\documentclass[draftclsnofoot]{IEEEtran}
\onecolumn
\usepackage{latexsym}
\usepackage{cite}
\usepackage{amssymb}
\usepackage{bbm}
\usepackage{bm}
\usepackage[usenames,dvipsnames]{color}
\usepackage{mathtools}
\usepackage{stmaryrd}
\usepackage{amsmath, amssymb,accents}
\usepackage{dsfont}
\usepackage{comment}
\usepackage[dvips]{graphics}
\DeclareMathOperator{\supp}{supp}
\usepackage{graphicx}
\usepackage{lipsum}
\usepackage{psfrag}
\usepackage{units}
\usepackage{setspace}
\usepackage{theorem}
\usepackage{float}
\usepackage{mathtools}
\allowdisplaybreaks

\newtheorem{definition}{Definition}

\newtheorem{theorem}{Theorem}
\newtheorem{lemma}{Lemma}
\newtheorem{remark}{Remark}

\newtheorem{corollary}{Corollary}
\newtheorem{proposition}{Proposition}

\newcommand\numberthis{\addtocounter{equation}{1}\tag{\theequation}}
{} %<- modify value to suit your needs
\begin{document}

\makeatletter
\newcommand{\vasti}{\bBigg@{3}}
\newcommand{\vast}{\bBigg@{4}}
\newcommand{\Vast}{\bBigg@{5}}
\makeatother
\newcommand{\be}{\begin{equation}}
\newcommand{\ee}{\end{equation}}
\newcommand{\ba}{\begin{align}}
\newcommand{\ea}{\end{align}}
\newcommand{\baa}{\begin{align*}}
\newcommand{\eaa}{\end{align*}}
\newcommand{\bea}{\begin{eqnarray}}
\newcommand{\eea}{\end{eqnarray}}
\newcommand{\beaa}{\begin{eqnarray*}}
\newcommand{\eeaa}{\end{eqnarray*}}
\newcommand{\p}[1]{\left(#1\right)}
\newcommand{\pp}[1]{\left[#1\right]}
\newcommand{\ppp}[1]{\left\{#1\right\}}
\newcommand{\ber}{$\ \mbox{Ber}$}
\newcommand{\mkv}{-\!\!\!\!\!\minuso\!\!\!\!\!-}
\newcommand{\fillater}[1] {{\Large \color{red} #1}}
\newcommand{\done}[1] {{\Large \color{blue} #1}}

\title{Wiretap Channels with Random States Non-Causally Available at the Encoder}

\author{Ziv Goldfeld, Paul Cuff and Haim H. Permuter

\thanks{%Manuscript received May 19, 2014; revised November 03, 2014; accepted November 26,
		%2016. Date of current version November 27, 2015. 
		The work of Z. Goldfeld and H. H. Permuter was supported by the Israel Science Foundation (grant no. 2012/14), the European Research Council under the European Union's Seventh Framework Programme (FP7/2007-2013) / ERC grant agreement n$^\circ$337752, and the Cyber Center and at Ben-Gurion University of the Negev. The work of P. Cuff was supported by the National Science Foundation (grant CCF-1350595) and the Air Force Office of Scientific Research (grant FA9550-15-1-0180).
		\newline This paper was presented in part at the 2016 IEEE International Conference on the Science of Electrical Engineers (ICSEE-2016), Eilat, Israel.
		\newline Z. Goldfeld and H. H. Permuter are with the Department of Electrical and Computer Engineering, Ben-Gurion University of the Negev, Beer-Sheva, Israel (gziv@post.bgu.ac.il, haimp@bgu.ac.il). Paul Cuff is with the Department of Electrical Engineering, Princeton University,
		Princeton, NJ 08544 USA (e-mail: cuff@princeton.edu).
		%\newline Copyright (c) 2016 IEEE. Personal use of this material is permitted. However, permission to use this material for any other purposes must be obtained from the IEEE by sending a request to pubs-permissions@ieee.org.
		}}
\maketitle

%%%%%%%%%%%%%%%%%%%%%%%%%%%%%%%%%%%%%%%%%%%%%%%%%%%%%%%%%%%%%%%%%%%%%%%%%%%%%%%%%%%%%%%%%%%%%%%%%%%%%%%%%%%%%%%%%%%
%%%%%%%%%%%%%%%%%%%%%%%%%%%%%%%%%%%%%%%%%%%%%%%%%%%%%%%%%%%%%%%%%%%%%%%%%%%%%%%%%%%%%%%%%%%%%%%%%%%%%%%%%%%%%%%%%%%
%%%%%%%%%%%%%%%%%%%%%%%%%%%%%%%%%%%%%%%%%%                         %%%%%%%%%%%%%%%%%%%%%%%%%%%%%%%%%%%%%%%%%%%%%%%%
%%%%%%%%%%%%%%%%%%%%%%%%%%%%%%%%%%%%%%%%%%         Abstract        %%%%%%%%%%%%%%%%%%%%%%%%%%%%%%%%%%%%%%%%%%%%%%%%
%%%%%%%%%%%%%%%%%%%%%%%%%%%%%%%%%%%%%%%%%%                         %%%%%%%%%%%%%%%%%%%%%%%%%%%%%%%%%%%%%%%%%%%%%%%%
%%%%%%%%%%%%%%%%%%%%%%%%%%%%%%%%%%%%%%%%%%%%%%%%%%%%%%%%%%%%%%%%%%%%%%%%%%%%%%%%%%%%%%%%%%%%%%%%%%%%%%%%%%%%%%%%%%%
%%%%%%%%%%%%%%%%%%%%%%%%%%%%%%%%%%%%%%%%%%%%%%%%%%%%%%%%%%%%%%%%%%%%%%%%%%%%%%%%%%%%%%%%%%%%%%%%%%%%%%%%%%%%%%%%%%%

\begin{abstract}

We study the state-dependent (SD) wiretap channel (WTC) with non-causal channel state information (CSI) at the encoder. This model subsumes all other instances of CSI availability as special cases, and calls for an efficient utilization of the state sequence for both reliability and security purposes. A lower bound on the secrecy-capacity, that improves upon the previously best known result published by Prabhakaran \emph{et al.}, is derived based on a novel superposition coding scheme. Our achievability gives rise to the exact secrecy-capacity characterization of a class of SD-WTCs that decompose into a product of two WTCs, where one is independent of the state and the other one depends only on the state. The results are derived under the strict semantic-security metric that requires negligible information leakage for all message distributions. %The proof of achievability relies on a stronger version of the soft-covering lemma for superposition codes. 
\end{abstract}


%%%%%%%%%%%%%%%%%%%%%%%%%%%%%%%%%%%%%%%%%%%%%%%%%%%%%%%%%%%%%%%%%%%%%%%%%%%%%%%%%%%%%%%%%%%%%%%%%%%%%%%%%%%%%%%%%%%
%%%%%%%%%%%%%%%%%%%%%%%%%%%%%%%%%%%%%%%%%%%%%%%%%%%%%%%%%%%%%%%%%%%%%%%%%%%%%%%%%%%%%%%%%%%%%%%%%%%%%%%%%%%%%%%%%%%
%%%%%%%%%%%%%%%%%%%%%%%%%%%%%%%%%%%%%%%%%%%%                         %%%%%%%%%%%%%%%%%%%%%%%%%%%%%%%%%%%%%%%%%%%%%%
%%%%%%%%%%%%%%%%%%%%%%%%%%%%%%%%%%%%%%%%%%%%       Introduction      %%%%%%%%%%%%%%%%%%%%%%%%%%%%%%%%%%%%%%%%%%%%%%
%%%%%%%%%%%%%%%%%%%%%%%%%%%%%%%%%%%%%%%%%%%%                         %%%%%%%%%%%%%%%%%%%%%%%%%%%%%%%%%%%%%%%%%%%%%%
%%%%%%%%%%%%%%%%%%%%%%%%%%%%%%%%%%%%%%%%%%%%%%%%%%%%%%%%%%%%%%%%%%%%%%%%%%%%%%%%%%%%%%%%%%%%%%%%%%%%%%%%%%%%%%%%%%%
%%%%%%%%%%%%%%%%%%%%%%%%%%%%%%%%%%%%%%%%%%%%%%%%%%%%%%%%%%%%%%%%%%%%%%%%%%%%%%%%%%%%%%%%%%%%%%%%%%%%%%%%%%%%%%%%%%%

\section{Introduction}\label{SEC:introduction}

\par Reliably transmitting a message over a noisy state-dependent (SD) channel with non-causal encoder channel state information (CSI) is a fundamental information-theoretic problem. Its formulation and the derivation of its capacity date back to Gelfand and Pinsker (GP) \cite{Gelfand_Pinsker}. A key virtue of the GP model is its generality. Namely, it is the most general instance of a SD point-to-point channel in which any or all of the terminals have non-causal access to CSI. Motivated by the above and by the importance of security in modern communication systems, we study the SD wiretap channel (WTC) with non-causal encoder CSI, which incorporates security in the presence of a wiretapper into the GP channel coding paradigm.

%The original motivation for the problem, as presented in \cite{Gelfand_Pinsker}, stems from the memory with stuck-at faults example \cite{Tsybakov_Memory_stuckat1974}. However, the implications of the result were much broader. One such prominent implication is that viewing the state sequence (known to the encoder) as a codeword of some other message naturally relates the GP scenario to the problem of broadcasting. It is therefore of no surprise that GP coding achieves the corner points of the best known inner bound on the capacity region of the broadcast channel \cite{Marton_BC1979}. 

% The capacity-achieving coding scheme from \cite{Gelfand_Pinsker} (subsequently referred to as GP coding) is based on binning an over-populating codebook and choosing from each bin a codeword that is correlated with the state-sequence. 

\par The study of secret communication over noisy channels was pioneered by Wyner, who introduced the degraded WTC and derived its secrecy-capacity \cite{Wyner_Wiretap1975}. Csisz{\'a}r and K{\"o}rner extended Wyner's result to the non-degraded WTC \cite{Csiszar_Korner_BCconfidential1978}. These two results formed the basis for the study of physical layer security and spawned a variety of works on related topics, among which are SD-WTCs. The interest in WTCs with random states relates to the observation that knowledge of the state sequence may be exploited as an additional source of randomness to boost secrecy performance. This oftentimes involves decorrelating the transmission and the state sequence so as to avoid leaking information that might compromise security. Reliable transmission over SD channels, on the other hand, favors coherent strategies that correlate the channel input and the state. Resolving the tension between these two different utilizations of the transmitter CSI is the main challenge in the considered communication scenario.

%Consequently, a key question is how to best exploit the state for secrecy purposes, while taking into account coding techniques designed for transmission over SD channels.

\par The first to consider a discrete and memoryless (DM) WTC with random states were Chen and Han Vinck \cite{SDWTC_Chen_HanVinck2006}, who studied the encoder CSI scenario. They established a lower bound on the secrecy-capacity based on a combination of wiretap coding with GP coding (see also \cite{SDWTC_2Sided_Liu2007} for the special case where the WTC is driven by a pair of states, one available to the encoder and the other one to the decoder). Their achievable rate, however, was shown to be suboptimal in general in a later work by Chia and El-Gamal \cite{SDWTC_Chia_ElGamal2012}. In that work, a coding scheme that uses both wiretap coding and secret key agreement\footnote{see also \cite{Khisti_Key_Agreement2011} for a related work focused solely on secret key agreement} was proposed for the scenario where the encoder has causal access to the state sequence, while the decoder has full CSI. Despite the restriction to use the state only in a causal manner, the authors of \cite{SDWTC_Chia_ElGamal2012} proved that their scheme can strictly outperform the adaptations of the non-causal schemes from \cite{SDWTC_Chen_HanVinck2006,SDWTC_2Sided_Liu2007} to the encoder and decoder CSI setup. Later related works include achievability results for the WTC with correlated sources \cite{Chen_WTC_Corr_Sources2014}, action-dependent SD-WTCs \cite{Han_Vinck_SDWTC_Actions2013} and WTCs with generalized feedback \cite{Bassi_WTC_GenFeed2015}. The benchmark result for the SD-WTC with non-causal encoder CSI considered here is the one derived by Prabhakaran \emph{et al.} \cite{Prabhakaran_SKSM2012}, via a two layered superposition coding scheme. As a consequence of the analysis in \cite{Prabhakaran_SKSM2012}, the inner layer of the superposition code therein is restricted to be independent of the state. 

%Other related directions of research include key-agreement over SD-WTCs by means of non-causal encoder CSI , and action-dependent SD-WTCs \cite{Han_Vinck_SDWTC_Actions2013}, where the encoder can affect the formation of the channel states by means of an action sequence (see also references therein).

%A more intricate coding scheme was proposed by Chia and El Gamal for the SD-WTC with causal encoder CSI and full decoder CSI \cite{SDWTC_Chia_ElGamal2012}. Their idea was to explicitly extract a cryptographic key from the random state, and protect a part of the confidential message via a one-time-pad with that key. The remaining portion of the confidential message is protected through a wiretap code (whenever wiretap coding is possible). 

%. This work was later generalized in \cite{SDWTC_2Sided_Liu2007} to a WTC that is driven by a pair of states, one available to the encoder and the other one to the decoder. However, as previously mentioned, since CSI at the encoder is the most general setup, the result of \cite{SDWTC_2Sided_Liu2007} is a special of \cite{SDWTC_Chen_HanVinck2006}.  

\par In this paper we propose a novel superposition-based coding scheme for the SD-WTC with non-causal encoder CSI, in which both layers are correlated with the state. The scheme results in a lower bound on the secrecy-capacity, which recovers the previously best known achievability formula from \cite{Prabhakaran_SKSM2012} (as well as all the preceding works) as a special case. The correlation between the inner layer of the superposition code and the state is fundamental as it allows our scheme to strictly outperform that of \cite{Prabhakaran_SKSM2012} for certain instances of the considered model. Our achievability formula also gives rise to some new secrecy-capacity results. In particular, we derive the semantic-security (SS) capacity of a class of SD-WTCs that decompose into a WTC that is independent of the state and another channel that generates two noisy versions of the state, each observed either by the legitimate receiver or by the eavesdropper.

We use an over-populated superposition codebook and encode the entire confidential message at the outer layer. The transmission is correlated with the state sequence by means of the likelihood encoder \cite{Cuff_Song_Likelihood2016}, while security is ensured by making the eavesdropper decode the inner layer codeword that contains no confidential information. Having done so, the eavesdropper is lacking the resources to extract any information about the secret message. Superposition-based code constructions for secrecy purposes have been considered before in the context of lossy source coding in \cite{Villard_Pinatanida_Secure_Source_Coding2010,Ramchandran_Superpos_Secrecy2013,Zaidi_Superpos_Secrecy_ISIT2016,Zaidi_Superpos_Secrecy_ITW2016}, where the eavesdropper was also allowed to decode a layer that contains no useful information

%by relation to the previous schemes can be strict, i.e., an example is fashioned where our scheme achieves strictly higher secrecy rates than \cite{SDWTC_Chen_HanVinck2006,SDWTC_2Sided_Liu2007}. The example is a specific instance of a class of SD-WTC whose channel transition probability decomposes into a WTC that is independent of the state and another channel that generates two noisy versions of the state, each observed either by the legitimate receiver or by the eavesdropper. We show that when the WTC's output to the eavesdropper is less noisy than the one observed by the legitimate user, our lower bound is tight - thus characterizing the secrecy-capacity.

%\par When specializing to the case where the decoder also knows the state sequence, our achievability is shown to be at least as good as the scheme from \cite{SDWTC_Chia_ElGamal2012}. In fact, \cite{SDWTC_Chia_ElGamal2012} provided two separate coding schemes and stated their achievability result as the maximum between the two. Recovering \cite{SDWTC_Chia_ElGamal2012} from our lower bound results in a compact and simplified (yet equivalent) characterization of their achievable formula. Thus, our superposition-based coding scheme encompasses a unification of the two schemes from \cite{SDWTC_Chia_ElGamal2012}. Interestingly, while both schemes from \cite{SDWTC_Chia_ElGamal2012} rely on generating the aforementioned cryptographic key, our code construction does not involve any explicit key generation/agreement phase. 

\par Our results are derived under the strict metric of SS. The SS criterion is a cryptographic benchmark that was adapted to suit the information-theoretic framework (of computationally unbounded adversaries) in \cite{Vardy_Semantic_WTC2012}. In that work, SS was shown to be equivalent to a negligible mutual information between the message and the eavesdropper's observations for all message distributions. In contrast to our stringent security requirement, all the aforementioned secrecy results were derived under the weak-secrecy metric, i.e., a vanishing \emph{normalized} mutual information with respect to a \emph{uniformly distributed} message. Nowadays, however, weak-secrecy is regarded as being insufficient, giving rise to the recent effort of upgrading information-theoretic secrecy results to strong-secrecy (by removing the normalization factor but keeping the uniformity assumption on the message). SS further strengthens both these; consequently, our achievability result outperforms the schemes from \cite{SDWTC_Chen_HanVinck2006,SDWTC_2Sided_Liu2007,Prabhakaran_SKSM2012} for the SD-WTC with non-causal encoder CSI, not only in terms of the achievable secrecy rate, but also in the upgraded sense of security it provides. %When CSI is also available at the decoder, our result implies that an upgrade to SS is possible, without inflicting any loss of rate compared to \cite{SDWTC_Chia_ElGamal2012}.

%\par While derivations of weak-secrecy largely rely on the groundwork laid by the early works Wyner \cite{Wyner_Wiretap1975} and Csisz{\'a}r and K{\"o}rner \cite{Csiszar_Korner_BCconfidential1978}, ensuring SS calls for stronger tools. In the spirit of our previous papers \cite{Goldfeld_WTCII_semantic2015} and \cite{Goldfeld_AVWTC_semantic2015}, the SS analysis relies on a stronger version of the soft-covering lemma (SCL) for superposition codebooks given in \cite[Corollary VII.8]{Cuff_Synthesis2013}. Namely, we show that a random superposition codebook achieves the soft-covering phenomenon with high probability. The probability of failure is doubly-exponentially small in the blocklength. The union bound combined with some additional distribution approximation arguments is then used to establish SS. Our code is also designed to produce an arbitrarily small \emph{maximal} error probability via the expurgation method (e.g., cf. \cite[Theorem 7.7.1]{Cover_Thomas}).

\par The remainder of this paper is organized as follows. Section \ref{SEC:preliminaries} provides notation and basic definitions and properties. In Section \ref{SEC:SDWTC} we describe the SD-WTC with non-causal encoder CSI and state the lower bound on its SS-capacity. Section \ref{SEC:Special_Cases} discusses our result and compares it to previous works, and also states some tight SS-capacity results. The proof of our main theorem is provided in Section \ref{SUBSUBSEC:SDWTC_lower_bound_proof}, while Section \ref{SEC:summary} summarizes the main achievements and insights of this work.

%%%%%%%%%%%%%%%%%%%%%%%%%%%%%%%%%%%%%%%%%%%%%%%%%%%%%%%%%%%%%%%%%%%%%%%%%%%%%%%%%%%%%%%%%%%%%%%%%%%%%%%%%%%%%%%%%%%
%%%%%%%%%%%%%%%%%%%%%%%%%%%%%%%%%%%%%%%%%%%%%%%%%%%%%%%%%%%%%%%%%%%%%%%%%%%%%%%%%%%%%%%%%%%%%%%%%%%%%%%%%%%%%%%%%%%
%%%%%%%%%%%%%%%%%%%%%%%%%%%%%%%%%%%%%%%%%%%%                         %%%%%%%%%%%%%%%%%%%%%%%%%%%%%%%%%%%%%%%%%%%%%%
%%%%%%%%%%%%%%%%%%%%%%%%%%%%%%%%%%%%%%%%%%%%      Preliminaries      %%%%%%%%%%%%%%%%%%%%%%%%%%%%%%%%%%%%%%%%%%%%%%
%%%%%%%%%%%%%%%%%%%%%%%%%%%%%%%%%%%%%%%%%%%%                         %%%%%%%%%%%%%%%%%%%%%%%%%%%%%%%%%%%%%%%%%%%%%%
%%%%%%%%%%%%%%%%%%%%%%%%%%%%%%%%%%%%%%%%%%%%%%%%%%%%%%%%%%%%%%%%%%%%%%%%%%%%%%%%%%%%%%%%%%%%%%%%%%%%%%%%%%%%%%%%%%%
%%%%%%%%%%%%%%%%%%%%%%%%%%%%%%%%%%%%%%%%%%%%%%%%%%%%%%%%%%%%%%%%%%%%%%%%%%%%%%%%%%%%%%%%%%%%%%%%%%%%%%%%%%%%%%%%%%%

\section{Notations and Preliminaries}\label{SEC:preliminaries}

\par In this paper, we use the following notations. As is customary, $\mathbb{N}$ is the set of natural numbers (which does not include 0), while  $\mathbb{R}$ denotes the reals. We further define $\mathbb{R}_+=\{x\in\mathbb{R}|x\geq 0\}$ and $\mathbb{R}_{++}=\{x\in\mathbb{R}|x> 0\}$. Given two real numbers $a,b$, we denote by $[a\mspace{-3mu}:\mspace{-3mu}b]$ the set of integers $\big\{n\in\mathbb{N}\big| \lceil a\rceil\leq n \leq\lfloor b \rfloor\big\}$. Calligraphic letters denote sets, e.g., $\mathcal{X}$, the complement of $\mathcal{X}$ is denoted by $\mathcal{X}^c$, while $|\mathcal{X}|$ stands for its cardinality. $\mathcal{X}^n$ denotes the $n$-fold Cartesian product of $\mathcal{X}$. An element of $\mathcal{X}^n$ is denoted by $x^n=(x_1,x_2,\ldots,x_n)$; whenever the dimension $n$ is clear from the context, vectors (or sequences) are denoted by boldface letters, e.g., $\mathbf{x}$. A substring of $\mathbf{x}\in\mathcal{X}^n$ is denoted by $x_i^j=(x_i,x_{i+1},\ldots,x_j)$, for $1\leq i\leq j \leq n$; when $i=1$, the subscript is omitted. We also define $x^{n\backslash i}=(x_1,\ldots,x_{i-1},x_{i+1},\ldots,x_n)$. Random variables are denoted by uppercase letters, e.g., $X$, with similar conventions for random vectors.

%For any $\mathcal{S}\subseteq[1:n]$, we use $\mathbf{x}^\mathcal{S}=(x_i)_{i\in\mathcal{S}}$ to denote the substring of $x^n$ defined by $\mathcal{S}$, with respect to the natural ordering of $\mathcal{S}$. For instance, if $\mathcal{S}=[i:j]$, where $1\leq i< j\leq n$, then $\mathbf{x}^\mathcal{S}=(x_i,x_{i+1},\ldots,x_j)$. 

Let $\big(\mathcal{X},\mathcal{F},\mathbb{P}\big)$ be a probability space, where $\mathcal{X}$ is the sample space, $\mathcal{F}$ is the $\sigma$-algebra and $\mathbb{P}$ is the probability measure. Random variables over $\big(\mathcal{X},\mathcal{F},\mathbb{P}\big)$ are denoted by uppercase letters, e.g., $X$, with conventions for random vectors similar to those for deterministic sequences. The probability of an event $\mathcal{A}\in\mathcal{F}$ is denoted by $\mathbb{P}(\mathcal{A})$, while $\mathbb{P}(\mathcal{A}\big|\mathcal{B}\mspace{2mu})$ denotes the conditional probability of $\mathcal{A}$ given $\mathcal{B}$. We use $\mathds{1}_\mathcal{A}$ to denote the indicator function of $\mathcal{A}$, while $p^{(U)}_\mathcal{A}$ denotes the uniform distribution over $\mathcal{A}$. The set of all probability mass functions (PMFs) on a finite set $\mathcal{X}$ is denoted by $\mathcal{P}(\mathcal{X})$, i.e., 
\begin{equation}
    \mathcal{P}(\mathcal{X})=\left\{P:\mathcal{X}\to[0,1]\Bigg| \sum_{x\in\mathcal{X}}P(x)=1]\right\}.
\end{equation}
In our notation for PMFs we oftentimes use subscripts to identify the involved random variable(s) and its possible conditioning. For example, for a discrete probability space $\big(\mathcal{X},\mathcal{F},\mathbb{P}\big)$ and two (correlated) random variables $X$ and $Y$ over that space, we use $p_X$, $p_{X,Y}$ and $p_{X|Y}$ to denote, respectively, the marginal PMF of $X$, the joint PMF of $(X,Y)$ and the conditional PMF of $X$ given $Y$. In particular, $p_{X|Y}$ represents the stochastic matrix whose elements are given by $p_{X|Y}(x|y)=\mathbb{P}\big(X=x|Y=y\big)$. Expressions such as $p_{X,Y}=p_Xp_{Y|X}$ are to be understood as $p_{X,Y}(x,y)=p_X(x)p_{Y|X}(y|x)$, for all $(x,y)\in\mathcal{X}\times\mathcal{Y}$. Accordingly, when three random variables $X$, $Y$ and $Z$ satisfy $p_{X|Y,Z}=p_{X|Y}$, they form a Markov chain, which we denote by $X\mkv Y\mkv Z$. We omit subscripts if the arguments of a PMF are lowercase versions of the random variables.% The support of a PMF $p$ and the expectation of a random variable $X$ are denoted by $\supp(P)$ and $\mathbb{E}\big[X\big]$, respectively. 

For a discrete measurable space $(\mathcal{X},\mathcal{F})$, a PMF $q\in\mathcal{P}(\mathcal{X})$ gives rise to a probability measure on $(\mathcal{X},\mathcal{F})$, which we denote by $\mathbb{P}_q$; accordingly, $\mathbb{P}_q\big(\mathcal{A})=\sum_{x\in\mathcal{A}}q(x)$ for every $\mathcal{A}\in\mathcal{F}$. We use $\mathbb{E}_q$ to denote an expectation taken with respect to $\mathbb{P}_q$. Similarly, we use $H_q$ and $I_q$ to indicate that an entropy or a mutual information term are calculated with respect to the PMF $q$. For a sequence of random variables $X^n$, if the entries of $X^n$ are drawn in an i.i.d. manner according to $p_X$, then for every $\mathbf{x}\in\mathcal{X}^n$ we have $p_{X^n}(\mathbf{x})=\prod_{i=1}^np_X(x_i)$ and we write $p_{X^n}(\mathbf{x})=p_X^n(\mathbf{x})$. Similarly, if for every $(\mathbf{x},\mathbf{y})\in\mathcal{X}^n\times\mathcal{Y}^n$ we have $p_{Y^n|X^n}(\mathbf{y}|\mathbf{x})=\prod_{i=1}^np_{Y|X}(y_i|x_i)$, then we write $p_{Y^n|X^n}(\mathbf{y}|\mathbf{x})=p_{Y|X}^n(\mathbf{y}|\mathbf{x})$. The conditional product PMF $p_{Y|X}^n$, given a specific sequence $\mathbf{x}\in\mathcal{X}^n$, is denoted by $p_{Y|X=\mathbf{x}}^n$.

The empirical PMF $\nu_{\mathbf{x}}$ of a sequence $\mathbf{x}\in\mathcal{X}^n$ is
\begin{equation}
	\nu_{\mathbf{x}}(x)\triangleq\frac{N(x|\mathbf{x})}{n},
\end{equation}
where $N(x|\mathbf{x})=\sum_{i=1}^n\mathds{1}_{\{x_i=x\}}$. We use $\mathcal{T}_\epsilon^{n}(p)$ to denote the set of letter-typical sequences of length $n$ with respect to the PMF $p\in\mathcal{P}(\mathcal{X})$ and the non-negative number $\epsilon$ \cite[Chapter 3]{Massey_Applied}, i.e., we have
\begin{equation}
	\mathcal{T}_\epsilon^{n}(p)=\Big\{\mathbf{x}\in\mathcal{X}^n\Big|\mspace{5mu}\big|\nu_{\mathbf{x}}(x)-p(x)\big|\leq\epsilon p(x),\ \forall x\in\mathcal{X}\Big\}.
\end{equation}

For a countable sample space $\Omega$ and $p,q\in\mathcal{P}(\Omega)$, the \emph{relative entropy} between $p$ and $q$ is
\begin{equation}
	\mathsf{D}(p||q)=\sum_{x\in\supp(p)}p(x)\log\left(\frac{p(x)}{q(x)}\right)\label{EQ:relative_entropy_def_discrete}
\end{equation}
and the \emph{total variation} between them is
\begin{equation}
	||p-q||_{\mathsf{TV}}=\frac{1}{2}\sum_{x\in\Omega}\big|p(x)-q(x)\big|.%=\sum_{\substack{x\in\Omega:\\p(x)>q(x)}}\big[p(x)-q(x)\big].\label{EQ:total_variation_def_discrete}
\end{equation}

Relative entropy dominates total variation through Pinsker's inequality \cite[Theorem 4.1]{Csiszar_Pinkser_Ineq1967}, which states that for any $p,q\in\mathcal{P}(\Omega)$
\begin{equation}
||p-q||_{\mathsf{TV}}\leq \sqrt{\frac{1}{2}\mathsf{D}(p||q)}.\label{EQ:Pinsker_Inequality}
\end{equation}
While no reverse Pinsker's inequality is known in general, a reverse asymptotic relation is sometimes valid (see \cite[Remark 1]{Goldfeld_BC_Cooperation_Secrecy2016}).

\begin{lemma}[Asymptotic Relation between Total Variation and Relative Entropy]\label{LEMMA:TV_divergence_relation}
Let $\Omega$ be a finite set and let
$\big\{p_n\big\}_{n\in\mathbb{N}}$ be a sequence of distributions with $p_n\in\mathcal{P}(\Omega^n)$. Let $q\in\mathcal{P}(\Omega)$ and assume $p_n\ll q^n$ for every $n\in\mathbb{N}$. Then\footnote{$f(n)\in O\big(g(n)\big)$ means that there exists $M>0$ such that $\big|f(n)\big|\leq M\big| g(n)\big|$, for any sufficiently large $n$.} 
\begin{equation}
\mathsf{D}(p_n||q^n)\in O\left(\left[n+\log\frac{1}{||p_n-q^n||_\mathsf{TV}}\right]||p_n-q^n||_\mathsf{TV}\right)\label{EQ:TV_relative_ent_relation}.
\end{equation}
\end{lemma}

\section{Wiretap Channels with Random States Non-Causally Available at the Encoder}\label{SEC:SDWTC}

We study the SD-WTC with non-causal encoder CSI, for which we establish a new and improved achievability formula that, in some cases, strictly outperforms the previously best known coding schemes for this scenario.

%The secrecy-capacity of a WTC with random states observed non-causally by some or all of the terminals is a highly challenging problem in information-theoretic security that have received noticeable attention throughout the years (see, e.g., \cite{Mitrpant_Gaussian_SDWTC2006,SDWTC_Chen_HanVinck2006,SDWTC_2Sided_Liu2007,SDWTC_Chia_ElGamal2012}). This interest in such secure communication scenarios stems from trying to understand how to optimally correlate the transmission with the state observation while exploiting the additional randomness offered by the knowledge of the state sequence to further enhance the secrecy rate. The optimal integration of these two ingredient is yet to be fully understood. 

%%%%%%%%%%%%%%%%%%%%%%%%%%%%%%%%%%%%%%%%%%%%%%%%%%%%%%%%%%%%%%%%%%%%%%%%%%%%%%%%%%%%%%%%%%%%%%%%%%%%%%%%%%%%%%%%%%%
%%%%%%%%%%%%%%%%%%%%%%%%          FIGURE: State-Dependent Wiretap Channel           %%%%%%%%%%%%%%%%%%%%%%%%%%%%%%%

\begin{figure}[!t]
	\begin{center}
	    \begin{psfrags}
	        \psfragscanon
	        \psfrag{A}[][][1]{\ \ \ $m$}
	        \psfrag{B}[][][0.9]{\ \ \ \ \ \ \ \ Encoder $f_n$}
	        \psfrag{D}[][][1]{\ \ \ \ \ \ \ \ \ \ \ \ }
	        \psfrag{C}[][][1]{\ \ \ $\mathbf{X}$}
	        \psfrag{E}[][][1]{\ \ \ \ \ \ \ \ \ \ $W^n_{Y,Z|X,S}$}
	        \psfrag{F}[][][1]{\ \ \ $\mathbf{Y}$}
	        \psfrag{G}[][][1]{\ \ \ $\mathbf{Z}$}
	        \psfrag{H}[][][0.9]{\ \ \ \ \ \ \ \ Decoder $\phi_n$}
	        \psfrag{I}[][][0.82]{\ \ \ \ \ \ \ \ \ Eavesdropper}
	        \psfrag{J}[][][1]{\ \ $\hat{m}$}
	        \psfrag{K}[][][1]{\ $m$}
	        \psfrag{L}[][][1]{\ \ \ \ $W^n_S$}
	        \psfrag{M}[][][1]{$\mathbf{S}$}
	        \includegraphics[scale = .37]{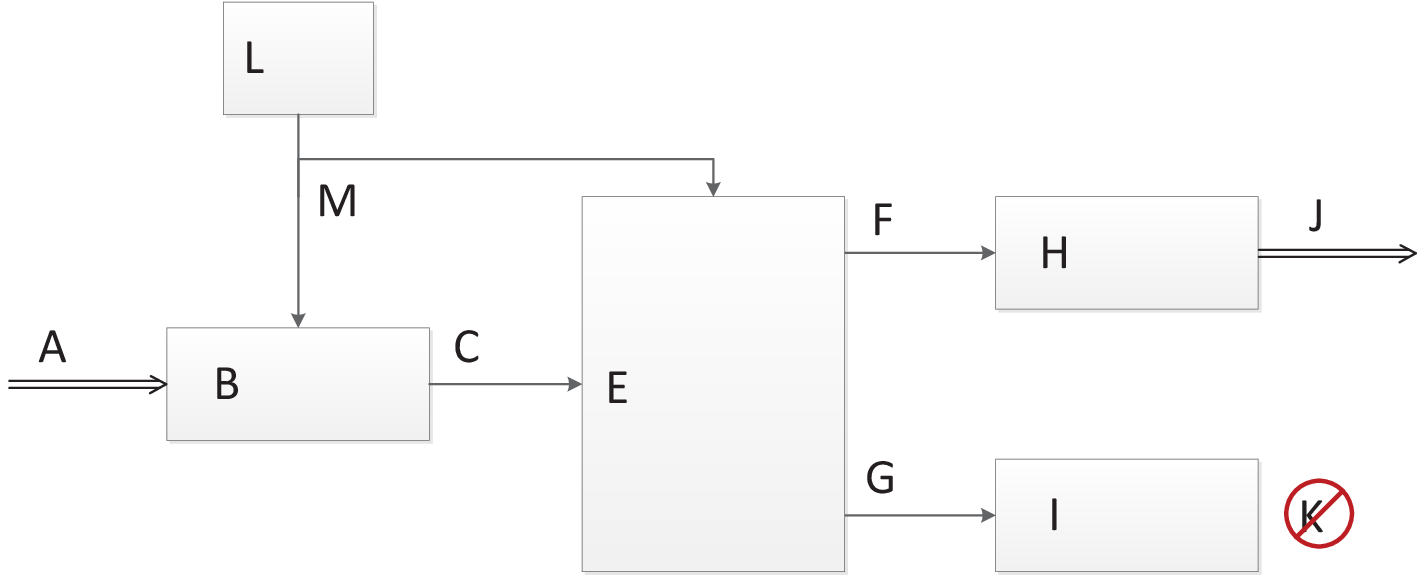}
	        \caption{The state-dependent wiretap channel with non-casual encoder channel state information.} \label{FIG:wiretap}
	        \psfragscanoff
	    \end{psfrags}
	\end{center}
\end{figure}
%%%%%%%%%%%%%%%%%%%%%%%%%%%%%%%%%%%%%%%%%%%%%%%%%%%%%%%%%%%%%%%%%%%%%%%%%%%%%%%%%%%%%%%%%%%%%%%%%%%%%%%%%%%%%

\subsection{Problem Setup}\label{SUBSEC:setup}

Let $\mathcal{S},\ \mathcal{X},\ \mathcal{Y}$ and $\mathcal{Z}$ be finite sets. The $\big(\mathcal{S},\mathcal{X},\mathcal{Y},\mathcal{Z},p_S,p_{Y,Z|X,S}\big)$ DMSD-WTC with non-causal encoder CSI is illustrated in Fig. \ref{FIG:wiretap}. A state sequence $\mathbf{s}\in\mathcal{S}^n$ is generated in an i.i.d. manner according to $p_S$ and is revealed in a non-causal fashion to the sender, who chooses a message $m$ from the set $\big[1:2^{nR}\big]$. The sender then maps the observed state sequence $\mathbf{s}$ and the chosen message $m$ onto a sequence $\mathbf{x}\in\mathcal{X}^n$ (the mapping may be random). The sequence $\mathbf{x}$ is transmitted over the DMSD-WTC with transition probability $p_{Y,Z|X,W}$. The output sequences $\mathbf{y}\in\mathcal{Y}^n$ and $\mathbf{z}\in\mathcal{Z}^n$ are observed by the receiver and the eavesdropper, respectively. Based on $\mathbf{y}$, the receiver produces an estimate $\hat{m}$ of $m$. The eavesdropper tries to glean whatever it can about the message from $\mathbf{z}$.

\begin{remark}[Most General Model]\label{REM:most_general_setting}
Before defining the setup and stating the result, we note that the considered model is the most general instance of a SD-WTC with non-causal CSI known at some or all of the terminals. The broadest model one may consider is when the SD-WTC $p_{\tilde{Y},\tilde{Z}|X,S_1,S_2,S_3}$ is driven by a triple of correlated state random variables $(S_1,S_2,S_3)\sim p_{S_1,S_2,S_3}$, where $S_1$ is known to the transmitter, $S_2$ is known to the receiver and $S_3$ is available at the eavesdropper's site. However, setting $S=S_1$, $Y=(\tilde{Y},S_2)$, $Z=(\tilde{Z},S_3)$ in a SD-WTC with non-causal encoder CSI and defining the channel's transition probability as
\begin{equation}
    p_{Y,Z|X,S}=p_{(\tilde{Y},S_2),(\tilde{Z},S_3)|X,S_1}=p_{S_2,S_3|S_1}p_{\tilde{Y},\tilde{Z}|X,S_1,S_2,S_3},
\end{equation}
one recovers this general SD-WTC from the model with non-causal encoder CSI only.
\end{remark}

\begin{definition}[Code]\label{DEF:SDWTC_code}
	An $(n,R)$-code $c_n$ for the SD-WTC with non-causal encoder CSI has a message set $\mathcal{M}_n\triangleq\big[1:2^{nR}\big]$, a stochastic encoder $f_n:\mathcal{M}_n\times\mathcal{S}^n\to\mathcal{P}(\mathcal{X}^n)$ and a decoder $\phi_n:\mathcal{Y}^n\to \hat{\mathcal{M}}_n$, where $\hat{\mathcal{M}}_n=\mathcal{M}_n\cup\{e\}$ and $e\notin\mathcal{M}_n$.
\end{definition}
	
For any message distribution $p_M\in\mathcal{P}(\mathcal{M}_n)$ and any $(n,R)$-code $c_n$, the induced joint PMF on $\mathcal{S}^n\times\mathcal{M}_n\times\mathcal{X}^n\times\mathcal{Y}^n\times\mathcal{Z}^n\times\hat{\mathcal{M}}_n$ is
\begin{equation}
	P^{(c_n)}(\mathbf{s},m,\mathbf{x},\mathbf{y},\mathbf{z},\hat{m})=p_S^n(\mathbf{s})p_M(m)f_n(\mathbf{x}|m,\mathbf{s})p^n_{Y,Z|X,S}(\mathbf{y},\mathbf{z}|\mathbf{x},\mathbf{s})\mathds{1}_{\big\{\hat{m}=\phi_n(\mathbf{y})\big\}}.\label{EQ:SDWTC_induced_PMF}
\end{equation}
The performance of $c_n$ is evaluated in terms of its rate $R$, the maximal decoding error probability and the SS-metric.

\begin{definition}[Maximal Error Probability]\label{DEF:SDWTC_error_probability} The maximal error probability of an $(n,R)$-code $c_n$ is
\begin{subequations}
	\begin{equation}
	    e(c_n)=\max_{m\in\mathcal{M}_n}e_m(c_n),\label{EQ:SDWTC_max_error_prob}
	\end{equation}
	where
	\begin{equation}
	    e_m(c_n)=\sum_{(\mathbf{s},\mathbf{x})\in\mathcal{S}^n\times\mathcal{X}^n}p^n_S(\mathbf{s})f_n(\mathbf{x}|m,\mathbf{s})\sum_{\substack{(\mathbf{y},\mathbf{z})\in\mathcal{Y}^n\times\mathcal{Z}^n:\\\phi_n(\mathbf{y})\neq m}}p_{Y,Z|X,S}^n(\mathbf{y},\mathbf{z}|\mathbf{x},\mathbf{s}).\label{EQ:SDWTC_message_error_prob}
    \end{equation}
\end{subequations}
\end{definition}

\begin{definition}[Information Leakage and SS Metric] The information leakage to the eavesdropper under the $(n,R)$-code $c_n$ and the message distribution $p_M\in\mathcal{P}(\mathcal{M}_n)$ is
\begin{equation}
\ell(p_M,c_n)=I_P(M;\mathbf{Z}),\label{EQ:SDWTC_info_leakage}
\end{equation}
where the subscript $P$ indicates that the underlying distribution is $P^{(c_n)}$ from \eqref{EQ:SDWTC_induced_PMF}. The SS metric with respect to $c_n$ is%\footnote{$\ell_\mathsf{Sem}(c_n)$ is actually the mutual-information-security (MIS) metric, which is equivalent to SS by \cite{Vardy_Semantic_WTC2012}. We use the representation in \eqref{EQ:SDWTC_SS_metric} rather than the formal definition of SS (see, e.g., \cite[Equation (4)]{Vardy_Semantic_WTC2012}) out of analytic convenience.}
\begin{equation}
\ell_\mathsf{Sem}(c_n)=\max_{p_M\in\mathcal{P}(\mathcal{M}_n)}\ell(p_M,c_n).\label{EQ:SDWTC_SS_metric}
\end{equation}
\end{definition}

%where the subscript $c_n$ denotes that the mutual information term is calculated with respect to the joint PMF of $M$ and $\mathbf{Z}$ induced by $c_n$, i.e., with respect to the marginal $P^{(c_n)}_{M,\mathbf{Z}}$ of \eqref{EQ:SDWTC_induced_PMF}. 

%\begin{remark}\label{REM:Message_PMF_function_of_code} SS requires that the code $c_n$ works well for all message PMFs. This means that the mutual information term in \eqref{EQ:SDWTC_SS_metric} is maximized over $p_M$ when the code $c_n$ is known. In other words, although not stated explicitly, $p_M$ may depend on $c_n$.
%\end{remark}

\begin{definition}[Achievability]\label{DEF:SDWTC_achievability}
A number $R\in\mathbb{R}_+$ is called an achievable SS-rate for the SD-WTC with non-causal encoder CSI if for every $\epsilon>0$ and sufficiently large $n$ there exists a CR $(n,R)$-code $c_n$ with
\begin{subequations}
\begin{align}
%\frac{1}{n}\log M_n&> R-\epsilon\label{EQ:SDWTC_achievability_rate}\\
e(c_n)&\leq\epsilon\label{EQ:SDWTC_achievability_reliability}\\
\ell_\mathsf{Sem}(c_n)&\leq\epsilon.\label{EQ:SDWTC_achievability_security}
\end{align}\label{EQ:SDWTC_achievability}
\end{subequations}
\end{definition}

\vspace{-10.5mm}

%\begin{remark}
%Our achievability proof shows that $\ell_\mathsf{Sem}(c_n)$ vanishes exponentially fast. This is a standard requirement in the cryptography community, commonly referred to as strong-SS (see, e.g., \cite[Section 3.2]{Vardy_Semantic_WTC2012}).
%\end{remark}

\begin{definition}[SS-Capacity]
The SS-capacity $C_{\mathsf{Sem}}$ of the SD-WTC with non-causal encoder CSI is the supremum of the set of achievable SS-rates.
\end{definition}

%\begin{remark}
%By Definition \ref{DEF:SS_codes}, for a sequence of WTC I codes to be semantically-secure, the SS metric from \eqref{EQ:WTCI_security_metric} must vanish exponentially fast. This is a standard requirement in the cryptography community, commonly referred to as strong-SS (see, e.g., \cite[Section 3.2]{Vardy_Semantic_WTC2012}). The coding scheme given in the direct proof of Theorem \ref{TM:WTCI_capacity_WTC} achieves this exponential decay of the SS-metric (see Section \ref{SUBSUBSEC:WTCI_proof_theorem1}). An exponential decay of the strong-secrecy metric was previously observed in \cite{Hayashi_Secrecy_Resolvability2006,Hayashi_Exponential_Resolvability2011,Hayashi_SS_BCConfidential2015}.
%\end{remark}
	
%%%%%%%%%%%%%%%%%%%%%%%%%%%%%%%%%%%%%%%%%%%%%%%%%%%%%%%%%%%%%%%%%%%%%%%%%%%%%%%%%%%%%%%%%%%%%%%%%%%%%%%%%%%%%%%%%%%%%
%%%%%%%%%%%%%%%%%                                                                                %%%%%%%%%%%%%%%%%%%%
%%%%%%%%%%%%%%%%%                                    MAIN RESULTS                                %%%%%%%%%%%%%%%%%%%%
%%%%%%%%%%%%%%%%%                                                                                %%%%%%%%%%%%%%%%%%%%
%%%%%%%%%%%%%%%%%%%%%%%%%%%%%%%%%%%%%%%%%%%%%%%%%%%%%%%%%%%%%%%%%%%%%%%%%%%%%%%%%%%%%%%%%%%%%%%%%%%%%%%%%%%%%%%%%%%%%

\subsection{Main Result}\label{SUBSEC:main_results}

%The main result of this work is a novel lower bound on the SS-capacity of the SD-WTC with non-causal encoder CSI. Our achievability formula is subsequently shown to strictly outperform the best previously known coding scheme for the considered scenario. It is also shown to be tight for certain instances of the SD-WTC of interest. 

The main result of this work is a novel lower bound on the SS-capacity of the SD-WTC with non-causal encoder CSI. To state it, let $\mathcal{U}$ and $\mathcal{V}$ be finite alphabets and for any $p_{U,V,X|S}:\mathcal{S}\to\mathcal{P}(\mathcal{U}\times\mathcal{V}\times\mathcal{X})$ define
\begin{equation}
R_\mathsf{A}\left(p_{U,V,X|S}\right)\triangleq\min\Big\{I(V;Y|U)-I(V;Z|U),I(U,V;Y)-I(U,V;S)\Big\},\label{EQ:SDWTC_lower_bound_prob}
\end{equation}
where the mutual information terms are calculated with respect to the joint distribution $p_Sp_{U,V,X|S}p_{Y,Z|X,S}$, i.e., such that $(U,V)\mkv (X,S)\mkv (Y,Z)$. 

\begin{theorem}[SD-WTC SS-Capacity Lower Bound]\label{TM:SDWTC_lower_bound}
The SS-capacity of the SD-WTC with non-causal encoder CSI is lower bounded by
\begin{equation}
C_{\mathsf{Sem}}\geq R_\mathsf{A}\triangleq\max_{\substack{p_{U,V,X|S}:\\I(U;Y)-I(U;S)\geq 0}}R_\mathsf{A}\left(p_{U,V,X|S}\right),\label{EQ:SDWTC_capacity_lower_bound}
\end{equation}\label{EQ:SDWTC_capacity_alt}
and one may restrict the cardinalities of $U$ and $V$ to $|\mathcal{U}|\leq |\mathcal{S}||\mathcal{X}|+5$ and $|\mathcal{V}|\leq |\mathcal{S}|^2|\mathcal{X}|^2+5|\mathcal{S}||\mathcal{X}|+3$.
\end{theorem}

The proof of Theorem \ref{TM:SDWTC_lower_bound} is given in Section \ref{SUBSUBSEC:SDWTC_lower_bound_proof} and is based on a superposition coding scheme for secrecy. The superposition codebook encodes the entire secret message in its \emph{outer layer}, carrying no confidential information in its inner layer. As explained in the following remark, the coding distribution is chosen so that the inner layer is better observable by the eavesdropper. This makes the eavesdropper `waste' channel resources on decoding it, leaving insufficient resources to extract information about the secret message. The outer codebook is designed to give a physical layer advantage to the legitimate parties, thus enabling wiretap coding through which the confidential message is protected. The transmission is correlated with the observed state sequence by means of the likelihood encoder \cite{Cuff_Song_Likelihood2016}. The SS analysis relies on the soft-covering for superposition codes (Lemma \ref{LEMMA:soft_covering}) and the expurgation technique (see, e.g., \cite[Theorem 7.7.1]{Cover_Thomas}).

\begin{remark}[Properties of Optimizing Distributions]
1) The underlying joint distribution in \eqref{EQ:SDWTC_alt_lower_bound_prob} is such that $(U,V)\mkv (X,S) \mkv(Y,Z)$ forms a Markov chain. However, since in all the mutual information terms from \eqref{EQ:SDWTC_alt_lower_bound_prob} the auxiliary random variable $V$ appears next to $U$ or conditioned on it, we may replace $V$ with $\tilde{V}=(U,V)$ without changing the region. Therefore, one may restrict the optimization domain of $C_{\mathsf{Sem}}$ to distributions with $U \mkv V \mkv (X,S)\mkv (Y,Z)$.\\
2) The allowed distributions $p_{U,V,X|S}$ in Theorem \ref{TM:SDWTC_lower_bound} are such that $I(U;Y)\geq I(U;S)$. We now argue that this restriction can be replaced with $I(U;Z)\geq I(U;Y)\geq I(U;S)$ without changing the region. Due to the Markov relation from 1), above, the second difference of mutual information terms can be expressed as $I(V;Y)-I(V;S)$. Rewriting the first bound as
\begin{equation}
    I(V;Y|U)-I(V;Z|U)=I(V;Y)-I(V;Z)+I(U;Z)-I(U;Y),
\end{equation}
we see that if $p_{U,V,X|S}$ is such that $I(U;Z)< I(U;Y)$, then taking $U=0$ achieves a higher rate.
\end{remark}

\begin{remark}[Interpretation of Achievable Rates]\label{REM:Alternative_interpretation}
To get some intuition on the structure of $R_\mathsf{A}$, notice that $I(V;Y|U)-I(V;Z|U)$ is the total rate of secrecy resources that are produced by the outer layer of the codebook. That is, the outer layer can achieve a secure communication rate of $I(V;Y|U)-\max\big\{I(V;Z|U),I(V;S|U)\big\}$, and it can produce a secret key at a rate of $\Big[I(V;S|U)-I(V;Z|U)\Big]^+$, where $[x]^+=\max\{0,x\}$. This is since some of the dummy bits needed to correlate the transmission with the state are secure for the same reason that a transmission is secure.

Also, the total amount of reliable (secured and unsecured) communication that this codebook allows is $I(U,V;Y)-I(U,V;S)$, including both the inner and outer layers. Therefore, one interpretation of our encoding scheme is that the secret key produced in the outer layer (if any) is applied to the non-secure communication in the inner layer.  In total, this achieves a secure communication rate that is the minimum of the total secrecy resources $I(V;Y|U)-I(V;Z|U)$ (i.e., secure communication and secret key) and the total communication rate $I(U,V;Y)-I(U,V;S)$, corresponding to the statement of $R_\mathsf{A}$. This effect happens naturally by the design of the superposition code, without explicitly extracting a key and applying a one-time pad.
\end{remark}

\begin{remark}[Cardinality Bounds]
The cardinality bounds on the auxiliary random variables $U$ and $V$ in Theorem \ref{TM:SDWTC_lower_bound} are established by standard application of the Eggleston-Fenchel-Carath{\'e}odory theorem \cite[Theorem 18]{Eggleston_Convexity1958} twice. The details are omitted.
\end{remark}

%%%%%%%%%%%%%%%%%%%%%%%%%%%%%%%%%%%%%%%%%%%%%%%%%%%%%%%%%%%%%%%%%%%%%%%%%%%%%%%%%%%%%%%%%%%%%%%%%%%%%%%%%%%%%%%%%%%%%
%%%%%%%%%%%%%%%%%                                                                                %%%%%%%%%%%%%%%%%%%%
%%%%%%%%%%%%%%%%%                                    MAIN RESULTS                                %%%%%%%%%%%%%%%%%%%%
%%%%%%%%%%%%%%%%%                                                                                %%%%%%%%%%%%%%%%%%%%
%%%%%%%%%%%%%%%%%%%%%%%%%%%%%%%%%%%%%%%%%%%%%%%%%%%%%%%%%%%%%%%%%%%%%%%%%%%%%%%%%%%%%%%%%%%%%%%%%%%%%%%%%%%%%%%%%%%%%

\subsection{Alternative Characterization of Achievable Result}\label{SUBSEC:Alternative_Achievability}
The achievable formula $R_\mathsf{A}$ can be restated in an alternative, yet equivalent, form. As before, let $\mathcal{U}$ and $\mathcal{V}$ be alphabets with cardinalities bounded as stated in Theorem \ref{TM:SDWTC_lower_bound}, and for any $p_{U,V,X|S}:\mathcal{S}\to\mathcal{P}(\mathcal{U}\times\mathcal{V}\times\mathcal{X})$ define
\begin{equation}
R_\mathsf{A}^\mathsf{Alt}\left(Q_{U,V,X|S}\right)\triangleq\min\left\{
\begin{aligned}
    I(V;Y|U)&-I(V;Z|U), \\
    I(U,V;Y)&-I(U,V;S), \\
    I(U,V;Y)&-I(U;S)-I(V;Z|U)
\end{aligned}
\right\},\label{EQ:SDWTC_alt_lower_bound_prob}
\end{equation}
where the mutual information terms are calculated with respect to the joint distribution $p_Sp_{U,V,X|S}p_{Y,Z|X,S}$. 

%that provides additional insight of our results. 
\begin{proposition}[Alternative Characterization of $R_\mathsf{A}$]\label{PROP:alternative_rate}
Setting
\begin{equation}
R_\mathsf{A}^\mathsf{Alt}\triangleq\max_{p_{U,V,X|S}}R_\mathsf{A}^{\mathsf{Alt}}\left(p_{U,V,X|S}\right),\label{EQ:SDWTC_capacity_alt_lower_bound}
\end{equation}
it holds that
\begin{equation}
    R_\mathsf{A}^\mathsf{Alt}= R_\mathsf{A}.\label{EQ:alternative_rate}
\end{equation}
\end{proposition}
The proof of Proposition \ref{PROP:alternative_rate} is relegated to Appendix \ref{APPEN:alternative_achievability_proof}. The challenge in the proof is showing that $R_\mathsf{A}^\mathsf{Alt}\leq R_\mathsf{A}$. This is since in $R_\mathsf{A}^\mathsf{Alt}$ the legitimate user may not be able to reliably decode the (inner) $U$ layer of the superposition codebook by itself. Decoding the $U$ layer in $R_\mathsf{A}^\mathsf{Alt}$ is possible, in general, with the assistance of the (outer) $V$ layer. This is evident from the second and third rate bounds in $R_\mathsf{A}^\mathsf{Alt}$, from which it is seen that even if an input distribution $p_{U,V,X|S}$ induces $I(U;Y)<I(U;S)$, it still might result in a positive achievable rate. In contrast, $R_\mathsf{A}$ only allows input distributions with $I(U;Y)\geq I(U;S)$, i.e., distributions that make $U$ decodable on its own by the legitimate user. Nonetheless, as the proof in Appendix \ref{APPEN:alternative_achievability_proof} shows that $R_\mathsf{A}=R_\mathsf{A}^\mathsf{Alt}$, it implies, in particular, that an optimal input distribution in $R_\mathsf{A}^\mathsf{Alt}$ always satisfies $I(U;Y)\geq I(U;S)$.

\section{Special Cases and Examples}\label{SEC:Special_Cases}

%%%%%%%%%%%%%%%%%%%%%%%%%%%%%%%%%%%%%%%%%%%%%%%%%%%%%%%%%%%%%%%%%%%%%%%%%%%%%%%%%%%%%%%%%%%%%%%%%%%%%%%%%%%%%%%%%%%
%%%%%%%%%%%%%%%%%%%%%%%%%                                                         %%%%%%%%%%%%%%%%%%%%%%%%%%%%%%%%%
%%%%%%%%%%%%%%%%%%%%%%%%%               COMPARISON TO CHIA & EL-GAMAL             %%%%%%%%%%%%%%%%%%%%%%%%%%%%%%%%%
%%%%%%%%%%%%%%%%%%%%%%%%%                                                         %%%%%%%%%%%%%%%%%%%%%%%%%%%%%%%%%
%%%%%%%%%%%%%%%%%%%%%%%%%%%%%%%%%%%%%%%%%%%%%%%%%%%%%%%%%%%%%%%%%%%%%%%%%%%%%%%%%%%%%%%%%%%%%%%%%%%%%%%%%%%%%%%%%%%

\subsection{Comparison to Previous Benchmark}\label{SUBSEC:comparison_prabhakaran}

The result of Theorem \ref{TM:SDWTC_lower_bound} recovers the previously best known achievable secrecy rate over the SD-WTC with non-causal encoder CSI by Prabhakarn \emph{et al.} in \cite{Prabhakaran_SKSM2012}. Theorem 2 of \cite{Prabhakaran_SKSM2012} established a tradeoff region between achievable secret message and secret key rate pairs. Specializing the result from \cite[Theorem 2]{Prabhakaran_SKSM2012} to the secret message only scenario (by nullifying the secret key rate) shows the achievability of 
\begin{subequations}
\begin{equation}
    R_\mathsf{PER}\triangleq\max_{p_U p_{V,X|U,S}}R_\mathsf{PER}\left(p_U p_{V,X|U,S}\right),\label{EQ:compare_prabhakaran_LB}
\end{equation}
where, for any $p_U\in\mathcal{P}(\mathcal{U})$ and $p_{V,X|U,S}:\mathcal{U}\times\mathcal{S}\to\mathcal{P}(\mathcal{V}\times\mathcal{X})$,
\begin{equation}
    R_\mathsf{PER}\left(p_U p_{V,X|U,S}\right)\triangleq\Big\{I(U,V;Y)-I(U,V;S),I(V;Y|U)-I(V;Z|U)\Big\},\label{EQ:compaer_PBK_final}
\end{equation}
\end{subequations}
and the mutual information terms are taken with respect to $p_Sp_Up_{V,X|U,S}p_{Y,Z|S,X}$, i.e., such $U$ and $S$ are independent and $(U,V) \mkv (S,X) \mkv (Y,Z)$ forms a Markov chain.

The difference between $R_\mathsf{PER}$ and $R_\mathsf{A}$ from Theorem \ref{TM:SDWTC_lower_bound} is that the former requires $U$ to be independent of $S$, while our formula allows correlation between $U$ and $S$ as long as $I(U;Y)\geq I(U;S)$. The independence of $U$ and $S$ in $R_\mathds{PER}$ essentially means that no GP coding is supported in the inner layer of the superposition code. Our scheme, on the other hand, supports GP coding in the inner layer as long as it is decodable by the legitimate receiver.

To compare our result to that of \cite[Theorem 2]{Prabhakaran_SKSM2012}, first note that Theorem \ref{TM:SDWTC_lower_bound} recovers $R_\mathsf{PER}$ by restricting $U$ to be independent of $S$ in $R_\mathsf{A}$. This choice of statistics is valid as it satisfies $I(U;S)=0$. Furthermore, there are instances of SD-WTC with non-causal encoder CSI for which $R_\mathsf{A}$ is strictly larger than $R_\mathsf{PER}$. In concurrent work with collaborators \cite{Goldfeld_Bunin_SNSK2017}, we construct a particular example of such a channel (see Section V-A therein). The main idea there is to consider a channel for which GP coding is necessary in order to attain capacity (e.g., the Memory with Stuck-at-Faults channel) and to force communication to happen in the inner layer of the codebook (by considering a strong eavesdropper). As the scheme from \cite{Prabhakaran_SKSM2012} does not allow inner layer GP coding, it turns out to be strictly below capacity. Our scheme, on the other hand, is optimal for the considered setup. This establishes the sub-optimality of $R_\mathsf{PER}$ and illuminates the fundamental role of the correlation between $U$ and $S$ for secure transmission over SD-WTCs with non-causal encoder CSI. 
	
Lastly, we note that the result from \cite{Prabhakaran_SKSM2012} was derived under the weak secrecy metric.\footnote{Weak secrecy refers to a vanishing \emph{normalized} mutual information $\frac{1}{n}I(M;\mathbf{Z})$ between a uniformly distributed confidential message and the eavesdropper's observation sequence.}. As our achievability ensures SS, Theorem \ref{TM:SDWTC_lower_bound} improves upon \cite[Theorem 2]{Prabhakaran_SKSM2012}, not only in the rate it achieves, but also in the sense of security it guarantees.

\begin{remark}[WTC with Correlated Sources]
Another related setup is that of the WTC with correlated sources \cite{Chen_WTC_Corr_Sources2014}, where the WTC $p_{Y,Z|X}$ is not SD and two correlated source sequences $(\mathbf{S},\mathbf{S}_1)\sim p_{S,S_1}^n$ are observed non-causally by the encoder and the legitimate receiver, respectively. In \cite{Chen_WTC_Corr_Sources2014},
\begin{equation}
\max_{\substack{p_{W|S}p_{T,X}:\\I(T;Y)\geq I(W;S)}}\Big\{
\big|I(T;Y)-\max\big\{I(T;Z),I(W;S)\big\}\big|+I(W;S_1)\Big\},\label{EQ:corr_sources_LB}
\end{equation}
where the joint distribution is 
$p_{S,S_1}p_{S|S}p_{T,X}p_{Y,Z|X}$, was established as a lower bound on the weak-secrecy capacity of that model. Setting $U=0$ and $V=(T,W)$ into $    R_\mathsf{PER}\left(p_U p_{V,X|U,S}\right)$ from \eqref{EQ:compaer_PBK_final} and maximizing over $p_{W|S}p_{T,X}$ recovers \eqref{EQ:corr_sources_LB}. As $R_\mathsf{A}$ from Theorem \ref{TM:SDWTC_lower_bound} captures $R_\mathsf{PER}$ as a special case, our result also subsumes that of \cite{Chen_WTC_Corr_Sources2014}
.
\end{remark}

%%%%%%%%%%%%%%%%%%%%%%%%%%%%%%%%%%%%%%%%%%%%%%%%%%%%%%%%%%%%%%%%%%%%%%%%%%%%%%%%%%%%%%%%%%%%%%%%%%%%%%%%%%%%%%%%%%%
%%%%%%%%%%%%%%%%%%%%%%%%%                                                         %%%%%%%%%%%%%%%%%%%%%%%%%%%%%%%%%
%%%%%%%%%%%%%%%%%%%%%%%%%                       EARLIER WORKS                     %%%%%%%%%%%%%%%%%%%%%%%%%%%%%%%%%
%%%%%%%%%%%%%%%%%%%%%%%%%                                                         %%%%%%%%%%%%%%%%%%%%%%%%%%%%%%%%%
%%%%%%%%%%%%%%%%%%%%%%%%%%%%%%%%%%%%%%%%%%%%%%%%%%%%%%%%%%%%%%%%%%%%%%%%%%%%%%%%%%%%%%%%%%%%%%%%%%%%%%%%%%%%%%%%%%%

\subsection{An Earlier Benchmark by Chen and Han Vinck}\label{SUBSEC:comparison_Han_Vicnk}

The benchmark result for the SD-WTC with non-causal encoder CSI prior to that from \cite{Prabhakaran_SKSM2012} is due to Chen and Han Vinck \cite{SDWTC_Chen_HanVinck2006}. Theorem 2 of \cite{SDWTC_Chen_HanVinck2006} shows that the weak-secrecy capacity of the considered SD-WTC is lower bounded by
\begin{subequations}
\begin{equation}
R_\mathsf{CHV}\triangleq\max_{p_{V,X|S}}R_\mathsf{CHV}\left(p_{V,X|S}\right),\label{EQ:compare_chen_LB}
\end{equation}
where for any $p_{V,X|S}:\mathcal{S}\to\mathcal{P}(\mathcal{V}\times\mathcal{X})$,
\begin{equation}
R_\mathsf{CHV}\left(p_{V,X|S}\right)\triangleq\min\Big\{I(V;Y)-I(V;Z),I(V;Y)-I(V;S)\Big\},\label{EQ:compare_chen_prob}
\end{equation}\label{EQ:compare_chen}%
\end{subequations}
and the mutual information terms are taken with respect to $p_Sp_{V,X|S}p_{Y,Z|X,S}$, i.e., such that $V\mkv(X,S)\mkv(Y,Z)$ forms a Markov Chain.

The code construction that achieves $R_\mathsf{CHV}$ combines GP coding and wiretap coding. Namely, a single-layered codebook is employed, in which the bins are large enough to simultaneously facilitate correlating the transmission with the state and confusing the eavesdropper. This construction is evident from the structure of the achievability formula by rewriting $R_\mathsf{CHV}\left(p_{V,X|S}\right)$ as 
\begin{equation}
R_\mathsf{CHV}\left(p_{V,X|S}\right)=I(V;Y)-\max\Big\{I(V;Z),I(V;S)\Big\}.
\end{equation}

This result was generalized in \cite[Theorem 1]{SDWTC_2Sided_Liu2007} to the case where the SD-WTC is governed by a pair of pairwise i.i.d. state sequences $(\mathbf{S},\mathbf{S}_1)$ with distribution $p^n_{S,S_1}$ (i.e., the SD-WTC's transition matrix is $p_{\tilde{Y},Z|X,S,S_1}$), the encoder is assumed to have non-causal access to $\mathbf{S}$, while the legitimate receiver has $\mathbf{S}_1$. However, as explained in Remark \ref{REM:most_general_setting}, this instance is a special case of the channel from \cite{SDWTC_Chen_HanVinck2006}, obtained by taking $Y=(\tilde{Y},S_1)$ and setting $p_{Y,Z|X,S}=p_{(\tilde{Y},S_1),Z|X,S}=p_{S_1|S}p_{\tilde{Y},Z|X,S,S_1}$. The achievability of $R_\mathsf{CHV}$ is recovered from Theorem \ref{TM:SDWTC_lower_bound} (and from $R_\mathsf{PER}$) by setting $U=0$. 

\begin{remark}[Sub Optimality of \cite{SDWTC_Chen_HanVinck2006}]
In \cite{SDWTC_Chia_ElGamal2012}, Chia and El Gamal showed that the Chen and Han Vinck result is sub-optimal in general for the considered SD-WTC. Specifically, \cite{SDWTC_Chia_ElGamal2012} considered a SD-WTC with causal encoder CSI and full decoder CSI. The coding scheme proposed in \cite{SDWTC_Chia_ElGamal2012} uses the state sequence to generate a cryptographic key, which is then used to one-time pad a part of the confidential message. The other part of the message is protected via a wiretap code (whenever wiretap coding is possible). Despite being restricted to exploit the state only in a causal manner, the aforementioned strategy was shown to achieve strictly higher rates than the one from \cite{SDWTC_Chen_HanVinck2006} for certain classes of SD-WTCs. 
\end{remark}

%%%%%%%%%%%%%%%%%%%%%%%%%%%%%%%%%%%%%%%%%%%%%%%%%%%%%%%%%%%%%%%%%%%%%%%%%%%%%%%%%%%%%%%%%%%%%%%%%%%%%%%%%%%%%%%%%%%
%%%%%%%%%%%%%%%%%%%%%%%%%                                                         %%%%%%%%%%%%%%%%%%%%%%%%%%%%%%%%%
%%%%%%%%%%%%%%%%%%%%%%%%%                        TIGHT RESULTS                    %%%%%%%%%%%%%%%%%%%%%%%%%%%%%%%%%
%%%%%%%%%%%%%%%%%%%%%%%%%                                                         %%%%%%%%%%%%%%%%%%%%%%%%%%%%%%%%%
%%%%%%%%%%%%%%%%%%%%%%%%%%%%%%%%%%%%%%%%%%%%%%%%%%%%%%%%%%%%%%%%%%%%%%%%%%%%%%%%%%%%%%%%%%%%%%%%%%%%%%%%%%%%%%%%%%%

\subsection{Tight SS-Capacity Results}\label{SUBSEC:tight_results}

\subsubsection{Reversely Less Noisy SD-WTC with Full Encoder and Noisy Decoder and Eavesdropper CSI}\label{SUBSUBSEC:LNWTC} Let $\mathcal{S}_1$ and $\mathcal{S}_2$ be finite sets and consider a SD-WTC $p_{\tilde{Y},\tilde{Z}|X,S}$ with non-causal encoder CSI, where $\tilde{Y}=(Y,S_1)$, $\tilde{Z}=(Z,S_2)$ and $p_{S_1,S_2,Y,Z|X,S}=p_{S_1,S_2|S}p_{Y,Z|X}$. Namely, the transition probability $p_{S_1,S_2,Y,Z|X,S}$ decomposes into a product of two WTCs, one being independent of the state, while the other one depends only on it. The legitimate receiver (respectively, the eavesdropper) observes not only the output $\mathbf{Y}$ (respectively, $\mathbf{Z}$) of the WTC $p^n_{Y,Z|X}$, but also $\mathbf{S}_1$ (respectively, $\mathbf{S}_2$) - a noisy version of the state sequence drawn according to the marginal of $p^n_{S_1,S_2|S}$. We characterize the SS-capacity of this setting when the WTC $p_{Y,Z|X}$ is reversely less noisy, i.e., when $I(U;Y)\leq I(U;Z)$, for every random variable $U$ with $U\mkv X\mkv (Y,Z)$. After the submission of this paper, the authors became aware of an independent derivation of this result under an average error probability and the weak-secrecy metric \cite{Bassi_WTC_GenFeed2015} as the performance criteria. In that work an achievable rate region based on secret key agreement was derived for the WTC with generalized feedback. Although being quite different from the setup considered herein, both problems capture the less noisy SD-WTC as a special case (in fact, this is also true for the slightly more general setup of the WTC with correlated sources \cite{Chen_WTC_Corr_Sources2014}). Each of the two achievability results (our region from Theorem \ref{TM:SDWTC_lower_bound} and the one from \cite[Theorem 1]{Bassi_WTC_GenFeed2015}) is tight for this instance.

To state the SS-capacity result, let $\mathcal{A}$ and $\mathcal{B}$ be finite sets and for any $p_X\in\mathcal{P}(\mathcal{X})$, $p_{A|S}:\mathcal{S}\to\mathcal{P}(\mathcal{A})$ and $p_{B|A}:\mathcal{A}\to\mathcal{P}(\mathcal{B})$ define
\begin{equation}
    R_\mathsf{RLN}\left(p_X,p_{A|S},p_{B|A}\right)=\min\Big\{I(A;S_1|B)-I(A;S_2|B),I(X;Y)-I(A;S|S_1)\Big\},\label{EQ:LNWTC_capacity_prob}
\end{equation}
where the mutual information terms are calculated with respect to the joint PMF $p_Sp_{A|S}p_{B|A}p_Xp_{S_1,S_2|S}p_{Y,Z|X}$, i.e., where $(X,Y,Z)$ is independent of $(S,S_1,S_2,A,B)$ and $A\mkv S\mkv (S_1,S_2)$ and $B\mkv A\mkv (S,S_1,S_2)$ form Markov chains (as well as the Markov relations implied by the channels).

\begin{corollary}[Reversely Less Noisy SD-WTC SS-Capacity]\label{CORR:LNWTC_capacity}
The SS-capacity of the reversely less noisy WTC with full encoder and noisy decoder and eavesdropper CSI is
\begin{equation}
    C_\mathsf{RLN}=\max_{p_X,p_{A|S},p_{B|A}}R_\mathsf{RLN}\left(p_X,p_{A|S},p_{B|A}\right).\label{EQ:LNWTC_capacity}
\end{equation}
\end{corollary}
A proof of Corollary \ref{CORR:LNWTC_capacity}, where the direct part is established based on Theorem \ref{TM:SDWTC_lower_bound}, is given in Appendix \ref{APPEN:LNWTC_capacity_proof}. Instead, one can derive an explicit achievability for \eqref{EQ:LNWTC_capacity} via a coding scheme based on a key agreement protocol through multiple blocks and one-time pad operations. To gain some intuition, an outline of the scheme for the simplified case where $S_2=0$ is described in the following remark. This scenario is fitting for intuitive purposes, since the absence of correlated observations with $S$ at the eavesdropper's site allows one to design a secure transmission strategy over a single block. Notwithstanding, a single-block-based coding scheme is feasible, even when $S_2$ is not a constant, via the superposition code construction given in the proof of Theorem \ref{TM:SDWTC_lower_bound}.

\begin{remark}[Explicit Achievability for Corollary \ref{CORR:LNWTC_capacity}]\label{REM:LNWTC_explicit_code}
It is readily verified that when $S_2=0$, setting $B=0$ in \eqref{EQ:LNWTC_capacity} is optimal. The resulting secrecy rate $\tilde{R}_\mathsf{RLN}\left(p_X,p_{A|S}\right)\triangleq\min\Big\{I(A;S_1),I(X;Y)-I(A;S|S_1)\Big\}$, for any fixed $p_X$ and $p_{A|S}$ as before, is achieved as follows: \footnote{A reminiscent coding scheme was employed in \cite{Khisti_Key_Generation2012} for the purpose of key generation (rather than the transmission of a confidential message) over the SD-WTC with non-causal encoder CSI. 
}
\begin{enumerate}
    \item Generate $2^{nR_A}$ $a$-codewords as i.i.d. samples from $p_A^n$.
    
    \item Partition the set of all $a$-codewords into $2^{nR_\mathsf{Bin}}$ equal sized bins. Accordingly, label each $a$-codeword as $\mathbf{a}(b,k)$, where $b\in\big[1:2^{nR_\mathsf{Bin}}\big]$ and $k\in\big[1:2^{n(R_A-R_\mathsf{Bin})}\big]$. 
    
    \item Generate a point-to-point codebook that comprises $2^{n(R+R_\mathsf{Bin})}$ codewords $\mathbf{x}(m,b)$, where $m\in\mathcal{M}_n$ and $b\in\big[1:2^{nR_\mathsf{Bin}}\big]$, drawn according to $p_X^n$.
    
    \item Upon observing the state sequence $\mathbf{s}\in\mathcal{S}^n$, the encoder searches the entire $a$-codebook for an $a$-codeword that is jointly-typical with $\mathbf{s}$, with respect to their joint PMF $p_Sp_{A|S}$. Such a codeword is found with high probability, provided that
    \begin{equation}
        R_A>I(A;S).
    \end{equation}
    Let $(b,k)\in\big[1:2^{nR_\mathsf{Bin}}\big]\times\big[1:2^{n(R_A-R_\mathsf{Bin})}\big]$ be the indices of the selected $a$-codeword. To sent the message $m\in\mathcal{M}_n$, the encoder one-time-pads $m$ with $k$ to get $\tilde{m}=m\oplus k\in\mathcal{M}_n$, and transmits $\mathbf{x}(\tilde{m},b)$ over the WTC. The one-time pad operation introduces the rate bound
    \begin{equation}
        R\leq R_A-R_\mathsf{Bin}.
    \end{equation}
    
    \item The legitimate receiver first decodes the $x$-codeword using its channel observation $\mathbf{y}$. Reliable decoding requires the total number of $x$-codewords to be less than the capacity of the sub-channel $p_{Y|X}$, i.e., 
    \begin{equation}
        R+R_\mathsf{Bin}<I(X;Y).
    \end{equation}
    Denoting the decoded indices by $(\hat{\tilde{m}},\hat{b})\in\mathcal{M}_n\times\big[1:2^{nR_\mathsf{Bin}}\big]$, the decoder then uses the noisy state observation $\mathbf{s}_1\in\mathcal{S}_1^n$ to isolate the exact $a$-codeword from the $\hat{b}$-th bin. Namely, it searches for a unique index $\hat{k}\in\big[1:2^{n(R_A-R_\mathsf{Bin})}\big]$, such that $\big(\mathbf{a}(\hat{b},\hat{k}),\mathbf{s}_1\big)$ are jointly-typical with respect to $p_{A,S_1}$ (the marginal of $p_Sp_{S_1|S}p_{A|S}$). The probability of error in doing so is arbitrarily small with the blocklength, provided that
    \begin{equation}
        R_A-R_\mathsf{Bin}<I(A;S_1).
    \end{equation}
    Having decoded $(\hat{\tilde{m}},\hat{b})$ and $\hat{k}$, the decoder declares $\hat{m}\triangleq \hat{\tilde{m}}\oplus\hat{k}$ as the decoded message.
    
    \item For the eavesdropper, note that although it has the correct $(\tilde{m},b)$ (due to the less noisy condition), it cannot decode $k$ since it has no observation that is correlated with $\mathbf{A}$, $\mathbf{S}$ and $\mathbf{S}_1$. Security of the protocol is, therefore, implied by the security of the one-time pad.
    
    \item Putting the aforementioned rate bounds together establishes the achievability of $\tilde{R}_\mathsf{RLN}\left(p_X,p_{A|S}\right)$.
\end{enumerate}
\end{remark}

\subsubsection{Semi-Deterministic SD-WTC with Non-Causal Encoder CSI}

Another observation is that $R_\mathsf{A}$ from Theorem \ref{TM:SDWTC_lower_bound} is tight when the main channel is deterministic, i.e., when $p_{Y,Z|X,S}=\mathds{1}_{\big\{Y=y(X,S)\big\}}p_{Z|X,S}$, for some function $y:\mathcal{S}\times\mathcal{X}\to\mathcal{Y}$. In fact, the achievability results from \cite{SDWTC_Chen_HanVinck2006,SDWTC_2Sided_Liu2007} are sufficient for achieving optimality in this case. We state this secrecy-capacity result merely because, to the best of our knowledge, it was not explicitly stated before. 

\begin{corollary}[Semi-Deterministic SD-WTC with Non-Causal Encoder CSI - SS-Capacity]\label{CORR:Semi_SDWTC_capacity}
The SS-capacity of the semi-deterministic SD-WTC with non-causal encoder CSI is
\begin{equation}
    C_\mathsf{Semi-Det}=\max_{p_{X|S}}\min\Big\{H(Y|Z),H(Y|S)\Big\},\label{EQ:Semi_SDWTC_capacity}
\end{equation}
where the entropy terms are calculated with respect to $p_Sp_{X|S}\mathds{1}_{\big\{Y=y(X,S)\big\}}p_{Z|X,S}$.
\end{corollary}
The achievability of $C_\mathsf{Semi-Det}$ follows by setting $U=0$ and $V=Y$ (which is a valid choice due to the deterministic nature of the main channel) in Theorem \ref{TM:SDWTC_lower_bound}. The converse is established by standard techniques - see Appendix \ref{APPEN:Semi_SDWTC_capacity_proof}.

Note that the SS-capacity is unaffected by whether or not the eavesdropper's channel is deterministic. Letting $Z=z(X,S)$, for some $z:\mathcal{S}\times\mathcal{X}\to\mathcal{Z}$ does not changes the result of Corollary \ref{CORR:Semi_SDWTC_capacity}.

\section{Proof of Theorem \ref{TM:SDWTC_lower_bound}}\label{SUBSUBSEC:SDWTC_lower_bound_proof}
Fix an $\epsilon>0$ and a conditional PMF $p_{U,V,X|S}:\mathcal{S}\to\mathcal{P}(\mathcal{U}\times\mathcal{V}\times\mathcal{X})$, which induces a joint single-letter distribution
\begin{equation}
    p\triangleq p_Sp_{U,V,X|S}p_{Y,Z|X,S},\label{EQ:single_letter_joint}
\end{equation}
such that $I(U;Y)-I(U;S)\geq 0$. Assume that $R<R_\mathsf{A}\left(p_{U,V,X|S}\right)$ and for any $n\in\mathbb{N}$, let $M\sim p^{(U)}_{\mathcal{M}_n}$ be a uniformly distributed message random variable. We first prove the existence of codes with an arbitrarily small \emph{average} error probability and a vanishing \emph{strong secrecy} metric.\footnote{Strong secrecy refers to the mutual information term $I(M;\mathbf{Z})$, where $M$ is uniformly distributed.} The expurgation method is then used to upgrade reliability to a vanishing \emph{maximal} error probability and upgrade strong secrecy to SS.

%let $M\sim P_M\in\mathcal{P}(\mathcal{M}_n)$, be the message distribution. We first show that for any $R\in\mathbb{R}_+$ with $R<R_\mathsf{A}\left(Q_{U,V,X|S}\right)$, there exists a semantically-secure sequence of $(n,R)$-codes with a vanishing \emph{average} error probability, i.e., when $M\sim p_{\mathcal{M}_n}^{(U)}$ - the uniform distribution over $\mathcal{M}_n$. Afterwards, the uniform message distribution assumption for the error probability analysis is dropped using the expurgation technique \cite[Theorem 7.7.1]{Cover_Thomas}, which allows upgrading reliability to achieve a vanishing \emph{maximal} error probability, while preserving SS.

%%%%%%%%%%%%%%%%%%%%%%%%%%%%%%%%%%%%%            Code Construction            %%%%%%%%%%%%%%%%%%%%%%%%%%%%%%%%%%%%%
%%%%%%%%%%%%%%%%%%%%%%%%%%%%%%%%%%%%%%%%%%%%%%%%%%%%%%%%%%%%%%%%%%%%%%%%%%%%%%%%%%%%%%%%%%%%%%%%%%%%%%%%%%%%%%%%%%%

\par\textbf{Codebook $\bm{\mathcal{C}_n}$:} We use a superposition codebook that encodes the confidential message in its outer layer. The codebook is drawn independently of the state sequence $\mathbf{S}$, but with sufficient redundancy to correlate the transmission with $\mathbf{S}$. 

Let $I$ and $J$ be two independent random variables uniformly distributed over $\mathcal{I}_n\triangleq\big[1:2^{nR_1}\big]$ and $\mathcal{J}_n\triangleq\big[1:2^{nR_2}\big]$, respectively.\footnote{For simplicity of notation we assume that $2^{nR}$, $2^{nR_1}$ and $2^{nR_2}$ are integers.} Let $\mathsf{C}_U^{(n)}\triangleq\big\{\mathbf{U}(i)\big\}_{i\in\mathcal{I}_n}$ be a random inner layer codebook, which is a set of random vectors of length $n$ that are i.i.d. according to $p_U^n$. An outcome of $\mathsf{C}_U^{(n)}$ is denoted by $\mathcal{C}_U^{(n)}\triangleq\big\{\mathbf{u}(i)\big\}_{i\in\mathcal{I}_n}$. 

To describe the outer layer codebook, fix $\mathcal{C}_U^{(n)}$ and for every $i\in\mathcal{I}_n$ let $\mathsf{C}_V^{(n)}(i)\triangleq\big\{\mathbf{V}(i,j,m)\big\}_{(j,m)\in\mathcal{J}_n\times\mathcal{M}_n}$ be a collection of i.i.d. random vectors of length $n$ with distribution $p^n_{V|U=\mathbf{u}(i)}$. A random outer layer codebook (with respect to an inner codebook $\mathcal{C}_U^{(n)}$) is defined as $\mathsf{C}_V^{(n)}\triangleq\big\{\mathsf{C}_V^{(n)}(i)\big\}_{i\in\mathcal{I}_n}$. A realization of $\mathsf{C}_V^{(n)}(i)$, for $i\in\mathcal{I}_n$, is denoted by  $\mathcal{C}_V^{(n)}(i)\triangleq\big\{\mathbf{v}(i,j,m)\big\}_{(j,m)\in\mathcal{J}_n\times\mathcal{M}_n}$, while $\mathcal{C}_V^{(n)}$ denotes a realization of $\mathsf{C}_V^{(n)}$. A random superposition codebook is $\mathsf{C}_n\triangleq\Big\{\mathsf{C}_U^{(n)},\mathsf{C}_V^{(n)}\Big\}$, while $\mathcal{C}_n=\Big\{\mathcal{C}_U^{(n)},\mathcal{C}_V^{(n)}\Big\}$ denotes a fixed codebook.

Let $\mathfrak{C}_n$ be the set of all possible outcomes of $\mathsf{C}_n$. The above codebook construction induces a PMF $\mu\in\mathcal{P}(\mathfrak{C}_n)$ over the codebook ensemble. For every $\mathcal{C}_n\in\mathfrak{C}_n$, we have
\begin{equation}
    \mu(\mathcal{C}_n)= \prod_{i\in\mathcal{I}_b}p^n_U\big(\mathbf{u}(i)\big) \prod_{\substack{\big(\hat{i},j,m\big)\\\in\mathcal{I}_n\times\mathcal{J}_n\times\mathcal{M}_m}}p^n_{V|U}\Big(\mathbf{v}\big(\hat{i},j,m\big)\Big|\mathbf{u}(\hat{i})\Big).\label{EQ:codebook_probability}
\end{equation}
The encoder and decoder are described next for any superposition codebook $\mathcal{C}_n\in\mathfrak{C}_n$.

%%%%%%%%%%%%%%%%%%%%%%%%%%%%%%%%%%%%%                 Encoding                %%%%%%%%%%%%%%%%%%%%%%%%%%%%%%%%%%%%%
%%%%%%%%%%%%%%%%%%%%%%%%%%%%%%%%%%%%%%%%%%%%%%%%%%%%%%%%%%%%%%%%%%%%%%%%%%%%%%%%%%%%%%%%%%%%%%%%%%%%%%%%%%%%%%%%%%%

\par\textbf{Encoder $\bm{f_{\mathcal{C}_n}}$:} The encoding phase is based on the likelihood-encoder \cite{Cuff_Song_Likelihood2016}, which, in turn, enables the approximation of the (rather cumbersome) induced joint distribution by a simpler distribution which is used for the analysis. 

To send $m\in\mathcal{M}_n$ upon observing the state sequence $\mathbf{s}\in\mathcal{S}^n$, the encoder randomly chooses $(i,j)\in\mathcal{I}_n\times\mathcal{J}_n$ according to
\begin{equation}
\hat{P}^{(\mathcal{C}_n)}(i,j|m,\mathbf{s})=\frac{p^n_{S|U,V}\big(\mathbf{s}\big|\mathbf{u}(i),\mathbf{v}(i,j,m)\big)}{\sum_{(i',j')\in\mathcal{I}_n\times\mathcal{J}_n}p^n_{S|U,V}\big(\mathbf{s}\big|\mathbf{u}(i'),\mathbf{v}(i',j',m)\big)},\label{EQ:main_proof_likelihood_enc}
\end{equation}
where $p_{S|U,V}$ is a conditional marginal distribution of $p$ from \eqref{EQ:single_letter_joint}. The channel input sequence is then generated by feeding the chosen $u$- and $v$-codewords along with the state sequence into a discrete and memoryless channel (DMC) $p_{X|U,V,S}$, i.e., it is sampled from the random vector $\mathbf{X}\sim p^n_{X|U=\mathbf{u}(i),V=\mathbf{v}(i,j,m),S=\mathbf{s}}$.

Accordingly, the (stochastic) encoding function $f_{\mathcal{C}_n}:\mathcal{M}_n\times\mathcal{S}^n\to\mathcal{P}(\mathcal{X}^n)$ is given by 
\begin{equation}
f_{\mathcal{C}_n}(\mathbf{x}|m,\mathbf{s})=\mspace{-12mu}\sum_{(i,j)\in\mathcal{I}_n\times\mathcal{J}_n}\mspace{-18mu}\hat{P}^{(\mathcal{C}_n)}(i,j|m,\mathbf{s})p_{X|U,V,S}^n\big(\mathbf{x}\big|\mathbf{u}(i),\mathbf{v}(i,j,m),\mathbf{s}\big),\mspace{-12mu}\quad\forall(
m,\mathbf{s},\mathbf{x})\in\mathcal{M}_n\times\mathcal{S}^n\times\mathcal{X}^n.\label{EQ:main_proof_encoder}
\end{equation}

%%%%%%%%%%%%%%%%%%%%%%%%%%%%%%%%%%%%%             Decoding            %%%%%%%%%%%%%%%%%%%%%%%%%%%%%%%%%%%%%%%%%%%%%%
%%%%%%%%%%%%%%%%%%%%%%%%%%%%%%%%%%%%%%%%%%%%%%%%%%%%%%%%%%%%%%%%%%%%%%%%%%%%%%%%%%%%%%%%%%%%%%%%%%%%%%%%%%%%%%%%%%%%

\par\textbf{Decoder $\bm{\phi_{\mathcal{C}_n}}$:} We define three decoding functions:
\begin{enumerate}
    \item $\phi_{\mathcal{C}_n}:\mathcal{Y}^n\to\hat{\mathcal{M}}_n$, which is the actual decoder of the message $m$.
    \item $\psi_{\mathcal{C}_n}^{(I)}:\mathcal{Y}^n\to\hat{\mathcal{I}}_n$, where $\hat{\mathcal{I}}_n\triangleq\mathcal{I}_n\cup\{e\}$.
    \item $\psi_{\mathcal{C}_n}^{(J)}:\mathcal{Y}^n\to\hat{\mathcal{J}}_n$, where $\hat{\mathcal{J}}_n\triangleq\mathcal{J}_n\cup\{e\}$.
\end{enumerate}
Here, $e$ is the same error symbol from the definition of $\hat{\mathcal{M}}_n$ for which we assume $e\notin\mathcal{M}_n\cup\mathcal{I}_n\cup\mathcal{J}_n$. The role of the functions $\psi_{\mathcal{C}_n}^{(I)}$ and $\psi_{\mathcal{C}_n}^{(J)}$ is to decode the indices $I$ and $J$, respectively. These functions will be used in the reliability analysis. Although, there is no reliability requirement on $(I,J)$, the subsequently chosen codebook rates enable their successful decoding.

Fix $\epsilon>0$. Upon observing $\mathbf{y}\in\mathcal{Y}^n$, the decoder searches for a unique triple $(\hat{i},\hat{j},\hat{m})\in\mathcal{I}_n\times\mathcal{J}_n\times\mathcal{M}_n$ such that
\begin{equation}
\Big(\mathbf{u}(\hat{i}),\mathbf{v}(\hat{i},\hat{j},\hat{m}),\mathbf{y}\Big)\in\mathcal{T}_\epsilon^{n}(p_{U,V,Y}).\label{EQ:main_proof_decoding_test}
\end{equation}
If such a unique triple is found, then set $\phi_{\mathcal{C}_n}(\mathbf{y})=\hat{m}$, $\psi_{\mathcal{C}_n}^{(I)}(\mathbf{y})=\hat{i}$ and $\psi_{\mathcal{C}_n}^{(J)}(\mathbf{y})=\hat{j}$; otherwise, $\phi_{\mathcal{C}_n}(\mathbf{y})=\psi_{\mathcal{C}_n}^{(I)}(\mathbf{y})=\psi_{\mathcal{C}_n}^{(J)}(\mathbf{y})=e$.

The triple $(\mathcal{M}_n,f_{\mathcal{C}_n},\phi_{\mathcal{C}_n})$ defined with respect to the codebook $\mathcal{C}_n$ constitutes an $(n,R)$-code $c_n$. The joint distribution $P^{(\mathcal{C}_n)}$ over $\mathcal{M}_n\times\mathcal{S}^n\times\mathcal{I}_n\times\mathcal{J}_n\times\mathcal{U}^n\times\mathcal{V}^n\times\mathcal{X}^n\times\mathcal{Y}^n\times\mathcal{Z}^n\times\hat{\mathcal{M}}_n$ induced by a fixed $\mathcal{C}_n$ is
\begin{align*}
P^{(\mathcal{C}_n)}(\mathbf{s},m,i,j,\mathbf{u},\mathbf{v},\mathbf{x},\mathbf{y},\mathbf{z},\hat{m})&=p_S^n(\mathbf{s})\frac{1}{|\mathcal{M}_n|}\hat{P}^{(\mathcal{C}_n)}(i,j|m,\mathbf{s})\mathds{1}_{\big\{\mathbf{u}=\mathbf{u}(i)\big\}\cap \big\{\mathbf{v}=\mathbf{v}(i,j,m)\big\}}\\&\mspace{120mu}\times p^n_{X|U,V,S}(\mathbf{x}|\mathbf{u},\mathbf{v},\mathbf{s})p^n_{Y,Z|X,S}(\mathbf{y},\mathbf{z}|\mathbf{x},\mathbf{s})\mathds{1}_{\big\{\phi_{\mathcal{C}_n}(\mathbf{y})=\hat{m}\big\}}.\numberthis\label{EQ:main_proof_induced_PMF}
\end{align*}
%If $P_M=p_{\mathcal{M}_n}^{(U)}$, i.e., the message distribution is uniform, we write $\bar{P}^{(\mathcal{C}_n)}$ instead of $P^{(\mathcal{C}_n)}$.
% When a random codebook $\mathsf{C}_n$ is used, we denote the corresponding random code by $\mathbb{C}_n$. 

%%%%%%%%%%%%%%%%%%%%%%%%%%           Approximating Distribution       %%%%%%%%%%%%%%%%%%%%%%%%%%%%%%%%%%%%%%%%%%%%%%
%%%%%%%%%%%%%%%%%%%%%%%%%%%%%%%%%%%%%%%%%%%%%%%%%%%%%%%%%%%%%%%%%%%%%%%%%%%%%%%%%%%%%%%%%%%%%%%%%%%%%%%%%%%%%%%%%%%%

\par \textbf{Approximating Distribution:} We next show that $P^{(\mathcal{C}_n)}$ is close in total variation to another distribution $Q^{(\mathcal{C}_n)}$, which we use for the reliability and security analyses. Let %For any $P_M\in\mathcal{P}(\mathcal{M}_n)$, we define $Q^{(\mathcal{C}_n)}$ by
\begin{align*}
Q^{(\mathcal{C}_n)}(m,i,j,\mathbf{u},\mathbf{v},\mathbf{s},\mathbf{x},\mathbf{y},\mathbf{z},\hat{m})&=\frac{1}{|\mathcal{M}_n||\mathcal{I}_n||\mathcal{J}_n|}\mathds{1}_{\big\{\mathbf{u}=\mathbf{u}(i)\big\}\cap \big\{\mathbf{v}=\mathbf{v}(i,j,m)\big\}}p^n_{S|U,V}(\mathbf{s}|\mathbf{u},\mathbf{v})\\&\mspace{120mu}\times p^n_{X|U,V,S}(\mathbf{x}|\mathbf{u},\mathbf{v},\mathbf{s})p^n_{Y,Z|X,S}(\mathbf{y},\mathbf{z}|\mathbf{x},\mathbf{s})\mathds{1}_{\big\{\phi_{\mathcal{C}_n}(\mathbf{y})=\hat{m}\big\}}.\numberthis\label{EQ:main_proof_target_PMF}
\end{align*}
For simplicity of notation, we sometimes abbreviate  $P^{(\mathcal{C}_n)}_{\mathbf{S},M,I,J,\mathbf{U},\mathbf{V},\mathbf{X},\mathbf{Y},\mathbf{Z},\hat{M}}$ and $Q^{(\mathcal{C}_n)}_{M,I,J,\mathbf{U},\mathbf{V},\mathbf{S},\mathbf{X},\mathbf{Y},\mathbf{Z},\hat{M}}$ as $P^{(\mathcal{C}_n)}$ and $Q^{(\mathcal{C}_n)}$, respectively. The following lemma states sufficient conditions for the expected value of the total variation between $P^{(\mathsf{C}_n)}$ and $Q^{(\mathsf{C}_n)}$ to converge exponentially fast to zero.% with double-exponential certainty (with respect to a random superposition codebook $\mathsf{B}_n$).

%As before, when $P_M=p_{\mathcal{M}_n}^{(U)}$, the notation $Q^{(\mathcal{C}_n)}$ replaces $Q^{(\mathcal{C}_n)}$. 

\begin{lemma}[Sufficient Conditions for Approximation]\label{LEMMA:good_approximation}
If $(R_1,R_2)\in\mathbb{R}_+^2$ satisfy
\begin{subequations}
	\begin{align}
		R_1&>I(U;S)\label{EQ:main_proof_approx_rate_bound1}\\
		R_1+R_2&>I(U,V;S),\label{EQ:main_proof_approx_rate_bound2}
	\end{align}\label{EQ:main_proof_approx_rate_bounds}%
\end{subequations}
then there exist $\alpha>0$, such that for any $n$ large enough
\begin{equation}
\mathbb{E}_\mu\Big|\Big|P^{(\mathsf{C}_n)}-Q^{(\mathsf{C}_n)}\Big|\Big|_{\mathsf{TV}}\leq e^{-n\alpha}.\label{EQ:main_proof_approx_soft_covering_expect}%+n\log\left(\frac{1}{\mu_S}\right)e^{-e^{n\alpha_2}}.\label{EQ:main_proof_approx_soft_covering_expect}
\end{equation}
%where $\mu_S=\min_{s\in\supp(p_S)}p_S(s)>0$.
\end{lemma}
%then there exist $\alpha _1,\alpha_2>0$, such that for any $n$ large enough
%\begin{equation}
%	\mathbb{P}\bigg(\max_{P_M\in\mathcal{P}(\mathcal{M}_n)}\Big|\Big|P^{(\mathsf{B}_n)}_{M,\mathbf{S},I,J,\mathbf{U},\mathbf{V},\mathbf{X},\mathbf{Y},\mathbf{Z},\hat{M}}-Q^{(\mathsf{B}_n)}_{M,\mathbf{S},I,J,\mathbf{U},\mathbf{V},\mathbf{X},\mathbf{Y},\mathbf{Z},\hat{M}}\Big|\Big|_{\mathsf{TV}}> e^{-n\alpha_1}\bigg)\leq e^{- e^{n\alpha_2}}.\label{EQ:main_proof_approx_soft_covering}
%\end{equation}

%In particular, for any such $n$ it also holds that

The proof of Lemma \ref{LEMMA:good_approximation} relies Lemmas \ref{LEMMA:soft_covering} and \ref{LEMMA:soft_covering_stronger} from Appendix \ref{APPEN:SCLs} and on some basic properties of total variation, see Appendix \ref{APPEN:good_approximation_proof} for details. Lemma \ref{LEMMA:good_approximation} is key in analyzing the performance of the proposed code. 

%%%%%%%%%%%%%%%%%%%%%%%%%%%%%%%%%%%%%           Error Analysis        %%%%%%%%%%%%%%%%%%%%%%%%%%%%%%%%%%%%%%%%%%%%%%
%%%%%%%%%%%%%%%%%%%%%%%%%%%%%%%%%%%%%%%%%%%%%%%%%%%%%%%%%%%%%%%%%%%%%%%%%%%%%%%%%%%%%%%%%%%%%%%%%%%%%%%%%%%%%%%%%%%%

\par \textbf{Average Error Probability Analysis:} For the reliability part, we first show that the average error probability can be made arbitrarily small. At the last step of this proof, the codebook is expurgated to attain a vanishing maximal error probability (in accordance with Definition \ref{DEF:SDWTC_achievability}). The main idea here is to use Lemma \ref{LEMMA:good_approximation} to move away from analyzing the error probability under $P^{(\mathcal{C}_n)}$ to an analysis with respect to $Q^{(\mathcal{C}_n)}$. Analyzing the latter involves only standard typicality arguments.

%We first establish reliability when $P_M=p_{\mathcal{M}_n}^{(U)}$. 

%A code $c_n$ (with respect to a fixed superposition codebook $\mathcal{C}_n$) and a uniformly distributed message induce the joint distribution $P^{(\mathcal{C}_n)}$ (see \eqref{EQ:main_proof_induced_PMF}). 

The average error of a code $c_n$, with an underlying superposition codebook $\mathcal{C}_n$, is denoted by $e_\mathsf{a}(\mathcal{C}_n)$ and is given by
\begin{equation}
e_a(\mathcal{C}_n)=\frac{1}{|\mathcal{M}_n|}\sum_{m\in\mathcal{M}_n}e_m(c_n)=\mathbb{P}_{P^{(\mathcal{C}_n)}}\big(\hat{M}\neq M\big),\label{EQ:main_proof_error_prob_P}
\end{equation}
where the subscript $P^{(\mathcal{C}_n)}$ on the RHS indicates that the probability measure is induced by the PMF $P^{(\mathcal{C}_n)}$ from \eqref{EQ:main_proof_induced_PMF}. 

We first show that a sufficient condition for the RHS of \eqref{EQ:main_proof_error_prob_P} to become arbitrarily small is that the average error probability induced by the $Q^{(\mathcal{C}_n)}$ PMF, i.e., $\mathbb{P}_{Q^{(\mathcal{C}_n)}}\big(\hat{M}\neq M\big)$,
is small. Recall the following property of total variation (see, e.g., \cite[Property (b)]{Cuff_Song_Likelihood2016}). Let $\mu,\nu$ be two measures on a measurable space $(\mathcal{X},\mathcal{F})$ and $g:\mathcal{X}\to\mathbb{R}$ be a non-negative measurable function bounded by $b\in\mathbb{R}$. It holds that
\begin{equation}
\big|\mathbb{E}_\mu g-\mathbb{E}_\nu g\big|\leq b\cdot\big|\big|\mu-\nu\big|\big|_\mathsf{TV}.\label{EQ:main_proof_TV_property}
\end{equation}
For every $n\in\mathbb{N}$, define $g_{\mathcal{C}_n}:\mathcal{M}_n\times\hat{\mathcal{M}}_n\to\mathbb{R}_+$ as
$g_{\mathcal{C}_n}(m,\hat{m})=\mathds{1}_{\{\hat{m}\neq m\}}$, and note that
\begin{subequations}
\begin{align}
\mathbb{E}_{P^{(\mathcal{C}_n)}}g_{\mathcal{C}_n}(M,\hat{M})&=\mathbb{P}_{P^{(\mathcal{C}_n)}}\big(\hat{M}\neq M\big)\\
\mathbb{E}_{Q^{(\mathcal{C}_n)}}g_{\mathcal{C}_n}(M,\hat{M})&=\mathbb{P}_{Q^{(\mathcal{C}_n)}}\big(\hat{M}\neq M\big).
\end{align}
\end{subequations}
The property from \eqref{EQ:main_proof_TV_property} gives that for any $\mathcal{C}_n$
\begin{equation}
    \Big|\mathbb{P}_{P^{(\mathcal{C}_n)}}\big(\hat{M}\neq M\big)-\mathbb{P}_{Q^{(\mathcal{C}_n)}}\big(\hat{M}\neq M\big)\Big|\leq \Big|\Big|P^{(\mathcal{C}_n)}_{M,\hat{M}}-Q^{(\mathcal{C}_n)}_{M,\hat{M}}\Big|\Big|_{\mathsf{TV}}\stackrel{(a)}\leq \Big|\Big|P^{(\mathcal{C}_n)}-Q^{(\mathcal{C}_n)}\Big|\Big|_{\mathsf{TV}},\label{EQ:main_proof_TV_property_applied}
\end{equation}
where (a) follows because for any $p_{X,Y},q_{X,Y}\in\mathcal{P}(\mathcal{X}\times\mathcal{Y})$ with marginals $p_X$ and $q_X$, respectively, it holds that $\big|\big|p_X-q_X\big|\big|_{\mathsf{TV}}\leq\big|\big|p_{X,Y}-q_{X,Y}\big|\big|_{\mathsf{TV}}$. Taking an expectation over the ensemble of superposition codebooks, after some algebra we obtain
\begin{align*}
\mathbb{E}_\mu\mathbb{P}_{Q^{(\mathsf{C}_n)}}\big(\hat{M}\neq M\big)-\mathbb{E}_\mu\Big|\Big|P^{(\mathsf{C}_n)}-Q^{(\mathsf{C}_n)}\Big|\Big|_{\mathsf{TV}}&\leq\mathbb{E}_\mu\mathbb{P}_{P^{(\mathsf{C}_n)}}\big(\hat{M}\neq M\big)\\
&\leq\mathbb{E}_\mu\mathbb{P}_{Q^{(\mathsf{C}_n)}}\big(\hat{M}\neq M\big)+\mathbb{E}_\mu\Big|\Big|\bar{P}^{(\mathsf{C}_n)}-Q^{(\mathsf{C}_n)}\Big|\Big|_{\mathsf{TV}}.\numberthis\label{EQ:main_proof_TV_error_prob_sandwich_expect}
\end{align*}

%We use the shorthand $\Big|\Big|P^{(\mathcal{C}_n)}-Q^{(\mathcal{C}_n)}\Big|\Big|_{\mathsf{TV}}$ for the total variation from the RHS of \eqref{EQ:main_proof_TV_property_applied} and rewrite it as
%\begin{align*}
%\mathbb{P}_{Q^{(\mathcal{C}_n)}}\big(\hat{M}\neq M\big)-\Big|\Big|P^{(\mathcal{C}_n)}-Q^{(\mathcal{C}_n)}\Big|\Big|_{\mathsf{TV}}&\leq\mathbb{P}_{P^{(\mathcal{C}_n)}}\big(\hat{M}\neq M\big)\\
%&\leq\mathbb{P}_{Q^{(\mathcal{C}_n)}}\big(\hat{M}\neq M\big)+\Big|\Big|P^{(\mathcal{C}_n)}-Q^{(\mathcal{C}_n)}\Big|\Big|_{\mathsf{TV}}.\numberthis\label{EQ:main_proof_TV_error_prob_sandwich}
%\end{align*}
%Taking an expectation over the ensemble of superposition codebooks, we obtain

Lemma \ref{LEMMA:good_approximation} states that $\mathbb{E}_\mu\Big|\Big|P^{(\mathsf{C}_n)}-Q^{(\mathsf{C}_n)}\Big|\Big|$ can be made arbitrarily small with $n$, provided that \eqref{EQ:main_proof_approx_rate_bounds} are satisfied. To show that the expected average error probability under $Q^{(\mathsf{C}_n)}$ also converges to 0 with $n$, consider the following arguments. For any codebook $\mathcal{C}_n\in\mathfrak{C}_n$ and $(\tilde{i},\tilde{j},\tilde{m})\in\mathcal{I}_n\times\mathcal{J}_n\times\mathcal{M}_n$, define the event
\begin{equation}
\mathcal{E}(\tilde{i},\tilde{j},\tilde{m},\mathcal{C}_n)=\Big\{\big(\mathbf{u}(\tilde{i}),\mathbf{v}(\tilde{i},\tilde{j},\tilde{m}),\mathbf{Y}\big)\in\mathcal{T}_\epsilon^{n}(p_{U,V,Y})\Big\},\label{EQ:main_proof_error_event}
\end{equation}
where $\mathbf{Y}\sim p^n_{Y|U=\mathbf{u}(\tilde{i}),V=\mathbf{v}(\tilde{i},\tilde{j},\tilde{m})}$ is the random sequence observed at the receiver when the transmitter sends $(\tilde{i},\tilde{j},\tilde{m})$ over the effective DMC $p^n_{Y|U,V}$ defined by
\begin{equation}
p_{Y|U,V}(y|u,v)=\mspace{-8mu}\sum_{(s,x,z)\in\mathcal{S}\times\mathcal{X}\times\mathcal{Z}}\mspace{-8mu}p_{S|U,V}(s|u,v)p_{X|S,U,V}(x|s,u,v)p_{Y,Z|X,S}(y,z|x,s),\quad\forall(u,v,y)\in\mathcal{U}\times\mathcal{V}\times\mathcal{Y}.
\end{equation}
Furthermore, the PMF $p_{U,V,Y}$ with respect to which the letter-typical set on the RHS of \eqref{EQ:main_proof_error_event} is defined is a marginal of $p$ from \eqref{EQ:single_letter_joint}. 

To upper bound the expected average error probability under $Q^{(\mathsf{C}_n)}$, for each $\mathcal{C}_n\in\mathfrak{C}_n$, we extend $Q^{(\mathcal{C}_n)}$ to the space $\mathcal{M}_n\times\mathcal{S}^n\times\mathcal{I}_n\times\mathcal{J}_n\times\mathcal{U}^n\times\mathcal{V}^n\times\mathcal{X}^n\times\mathcal{Y}^n\times\mathcal{Z}^n\times\hat{\mathcal{M}}_n\times\hat{\mathcal{I}}_n\times\hat{\mathcal{J}}_n$ by
\begin{equation}
Q^{(\mathcal{C}_n)}(m,i,j,\mathbf{u},\mathbf{v},\mathbf{s},\mathbf{x},\mathbf{y},\mathbf{z},\hat{m},\hat{i},\hat{j})=Q^{(\mathcal{C}_n)}(m,i,j,\mathbf{u},\mathbf{v},\mathbf{s},\mathbf{x},\mathbf{y},\mathbf{z},\hat{m})\mathds{1}_{\big\{\psi_{\mathcal{C}_n}^{(I)}(\mathbf{y})=\hat{i}\big\}\cap\big\{\psi_{\mathcal{C}_n}^{(J)}(\mathbf{y})=\hat{j}\big\}},\numberthis\label{EQ:main_proof_extended_gamma}
\end{equation}
thus allowing us to account for errors in decoding $I$ and $J$ as well. We have the following upper bound:
\begin{align*}
\mathbb{E}_\mu\mathbb{P}_{Q^{(\mathsf{C}_n)}}\big(\hat{M}\neq M\big)&\stackrel{(a)}\leq\mathbb{E}_\mu\mathbb{P}_{Q^{(\mathsf{C}_n)}}\Big((\hat{M},\hat{I},\hat{J})\neq (M,I,J)\Big)\\
&\stackrel{(b)}\leq\mathbb{E}_\mu\mathbb{P}_{Q^{(\mathsf{C}_n)}}\Big((\hat{M},\hat{I},\hat{J})\neq (1,1,1)\Big|(M,I,J)=(1,1,1)\Big)\\
&\begin{multlined}[b][.8\linewidth]\stackrel{(c)}=\mathbb{E}_\mu\mathbb{P}_{Q^{(\mathsf{C}_n)}}\vast(\mathcal{E}(1,1,1,\mathsf{C}_n)^c\cup\left\{\bigcup_{\tilde{i}\neq 1}\mathcal{E}(\tilde{i},1,1,\mathsf{C}_n)\right\}\\\cup\left\{\bigcup_{(\tilde{j},\tilde{m})\neq(1,1)}\mathcal{E}(1,\tilde{j},\tilde{m},\mathsf{C}_n)\right\}\cup\left\{\bigcup_{(\tilde{i},\tilde{j},\tilde{m})\neq(1,1,1)}\mathcal{E}(\tilde{i},\tilde{j},\tilde{m},\mathsf{C}_n)\right\}\vast)\end{multlined}\\
&\begin{multlined}[b][.8\linewidth]\stackrel{(d)}\leq \underbrace{\mathbb{P}_{p^n_{U,V,Y}}\Big((\mathbf{U},\mathbf{V},\mathbf{Y})\in\mathcal{T}_\epsilon^{n}(p_{U,V,Y})\Big)}_{P_1}+\underbrace{\sum_{\tilde{i}\neq 1}\mathbb{P}_{p^n_{U,V}\times p^n_Y}\Big((\mathbf{U},\mathbf{V},\mathbf{Y})\in\mathcal{T}_\epsilon^{n}(p_{U,V,Y})\Big)}_{P_2}\\+\underbrace{\sum_{(\tilde{j},\tilde{m})\neq (1,1)}\mathbb{P}_{p^n_{U,V}\times p^n_{Y|U}}\Big((\mathbf{U},\mathbf{V},\mathbf{Y})\in\mathcal{T}_\epsilon^{n}(p_{U,V,Y})\Big)}_{P_3}\\+\underbrace{\sum_{(\tilde{i},\tilde{j},\tilde{m})\neq(1,1,1)}\mathbb{P}_{p^n_{U,V}\times p^n_Y}\Big((\mathbf{U},\mathbf{V},\mathbf{Y})\in\mathcal{T}_\epsilon^{n}(p_{U,V,Y})\Big)}_{P_4},\end{multlined}
\end{align*}
where:\\
(a) is because the probability of error in decoding $M$ is upper bounded by the probability of error in decoding $(I,J,M)$;\\
(b) follows by the symmetry of the code under $Q^{(\mathcal{C}_n)}$ with respect to $(i,j,m)$;\\
(c) is the definition of the decoding rules $\phi_{\mathcal{C}_n}$, $\psi_{\mathcal{C}_n}^{(I)}$ and $\psi_{\mathcal{C}_n}^{(J)}$;\\
(d) uses the union bound and takes the expectation over the ensemble of codebooks.

By the law of large numbers $P_1\to 0$ as $n\to\infty$, while $P_2$, $P_3$ and $P_4$ also converge to 0 as $n$ grows if
\begin{subequations}
\begin{align}
R+R_2<I(V;Y|U)\label{EQ:main_proof_reliability_bounds1}\\
R+R_1+R_2<I(U,V;Y)\label{EQ:main_proof_reliability_bound2}.
\end{align}\label{EQ:main_proof_reliability_bounds}%
\end{subequations}
Specifically, \eqref{EQ:main_proof_reliability_bounds1} implies that $P_3\to 0$ as $n\to\infty$, while \eqref{EQ:main_proof_reliability_bound2} ensures that both $P_2\to 0$ and $P_4\to 0$ as $n\to\infty$. A sufficient condition for the former is
\begin{equation}
R_1<I(U,V;Y).\label{EQ:main_proof_reliability_bound_redundant}
\end{equation}
However, \eqref{EQ:main_proof_reliability_bound_redundant} is redundant having \eqref{EQ:main_proof_reliability_bound2}. Concluding, as long as \eqref{EQ:main_proof_approx_rate_bounds} and \eqref{EQ:main_proof_reliability_bounds} simultaneously hold, we have
\begin{equation}
\mathbb{E}_\mu e_a(\mathsf{C}_n)\xrightarrow[n\to\infty]{}0.\label{EQ:main_proof_reliability_expect}
\end{equation}

%%%%%%%%%%%%%%%%%%%%%%%%%%%%%%%%%%%%%         Security Analysis       %%%%%%%%%%%%%%%%%%%%%%%%%%%%%%%%%%%%%%%%%%%%%%
%%%%%%%%%%%%%%%%%%%%%%%%%%%%%%%%%%%%%%%%%%%%%%%%%%%%%%%%%%%%%%%%%%%%%%%%%%%%%%%%%%%%%%%%%%%%%%%%%%%%%%%%%%%%%%%%%%%%

\par \textbf{Security Analysis:} The security analysis shows that under proper conditions the induced conditional distribution of $\mathbf{Z}$ given $(M,\mathbf{U})$ approximates the product distribution $p^n_{Z|U}$. To demonstrate this, we once again rely on the approximation of $P^{(\mathcal{C}_n)}$ through $Q^{(\mathcal{C}_n)}$. It is first shown that if strong secrecy is achieved under $Q^{(\mathcal{C}_n)}$, then it is also achieved under $P^{(\mathcal{C}_n)}$. Strong secrecy is then upgraded to SS through expurgation. Having that, it remains to be shown that security is attainable under $Q^{(\mathcal{C}_n)}$. The following lemma justifies that strong secrecy under $Q^{(\mathcal{C}_n)}$ implies strong secrecy under $P^{(\mathcal{C}_n)}$.

% The underlying idea behind the rate bound we derive for SS is to allow the eavesdropper to decode the inner layer codeword, thus making him waste channel resources on decoding a codeword that carries no confidential information. The remaining resources are insufficient for extracting any information on the outer layer codeword, which, in turn, results in our code being semantically-secure.  

\begin{lemma}[SS via Approximating Distribution]\label{LEMMA:ss_p_gamma}
Let $\mathcal{C}_n\in\mathfrak{C}_n$ be a superposition codebook for which there exists a $\beta_1>0$, such that for all sufficiently large $n$
\begin{equation}
\Big|\Big|P^{(\mathcal{C}_n)}_{M,\mathbf{Z}}-Q^{(\mathcal{C}_n)}_{M,\mathbf{Z}}\Big|\Big|_{\mathsf{TV}}\leq e^{-n\beta_1}.\label{EQ:lemma_ss_if}
\end{equation}
Then, there exists a $\beta_2>0$, such that for any $n$ large enough (possibly larger than the values of $n$ needed for \eqref{EQ:lemma_ss_if} to become valid)
\begin{equation}
\Big|I_{P^{(\mathcal{C}_n)}}(M;\mathbf{Z})-I_{Q^{(\mathcal{C}_n)}}(M;\mathbf{Z})\Big|\leq e^{-n\beta_2}.\label{EQ:lemma_ss_then}
\end{equation}
%where the subscripts $P$ and $Q$ indicate that a mutual information term is calculated with respect to $P^{(\mathcal{C}_n)}$ or $Q^{(\mathcal{C}_n)}$, respectively.
\end{lemma}

The proof of Lemma \ref{LEMMA:ss_p_gamma} is relegated to Appendix \ref{APPEN:ss_p_gamma_proof}. As subsequently shown, the existence of a codebook $\mathcal{C}_n$ that satisfies \eqref{EQ:lemma_ss_if} follows by Lemma \ref{LEMMA:good_approximation}. For such a $\mathcal{C}_n$, we have% that for any $P_M\in\mathcal{P}(\mathcal{M}_n)$
\begin{equation}
I_{P^{(\mathcal{C}_n)}}(M;\mathbf{Z})\leq I_{Q^{(\mathcal{C}_n)}}(M;\mathbf{Z})+e^{-n\beta_2},\label{EQ:main_proof_SSP_UB1}
\end{equation}
for $n$ sufficiently large. %In particular, \eqref{EQ:main_proof_SSP_UB1} also holds for $P_M^\star$, the maximizer of $I_P(M;\mathbf{Z})$ (that exists due to concavity). Further increasing the RHS by maximizing it over all $P_M\in\mathcal{P}(\mathcal{M}_n)$ as well, gives
%\begin{equation}
%\ell_{\mathsf{Sem}}(c_n)=\max_{P_M\in\mathcal{P}(\mathcal{M}_n)}I_P(M;\mathbf{Z})\leq \max_{P_M\in\mathcal{P}(\mathcal{M}_n)}I_Q(M;\mathbf{Z})+e^{-n\beta_2},\label{EQ:main_proof_SSP_UB2}
%\end{equation}
%where $c_n$ is the code associated with $\mathcal{C}_n$. Thus, finding a codebook for which the RHS of \eqref{EQ:main_proof_SSP_UB2} can be made arbitrarily small implies SS.

With that in mind, we now focus on the mutual information term from the RHS of \eqref{EQ:main_proof_SSP_UB1}. For any $\mathcal{C}_n\in\mathfrak{C}_n$, we have
\begin{align*}
    I_{Q^{(\mathcal{C}_n)}}(M;\mathbf{Z})&\leq I_{Q^{(\mathcal{C}_n)}}(M;I,\mathbf{U},\mathbf{Z})\\
                          &\stackrel{(a)}=I_{Q^{(\mathcal{C}_n)}}(M;\mathbf{Z}|I,\mathbf{U})\\
                          &=\mathsf{D}\Big(Q^{(\mathcal{C}_n)}_{M,\mathbf{Z}|I,\mathbf{U}}\Big|\Big|p_{\mathcal{M}_n}^{(U)}Q^{(\mathcal{C}_n)}_{\mathbf{Z}|I,\mathbf{U}}\Big|Q^{(\mathcal{C}_n)}_{I,\mathbf{U}}\Big)\\
                          &\stackrel{(b)}=\mathsf{D}\Big(Q^{(\mathcal{C}_n)}_{\mathbf{Z}|M,I,\mathbf{U}}\Big|\Big|Q^{(\mathcal{C}_n)}_{\mathbf{Z}|I,\mathbf{U}}\Big|p_{\mathcal{M}_n}^{(U)}Q^{(\mathcal{C}_n)}_{I,\mathbf{U}}\Big)\\
                          &\stackrel{(c)}\leq \mathsf{D}\Big(Q^{(\mathcal{C}_n)}_{\mathbf{Z}|M,I,\mathbf{U}}\Big|\Big|p^n_{Z|U}\Big|p_{\mathcal{M}_n}^{(U)}Q^{(\mathcal{C}_n)}_{I,\mathbf{U}}\Big),\numberthis\label{EQ:main_proof_SS_gamma_MI_UB}
\end{align*}
where (a) is because $M$ and $(I,\mathbf{U})$ are independent under $Q^{(\mathcal{C}_n)}$, (b) is by the relative entropy chain rule and because $Q^{(\mathcal{C}_n)}_{M|I,\mathbf{U}}=p_{\mathcal{M}_n}^{(U)}$, while (c) follows from
\begin{equation}
    \mathsf{D}\Big(Q^{(\mathcal{C}_n)}_{\mathbf{Z}|M,I,\mathbf{U}}\Big|\Big|Q^{(\mathcal{C}_n)}_{\mathbf{Z}|I,\mathbf{U}}\Big|p_{\mathcal{M}_n}^{(U)}Q^{(\mathcal{C}_n)}_{I,\mathbf{U}}\Big)=\mathsf{D}\Big(Q^{(\mathcal{C}_n)}_{\mathbf{Z}|M,I,\mathbf{U}}\Big|\Big|p^n_{Z|U}\Big|p_{\mathcal{M}_n}^{(U)}Q^{(\mathcal{C}_n)}_{I,\mathbf{U}}\Big)-\mathsf{D}\Big(Q^{(\mathcal{C}_n)}_{\mathbf{Z}|I,\mathbf{U}}\Big|\Big|p^n_{Z|U}\Big|Q^{(\mathcal{C}_n)}_{I,\mathbf{U}}\Big)\label{EQ:similar_reasoning}
\end{equation}
and the non-negativity or relative entropy. Although, the inequality from \eqref{EQ:main_proof_SS_gamma_MI_UB} is true for any $p_{Z|U}:\mathcal{U}\to\mathcal{P}(\mathcal{Z})$, by $p_{Z|U}$ we refer to the conditional marginal of the single-letter distribution $p$ from \eqref{EQ:single_letter_joint}.

Recall that $Q^{(\mathcal{C}_n)}_{I,\mathbf{U}}=p_{\mathcal{I}_n}^{(U)}\mathds{1}_{\big\{\mathbf{U}=\mathbf{u}(I)\big\}}$ and apply an expectation over the codebook ensemble on both sides of \eqref{EQ:main_proof_SS_gamma_MI_UB}. This gives
\begin{align*}
    \mathbb{E}_\mu I_{Q^{(\mathsf{C}_n)}}(M;\mathbf{Z})&\leq \mathbb{E}_\mu\mathsf{D}\Big(Q^{(\mathsf{C}_n)}_{\mathbf{Z}|M,I,\mathbf{U}}\Big|\Big|p^n_{Z|U}\Big|p_{\mathcal{M}_n\times\mathcal{I}_n}^{(U)}Q^{(\mathsf{C}_n)}_{\mathbf{U}|I}\Big)\\
    &\stackrel{(a)}=\mathbb{E}_\mu\mathsf{D}\Big(Q^{(\mathsf{C}_n)}_{\mathbf{Z}|M=1,I=1,\mathbf{U}}\Big|\Big|p^n_{Z|U}\Big|Q^{(\mathsf{C}_n)}_{\mathbf{U}|I=1}\Big)\\
    &=\mathbb{E}_\mu\left[\sum_{\mathbf{u}\in\mathcal{U}^n}Q^{(\mathsf{C}_n)}_{\mathbf{U}|I}(\mathbf{u}|1)\mathsf{D}\Big(Q^{(\mathsf{C}_n)}_{\mathbf{Z}|M=1,I=1,\mathbf{U}=\mathbf{u}}\Big|\Big|p^n_{Z|U=\mathbf{u}}\Big)\right]\\
    &=\sum_{\mathbf{u}\in\mathcal{U}^n}\mathbb{E}_\mu\left[\mathds{1}_{\big\{\mathbf{U}(1)=\mathbf{u}\big\}}\mathsf{D}\Big(Q^{(\mathsf{C}_n)}_{\mathbf{Z}|M=1,I=1,\mathbf{U}=\mathbf{u}}\Big|\Big|p^n_{Z|U=\mathbf{u}}\Big)\right]\\
    &\stackrel{(b)}=\sum_{\mathbf{u}\in\mathcal{U}^n}\mathbb{E}_{\mathsf{C}_U^{(n)}}\Bigg[\mathbb{E}_{\mathsf{C}_V^{(n)}|\mathsf{C}_U^{(n)}}\bigg\{\mathds{1}_{\big\{\mathbf{U}(1)=\mathbf{u}\big\}}\mathsf{D}\Big(Q^{(\mathsf{C}_n)}_{\mathbf{Z}|M=1,I=1,\mathbf{U}=\mathbf{u}}\Big|\Big|p^n_{Z|U=\mathbf{u}}\Big)\bigg|\mathsf{C}_U^{(n)}\bigg\}\Bigg]\numberthis\label{EQ:new_security_analysis_UB1}%\\
    %&\stackrel{(c)}=\sum_{\mathbf{u}\in\mathcal{U}^n}\mathbb{E}_{\mathsf{C}_U^{(n)}}\Bigg[\mathds{1}_{\big\{\mathbf{U}(1)=\mathbf{u}\big\}}\mathbb{E}_{\mathsf{C}_V^{(n)}|\mathsf{C}_U^{(n)}}\bigg\{\mathsf{D}\Big(Q^{(\mathsf{C}_n)}_{\mathbf{Z}|M=1,I=1,\mathbf{U}=\mathbf{u}}\Big|\Big|p^n_{Z|U=\mathbf{u}}\Big)\bigg|\mathsf{C}_U^{(n)}\bigg\}\Bigg]\numberthis\label{EQ:new_security_analysis_UB1}
\end{align*}
where (a) is a consequence of symmetry, while (b) is the law of total expectation. In step (b) above we switched from the notation $\mathbb{E}_\mu$ that emphasizes the distribution of the random codebook $\mathsf{C}_n=\Big\{\mathsf{C}_U^{(n)},\mathsf{C}_V^{(n)}\Big\}$, to a notation that states the random variables themselves (and their possible conditioning).

The inner (conditional) expectation from the RHS of \eqref{EQ:new_security_analysis_UB1} is evaluated next. To do so, we present an argument for decorrelating the relative entropy inside the expectation and the inner layer random codebook $\mathsf{C}_U^{(n)}$. This will enable removing the conditioning from the inner expectation, which will simplify the term and adjust it to the framework of the SCL from \cite[Corollary VII.5]{Cuff_Synthesis2013}. Applying this SCL will, in turn, imply the desired strong secrecy.

Fix $\mathbf{u}\in\mathcal{U}^n$, an inner layer codebook $\mathsf{C}_U^{(n)}=\mathcal{C}_U^{(n)}$, and consider the quantity
\begin{equation}
    \mathbb{E}_{\mathsf{C}_V^{(n)}|\mathsf{C}_U^{(n)}=\mathcal{C}_U^{(n)}}\bigg\{\mathds{1}_{\big\{\mathbf{u}=\mathbf{u}(1)\big\}}\mathsf{D}\Big(Q^{(\mathsf{C}_n)}_{\mathbf{Z}|M=1,I=1,\mathbf{U}=\mathbf{u}(1)}\Big|\Big|p^n_{Z|U=\mathbf{u}(1)}\Big)\bigg|\mathsf{C}_U^{(n)}=\mathcal{C}_U^{(n)}\bigg\}.\label{EQ:conditional_expectation1}
\end{equation}
For each $\mathbf{u}\in\mathcal{U}^n$, let $\tilde{\mathsf{C}}_V^{(n)}(\mathbf{u})\triangleq\big\{\tilde{\mathbf{V}}(\mathbf{u},j)\big\}_{j\in\mathcal{J}_n}$ be a collection of i.i.d. random vectors of length $n$, each distributed according to $p^n_{V|U=\mathbf{u}}$ independently of $\mathsf{C}_n$. The collection $\tilde{\mathsf{C}}_V^{(n)}\triangleq\left\{\tilde{\mathsf{C}}_V^{(n)}(\mathbf{u})\right\}_{\mathbf{u}\in\mathcal{U}^n}$ is distributed according to 
\begin{equation}
    \tilde{\mu}\big(\tilde{\mathcal{C}}_V^{(n)}\big)=\prod_{\mathbf{u}\in\mathcal{U}^n}\prod_{j\in\mathcal{J}_n}p^n_{V|U}\big(\tilde{\mathbf{v}}(\mathbf{u},j)\big|\mathbf{u}\big),\label{EQ:tilde_mu_PMF}
\end{equation}
where, as before, $\tilde{\mathcal{C}}_V^{(n)}(\mathbf{u})\triangleq\big\{\tilde{\mathbf{v}}(\mathbf{u},j)\big\}_{j\in\mathcal{J}_n}$ stands for an outcome of $\tilde{\mathsf{C}}_V^{(n)}(\mathbf{u})$. For each $\mathbf{u}\in\mathcal{U}^n$ define a conditional PMF
\begin{equation}
\Pi^{\big(\tilde{\mathcal{C}}_V^{(n)}\big)}(j,\mathbf{v},\mathbf{z}|\mathbf{u})\triangleq\frac{1}{|\mathcal{J}_n|}\mathds{1}_{\big\{\mathbf{v}=\tilde{\mathbf{v}}(\mathbf{u},j)\big\}}p^n_{Z|U,V}(\mathbf{z}|\mathbf{u},\mathbf{v}).\label{EQ:main_proof_pi_PMF}
\end{equation}

Let $\mathsf{C}_V^{(n)}(1,1)\triangleq\big\{\mathbf{V}(1,j,1)\big\}_{j\in\mathcal{J}_n}$ be the collection of outer layer codewords from the codebook $\mathsf{C}_V^{(n)}(1)$ that correspond to $m=1$. Note that the random distribution $Q^{(\mathcal{C}_n)}_{\mathbf{Z}|M=1,I=1,\mathbf{U}=\mathbf{u}(1)}$ is a function of the collection $\mathsf{C}_V^{(n)}(1,1)$ only. Furthermore, whenever $\mathcal{C}_V^{(n)}(1,1)=\tilde{\mathcal{C}}_V^{(n)}\big(\mathbf{u}(1)\big)$, 
the distributions $Q^{(\mathsf{C}_n)}_{\mathbf{Z}|M=1,I=1,\mathbf{U}=\mathbf{u}(1)}$ and $\Pi^{\big(\tilde{\mathcal{C}}_V^{(n)}(\mathbf{u})\big)}_\mathbf{Z}$ are equal as PMFs on $\mathcal{Z}^n$. Since the set of possible outcomes of $\mathsf{C}_V^{(n)}(1,1)$ coincides with that of $\tilde{\mathsf{C}}_V^{(n)}\big(\mathbf{u}(1)\big)$, we may rewrite the conditional expectation from \eqref{EQ:conditional_expectation1} as
\begin{align*}
    \mathbb{E}_{\mathsf{C}_V^{(n)}|\mathsf{C}_U^{(n)}=\mathcal{C}_U^{(n)}}\bigg\{\mathds{1}_{\big\{\mathbf{u}=\mathbf{u}(1)\big\}}\mathsf{D}\Big(Q^{(\mathsf{C}_n)}_{\mathbf{Z}|M=1,I=1,\mathbf{U}=\mathbf{u}(1)}\Big|\Big|&p^n_{Z|U=\mathbf{u}(1)}\Big)\bigg|\mathsf{C}_U^{(n)}=\mathcal{C}_U^{(n)}\bigg\}\\
    &=\mathds{1}_{\big\{\mathbf{u}=\mathbf{u}(1)\big\}}\mathbb{E}_{\tilde{\mu}}\mathsf{D}\Big(\Pi^{\big(\tilde{\mathsf{C}}_V^{(n)}\big)}_{\mathbf{Z}|\mathbf{U}=\mathbf{u}}\Big|\Big|p^n_{Z|U=\mathbf{u}}\Big).\numberthis\label{EQ:conditional_expectation2}
\end{align*}
This essentially follows by the independence of the measures $\mu$ and $\tilde{\mu}$. Inserting \eqref{EQ:conditional_expectation2} into the RHS of \eqref{EQ:new_security_analysis_UB1}, we obtain
\begin{align*}
    \mathbb{E}_\mu I_{Q^{(\mathsf{C}_n)}}(M;\mathbf{Z})&\leq\sum_{\mathbf{u}\in\mathcal{U}^n}\mathbb{E}_{\mathsf{C}_U^{(n)}}\Bigg[\mathds{1}_{\big\{\mathbf{U}(1)=\mathbf{u}\big\}}\mathbb{E}_{\mathsf{C}_V^{(n)}|\mathsf{C}_U^{(n)}}\mathsf{D}\Big(Q^{(\mathsf{C}_n)}_{\mathbf{Z}|M=1,I=1,\mathbf{U}=\mathbf{u}}\Big|\Big|p^n_{Z|U=\mathbf{u}}\Big)\bigg|\mathsf{C}_U^{(n)}\Bigg]\\
    &=\sum_{\mathbf{u}\in\mathcal{U}^n}\mathbb{E}_{\mathsf{C}_U^{(n)}}\Bigg[\mathds{1}_{\big\{\mathbf{U}(1)=\mathbf{u}\big\}}\mathbb{E}_{\tilde{\mu}}\mathsf{D}\Big(\Pi^{\big(\tilde{\mathcal{C}}_V^{(n)}\big)}_{\mathbf{Z}|\mathbf{U}=\mathbf{u}}\Big|\Big|p^n_{Z|U=\mathbf{u}}\Big)\Bigg]\\
    &\stackrel{(a)}=\sum_{\mathbf{u}\in\mathcal{U}^n}q_U^n(\mathbf{u})\mathbb{E}_{\tilde{\mu}}\mathsf{D}\Big(\Pi^{\big(\tilde{\mathcal{C}}_V^{(n)}\big)}_{\mathbf{Z}|\mathbf{U}=\mathbf{u}}\Big|\Big|p^n_{Z|U=\mathbf{u}}\Big)\\
    &\stackrel{(b)}=\mathbb{E}_{\tilde{\mu}}\mathsf{D}\Big(q_U^n\Pi^{\big(\tilde{\mathcal{C}}_V^{(n)}\big)}_{\mathbf{Z}|\mathbf{U}}\Big|\Big|p^n_{U,Z}\Big)\numberthis\label{EQ:new_security_analysis_UB2}
\end{align*}
where (a) is because $\mathbf{U}(1)\sim q_U^n$, while (b) follows by the relative entropy chain rule. The expected value of the relative entropy on the RHS of \eqref{EQ:new_security_analysis_UB2} falls within the framework of \cite[Corollary VII.5]{Cuff_Synthesis2013} and it converges exponentially fast to zero as $n\to\infty$, provided\footnote{The original statement from \cite[Corollary VII.5]{Cuff_Synthesis2013} deals with total variation rather than with relative entropy. Nonetheless, the result applies here as well due to Lemma \ref{LEMMA:TV_divergence_relation}. Namely, because over finite probability spaces an exponential decay of total variation implies an exponential decay of the corresponding relative entropy.}
\begin{equation}
    R_2>I(V;Z|U).\label{EQ:main_proof_SS_RB_final}
\end{equation}

\par \textbf{Code Extraction:} Summarizing the results up to this point, we have that as long as \eqref{EQ:main_proof_approx_rate_bounds}, \eqref{EQ:main_proof_reliability_bounds} and \eqref{EQ:main_proof_SS_RB_final} are simultaneously satisfied, then $\mathbb{E}_\mu e_a(\mathsf{C}_n)\xrightarrow[n\to\infty]{}0$ and, for sufficiently large $n$,
\begin{equation}
    \mathbb{E}_\mu I_{Q^{(\mathsf{C}_n)}}(M;\mathbf{Z})\leq e^{-n\tilde{\gamma}}\label{EQ:expected_results_leakage_Q}
\end{equation}
also hold true for some $\tilde{\gamma}>0$ independent of $n$.

The Selection Lemma from \cite[Lemma 5]{Goldfeld_WTCII_semantic2015} implies the existence of a sequence of superposition codebooks $\big\{\mathcal{C}_n\big\}_{n\in\mathbb{N}}$ (giving rise to a sequence of $(n,R)$-codes $\big\{c_n\big\}_{n\in\mathbb{N}}$), for which
\begin{subequations}
\begin{align}
    e_a(\mathcal{C}_n)&\xrightarrow[n\to\infty]{}0\label{EQ:existance_results_error_prob}\\
    I_{Q^{(\mathcal{C}_n)}}(M;\mathbf{Z})&\leq e^{-n\gamma},\label{EQ:existance_results_leakage_Q}
\end{align}\label{EQ:existance_results}%
\end{subequations}
where \eqref{EQ:existance_results_leakage_Q} holds for $n$ large enough and some $\gamma>0$. Through the relation from \eqref{EQ:main_proof_SSP_UB1}, we further deduce that there exists $\delta>0$ such that for sufficiently large $n$
\begin{equation}
I_{P^{(\mathcal{C}_n)}}(M;\mathbf{Z})\leq e^{-n\delta}.\label{EQ:security_strong_sec_final}
\end{equation}

It is left to upgrade the vanishing average error probability and strong secrecy metric to a vanishing maximal error probability and SS. This is done by expurgating the superposition codebook \cite[Theorem 7.7.1]{Cover_Thomas} (see also \cite{Expurgation_Renner2011}). Let $n$ be sufficiently large, so that
\begin{subequations}
\begin{align}
    e_a(\mathcal{C}_n)&=\frac{1}{\mathcal{M}_n}\sum_{m\in\mathcal{M}_n}\mathbb{P}_{P^{(\mathcal{C}_n)}}\big(\tilde{M}\neq m |M=m\big)\leq\frac{\epsilon}{3}\label{EQ:error_epsilon}\\
    I_{P^{(\mathcal{C}_n)}}(M;\mathbf{Z})&=\frac{1}{\mathcal{M}_n}\sum_{m\in\mathcal{M}_n}\mathsf{D}\Big(P^{(\mathcal{C}_n)}_{\mathbf{Z}|M=m}\Big|\Big|P^{(\mathcal{C}_n)}_\mathbf{Z}\Big)\leq \frac{\epsilon}{3}.\label{EQ:leakage_epsilon}
\end{align}\label{EQ:epsilon}%
\end{subequations}
The fraction of messages that induce an error probability greater than $\epsilon$ is less than $\frac{1}{3}$. Similarly, the fraction of messages with relative entropy greater than $\epsilon$ is less than $\frac{1}{3}$. Therefore, the fraction of offending messages is less than $\frac{2}{3}$. By removing them one obtains a new sequence of codes that is $\big\{\mathcal{C}^\star_n\big\}_{n\in\mathbb{N}}$, such that for every large enough~$n$
\begin{subequations}
\begin{align}
    \max_{m\in\mathcal{M}_n}\mathbb{P}_{P^{(\mathcal{C}^\star_n)}}\big(\tilde{M}\neq m |M=m\big)&\leq\epsilon\label{EQ:error_epsilon_max}\\
    \max_{m\in\mathcal{M}_n}\mathsf{D}\Big(P^{(\mathcal{C}^\star_n)}_{\mathbf{Z}|M=m}\Big|\Big|P^{(\mathcal{C}^\star_n)}_\mathbf{Z}\Big)&\leq \epsilon.\label{EQ:leakage_epsilon_max}
\end{align}\label{EQ:epsilon_max}%
\end{subequations}
The rate of the $n$-th code in the new sequence $\big\{\mathcal{C}^\star_n\big\}_{n\in\mathbb{N}}$ is $R-\frac{\log(3)}{n}$, and the loss is negligible for large $n$.

\eqref{EQ:error_epsilon_max} is the small maximal error probability requirement from \eqref{EQ:SDWTC_achievability_reliability}. It remains to be shown that \eqref{EQ:leakage_epsilon_max} implies SS. Recall that $P^{(\mathcal{C}_n^\star)}$ is the induced probability distribution with respect to a uniformly distributed message, i.e., $P^{(\mathcal{C}_n^\star)}=p_{\mathcal{M}_n}^{(U)}$. For any $q\in\mathcal{P}(\mathcal{M}_n)$, let $P^{(\mathcal{C}_n^\star,q)}$
be the induced probability distribution when $M\sim q$. Namely, $P^{(\mathcal{C}_n^\star,q)}$ is given by \eqref{EQ:main_proof_induced_PMF}, but with $q(m)$ instead of $\frac{1}{|\mathcal{M}_n|}$. For any $q\in\mathcal{P}(\mathcal{M}_n)$, consider the following:
\begin{align*}
    I_{P^{(\mathcal{C}^\star_n,q)}}(M;\mathbf{Z})&=\sum_{m\in\mathcal{M}_n}q(m)\mathsf{D}\Big(P^{(\mathcal{C}^\star_n,q)}_{\mathbf{Z}|M=m}\Big|\Big|P^{(\mathcal{C}^\star_n,q)}_\mathbf{Z}\Big)\\
    &\stackrel{(a)}=\sum_{m\in\mathcal{M}_n}q(m)\bigg[\mathsf{D}\Big(P^{(\mathcal{C}^\star_n,q)}_{\mathbf{Z}|M=m}\Big|\Big|P^{(\mathcal{C}^\star_n)}_\mathbf{Z}\Big)-\mathsf{D}\Big(P^{(\mathcal{C}^\star_n,q)}_{\mathbf{Z}}\Big|\Big|P^{(\mathcal{C}^\star_n)}_\mathbf{Z}\Big)\bigg]\\
    &\leq\sum_{m\in\mathcal{M}_n}q(m)\max_{\tilde{m}\in\mathcal{M}_n}\mathsf{D}\Big(P^{(\mathcal{C}^\star_n,q)}_{\mathbf{Z}|M=\tilde{m}}\Big|\Big|P^{(\mathcal{C}^\star_n)}_\mathbf{Z}\Big)\\
    &\stackrel{(b)}=\max_{m\in\mathcal{M}_n}\mathsf{D}\Big(P^{(\mathcal{C}^\star_n)}_{\mathbf{Z}|M=m}\Big|\Big|P^{(\mathcal{C}^\star_n)}_\mathbf{Z}\Big)\\
    &\leq \epsilon,\numberthis\label{EQ:SS_derived}
\end{align*}
where (a) follows by a similar reasoning as step (c) in the derivation of \eqref{EQ:main_proof_SS_gamma_MI_UB} (see \eqref{EQ:similar_reasoning}), while (b) is because  $P^{(\mathcal{C}^\star_n,q)}_{\mathbf{Z}|M=m}=P^{(\mathcal{C}^\star_n)}_{\mathbf{Z}|M=m}$, for any $q\in\mathcal{P}(\mathcal{M}_n)$. Maximizing both sides of \eqref{EQ:SS_derived} over all $q\in\mathcal{P}(\mathcal{M}_n)$ establishes the SS requirement from \eqref{EQ:SDWTC_achievability_security}.

Finally, we apply Fourier-Motzkin Elimination on \eqref{EQ:main_proof_approx_rate_bounds}, \eqref{EQ:main_proof_reliability_bounds} and \eqref{EQ:main_proof_SS_RB_final}, to eliminate $R_1$ and $R_2$. Doing so shows that any $R<R_\mathsf{A}\left(p_{U,V,X|S}\right)$ is achievable. Maximizing over all $p_{U,V,X|S}$ establishes Theorem \ref{TM:SDWTC_lower_bound}. 

\begin{remark}[Alternative Security Analysis]
The security analysis shows that under the conditions \eqref{EQ:main_proof_approx_rate_bounds} and \eqref{EQ:main_proof_SS_RB_final}, the induced conditional distribution of $\mathbf{Z}$ given $\mathbf{U}$ and $M$ approximates a product distribution $p^n_{Z|U}$, on average over the messages. Since the inner layer codebook (which is encoded by $U$) carries no confidential information, this implies a vanishing information leakage. An alternative approach to establish this is to make the induced conditional distribution of $\mathbf{Z}$ given $M$ (without the conditioning on $\mathbf{U}$) be a good approximation of $p_Z^n$. This also implies security because
\begin{equation}
I_{P^{(\mathcal{C}_n)}}(M;\mathbf{Z})\leq\frac{1}{|\mathcal{M}_n|}\sum_{m\in\mathcal{M}_n}\mathsf{D}\Big(P^{(\mathcal{C}_n)}_{\mathbf{Z}|M=m}\Big|\Big|p^n_Z\Big).\label{EQ:SDWTC_main_proof_altSS}
\end{equation}
The SCL for superposition codebooks implies that the RHS of \eqref{EQ:SDWTC_main_proof_altSS} decays exponentially fast to 0, provided that
\begin{subequations}
    \begin{align}
        R_1&>I(U;Z)\label{EQ:main_proof_altSS_RB1}\\
        R_1+R_2&>I(U,V;Z).\label{EQ:main_proof_altSS_RB2}
    \end{align}\label{EQ:main_proof_altSS_RB}
\end{subequations}
Replacing \eqref{EQ:main_proof_SS_RB_final} with \eqref{EQ:main_proof_altSS_RB} and combining it with \eqref{EQ:main_proof_approx_rate_bounds} and \eqref{EQ:main_proof_reliability_bounds}, achieves any $R$ with
\begin{equation}
R\leq \tilde{R}_\mathsf{A}\left(p_{U,V,X|S}\right)\triangleq\min\Big\{I(U,V;Y)-I(U,V;Z),I(V;Y|U),I(U,V;Y)-I(U,V;S)\Big\}.\label{EQ:main_proo_altSS_totalRB}
\end{equation}
Seemingly, the best secrecy rates our scheme achieves is the maximum between the RHS of \eqref{EQ:main_proo_altSS_totalRB} and $\tilde{R}_\mathsf{A}\left(p_{U,V,X|S}\right)$ from \eqref{EQ:SDWTC_lower_bound_prob}. However, a closer examination of the expressions in $\tilde{R}_\mathsf{A}\left(p_{U,V,X|S}\right)$ reveals that when optimizing over all $p_{U,V,X|S}$, $\tilde{R}_\mathsf{A}\left(p_{U,V,X|S}\right)$ is actually redundant. To see this, notice that for any $p_{U,V,X|S}$, such that $\tilde{R}_\mathsf{A}\left(p_{U,V,X|S}\right)\geq R_\mathsf{A}\left(p_{U,V,X|S}\right)$, taking $p_{\tilde{U},\tilde{V},\tilde{X}|S}$ with $\tilde{U}=0$, $\tilde{V}=(U,V)_p$ and $p_{\tilde{X}|S,\tilde{U},\tilde{V}}=p_{X|S,U,V}$, where the subscript $p$ in the definition of $\tilde{V}$ denotes that the random variables are distributed according to $p$, gives
\begin{equation}
R_\mathsf{A}\big(p_{\tilde{U},\tilde{V},\tilde{X}|S}\big)=\min\Big\{I(U,V;Y)-I(U,V;Z),I(U,V;Y)-I(U,V;S)\Big\}\geq \tilde{R}_\mathsf{A}\left(p_{U,V,X|S}\right).\label{EQ:main_proof_altSS_redundant}
\end{equation}
This implies that $R_\mathsf{A}$ is at least as high as the maximal $\tilde{R}_\mathsf{A}\left(p_{U,V,X|S}\right)$.%the approach for establishing SS given in the proof of Theorem \ref{TM:SDWTC_lower_bound} is superior to the alternative path discussed in this remark. The interpretation of this conclusion is that it is always better to let the eavesdropper decode $\mathbf{U}$, since this makes it `waste' channel resources on decoding a layer of the codebook that carries no confidential information. After doing so, the eavesdropper is lacking the required resources to extract any information about $M$ (regardless of its distribution) and SS follows.
\end{remark}

\begin{remark}[SS via Strong Soft-Covering]
The above proof establishes SS via expurgation. The random coding argument first produces a sequence of codes that attain strong secrecy. Then, the messages with the highest information leakage are eliminated to obtain SS. Another possible approach is to derive SS directly from the random coding argument using a pair of strong SCLs. Namely, thought Lemma \ref{LEMMA:soft_covering} it can be shown that the probability that the the approximation from \eqref{EQ:main_proof_approx_soft_covering_expect} fails is doubly-exponentially small. Having that, the heterogeneous strong SCL from \cite[Lemma 1]{Goldfeld_AVWTC_semantic2015} can be used to argue that $P^{(\mathsf{C}_n)}_{\mathbf{Z}|M=m,\mathbf{U}}$ is close in total variation to $p_{Z|U}^n$, for each $m\in\mathcal{M}_n$ (rather than on average as argued above). The continuity of mutual information over discrete probability spaces with respect to total variation would then imply SS with (doubly-exponentially) high probability, with respect to the random coding ensemble. Although this approach is not necessary here, we note it because it applies in various scenarios where the expurgation argument fails. Such scenarios include compound or arbitrarily varying settings, as well as cases where instead of (or in addition to) a secret message transmission, the legitimate parties aim to agree upon a semantically secured secret key. A key is typically required to be approximately uniform; however, expurgation can undesirably alter the distribution of the key. Strong soft-covering arguments, on the other hand, enable SS proofs in all these aforementioned instances (see \cite{Goldfeld_WTCII_semantic2015,Goldfeld_AVWTC_semantic2015,Goldfeld_Bunin_SNSK2017}).
\end{remark}
%%%%%%%%%%%%%%%%%%%%%%%%%%%%%%%%%%%%%%%%%%%%%%%%%%%%%%%%%%%%%%%%%%%%%%%%%%%%%%%%%%%%%%%%%%%%%%%%%%%%%%%%%%%%%%%%%%% 
%%%%%%%%%%%%%%%%%%%%%%%%%%%%%%%%%%%%%%%%%%%%%%%%%%%%%%%%%%%%%%%%%%%%%%%%%%%%%%%%%%%%%%%%%%%%%%%%%%%%%%%%%%%%%%%%%%%
%%%%%%%%%%%%%%%%%%%%%%%%%%%%%%%%%%%%                                         %%%%%%%%%%%%%%%%%%%%%%%%%%%%%%%%%%%%%%
%%%%%%%%%%%%%%%%%%%%%%%%%%%%%%%%%%%%                  SUMMARY                %%%%%%%%%%%%%%%%%%%%%%%%%%%%%%%%%%%%%%
%%%%%%%%%%%%%%%%%%%%%%%%%%%%%%%%%%%%                                         %%%%%%%%%%%%%%%%%%%%%%%%%%%%%%%%%%%%%%
%%%%%%%%%%%%%%%%%%%%%%%%%%%%%%%%%%%%%%%%%%%%%%%%%%%%%%%%%%%%%%%%%%%%%%%%%%%%%%%%%%%%%%%%%%%%%%%%%%%%%%%%%%%%%%%%%%%
%%%%%%%%%%%%%%%%%%%%%%%%%%%%%%%%%%%%%%%%%%%%%%%%%%%%%%%%%%%%%%%%%%%%%%%%%%%%%%%%%%%%%%%%%%%%%%%%%%%%%%%%%%%%%%%%%%%

\section{Summary and Concluding Remarks}\label{SEC:summary}

This paper studied SD-WTCs with non-causal encoder CSI. A novel lower bound on the SS-capacity was derived. The coding scheme that achieves the lower bound is based on a superposition codebook, which encodes the confidential message in the outer layer. The superposition codebook was constructed with sufficient redundancy to facilitate correlating both layers and the transmission itself with the observed state sequence. The correlation is attained by means of the likelihood encoder \cite{Cuff_Song_Likelihood2016}. SS is ensured via distribution approximation arguments and the expurgation technique. The structure of the rate bounds for secrecy implies that the eavesdropper can decode the inner layer codeword. Since no confidential information is encoded in the inner layer, this doesn't compromise security. The gain from doing so is that decoding the inner layer exhausts the eavesdropper's channel resources. Consequently, this prevents him from inferring any information on the outer layer, which contains the confidential~message. 

Our result was compared to several previous achievability results from the literature. Notably, a comparison to the best past achievable scheme for the SD-WTC with non-causal encoder CSI from \cite{Prabhakaran_SKSM2012} revealed that our scheme not only captures it as a special case, but also strictly outperforms it in some cases. Finally, the SS-capacity of the reversely less noisy SD-WTC was characterized. It was also shown that our scheme is tight for the semi-deterministic SD-WTD, where $Y=y(X,S)$ is the deterministic output observed by the legitimate receiver. This SS-capacity result, however, can also be retrieved from \cite{Prabhakaran_SKSM2012}, and even from the simpler achievable regions found in~\cite{SDWTC_Chen_HanVinck2006,SDWTC_2Sided_Liu2007}.

%%%%%%%%%%%%%%%%%%%%%%%%%%%%%%%%%%%%%%%%%%%%%%%%%%%%%%%%%%%%%%%%%%%%%%%%%%%%%%%%%%%%%%%%%%%%%%%%%%%%%%%%%%%%%%%%%%%
%%%%%%%%%%%%%%%%%%%%%%%%%%%%%%%%%%%%%%%%%%%%%%%%%%%%%%%%%%%%%%%%%%%%%%%%%%%%%%%%%%%%%%%%%%%%%%%%%%%%%%%%%%%%%%%%%%%
%%%%%%%%%%%%%%%%%%%%%%%%%                                                         %%%%%%%%%%%%%%%%%%%%%%%%%%%%%%%%%
%%%%%%%%%%%%%%%%%%%%%%%%%                        APPENDICES                       %%%%%%%%%%%%%%%%%%%%%%%%%%%%%%%%%
%%%%%%%%%%%%%%%%%%%%%%%%%                                                         %%%%%%%%%%%%%%%%%%%%%%%%%%%%%%%%%
%%%%%%%%%%%%%%%%%%%%%%%%%%%%%%%%%%%%%%%%%%%%%%%%%%%%%%%%%%%%%%%%%%%%%%%%%%%%%%%%%%%%%%%%%%%%%%%%%%%%%%%%%%%%%%%%%%%
%%%%%%%%%%%%%%%%%%%%%%%%%%%%%%%%%%%%%%%%%%%%%%%%%%%%%%%%%%%%%%%%%%%%%%%%%%%%%%%%%%%%%%%%%%%%%%%%%%%%%%%%%%%%%%%%%%%

\appendices

%%%%%%%%%%%%%%%%%%%%%%%%%%%%%%%%%%%%%%%%%%%%%%%%%%%%%%%%%%%%%%%%%%%%%%%%%%%%%%%%%%%%%%%%%%%%%%%%%%%%%%%%%%%%%%%%%%%
%%%%%%%%%%%%%%%%%%%%%%%%%%%%%%%%%%%%%%%%                                   %%%%%%%%%%%%%%%%%%%%%%%%%%%%%%%%%%%%%%%%
%%%%%%%%%%%%%%%%%%%%%%%%%%%%%%%%%%%%%%%%      Strong Soft-Covering         %%%%%%%%%%%%%%%%%%%%%%%%%%%%%%%%%%%%%%%%
%%%%%%%%%%%%%%%%%%%%%%%%%%%%%%%%%%%%%%%%                                   %%%%%%%%%%%%%%%%%%%%%%%%%%%%%%%%%%%%%%%%
%%%%%%%%%%%%%%%%%%%%%%%%%%%%%%%%%%%%%%%%%%%%%%%%%%%%%%%%%%%%%%%%%%%%%%%%%%%%%%%%%%%%%%%%%%%%%%%%%%%%%%%%%%%%%%%%%%%

\section{Soft-Covering Lemmas}\label{APPEN:SCLs}
		
%%%%%%%%%%%%%%%%%%%%%%%%%%%%%%%%%%%%%%%%%%%%%%%%%%%%%%%%%%%%%%%%%%%%%%%%%%%%%%%%%%%%%%%%%%%%%%%%%%%%%%%%%%%%%%%%%%%
%%%%%%%%%%%%%%%%%%%%%%%%%%%%%%%%%%%%%%%%     Superposition Strong SLC      %%%%%%%%%%%%%%%%%%%%%%%%%%%%%%%%%%%%%%%%
%%%%%%%%%%%%%%%%%%%%%%%%%%%%%%%%%%%%%%%%%%%%%%%%%%%%%%%%%%%%%%%%%%%%%%%%%%%%%%%%%%%%%%%%%%%%%%%%%%%%%%%%%%%%%%%%%%%
\subsection{Strong Soft-Covering Lemma for Superposition Codes}\label{SEC:soft_covering}
		
The SS analysis for the SD-WTC with non-causal encoder CSI relies on a SCL for superposition codes. Here, we give a strong version of this lemma (in the spirit of \cite{Goldfeld_WTCII_semantic2015,Goldfeld_AVWTC_semantic2015}). The proof of Theorem \ref{TM:SDWTC_lower_bound} only uses a classic soft-covering statement (i.e., convergence of expected value); the reason the stronger version is presented is twofold. First, the SS derivation in the proof of Theorem \ref{TM:SDWTC_lower_bound} can be preformed directly using the stronger version. Second, we believe that the sharp claim of Lemma \ref{LEMMA:soft_covering} could prove useful for other research problems.

%%%%%%%%%%%%%%%%%%%%%%%%%%%%%%%%%%%%%%%%%%%%%%%%%%%%%%%%%%%%%%%%%%%%%%%%%%%%%%%%%%%%%%%%%%%%%%%%%%%%%%%%%%%%%%%%%%%
%%%%%%%%%%%%%%%%%%%%          Figure - Soft-Covering Setup for Superposition Codes         %%%%%%%%%%%%%%%%%%%%%%%%
		
\begin{figure}[t!]
	\begin{center}
		\begin{psfrags}
			\psfragscanon
			\psfrag{A}[][][1]{$\mspace{-10mu}I$}
			\psfrag{B}[][][1]{\ \ \ \ \ \ \ \ \ \ \ \ \ \ $\mathcal{B}^{(n)}_U$}
			\psfrag{C}[][][1]{$\mspace{-10mu}J$}
			\psfrag{D}[][][1]{\ \ \ \ \ \ \ \ \ \ \ \ \ \ \  $\mathcal{B}^{(n)}_V\mspace{-6mu}=\mspace{-4mu}\big\{\mathcal{B}^{(n)}_V\mspace{-2mu}(i)\big\}$}
			\psfrag{E}[][][1]{\ \ \ \ \ \ \ \ \ $\mathbf{U}(I)$}
			\psfrag{F}[][][1]{\ \ \ \ \ \ \ \ \ \ \ $\mathbf{V}\big(I,J\big)$}
			\psfrag{G}[][][1]{\ \ \ \ \ \ \ $p^n_{W|U,V}$}
			\psfrag{X}[][][0.8]{\ \ \ \ \ \ \ \ \ \ \ \ \ \ \ \ \ \ Inner Codebook}
			\psfrag{Y}[][][0.8]{\ \ \ \ \ \ \ \ \ \ \ \ \ \ \ \ \ \ Outer Codebook}
			\psfrag{H}[][][1]{\ \ \ \ \ \ \ \ \ \ \ \ \ $\mathbf{W}\sim P^{(\mathcal{B}_n)}_{\mathbf{W}}$}
			\includegraphics[scale = .53]{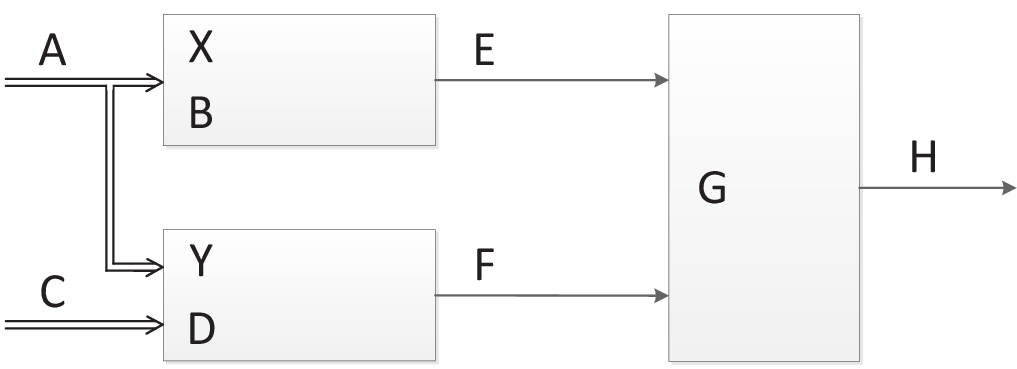}
			\caption{Superposition soft-covering setup with the goal of making $P^{(\mathcal{B}_n)}_{\mathbf{W}}\approx p_W^n$, where $\mathcal{B}_n=\Big\{\mathcal{B}^{(n)}_U,\mathcal{B}^{(n)}_V\Big\}$ is a fixed superposition codebook.} \label{FIG:soft_covering_superpos}
			\psfragscanoff
		\end{psfrags}
	\end{center}
\end{figure}
		
%%%%%%%%%%%%%%%%%%%%%%%%%%%%%%%%%%%%%%%%%%%%%%%%%%%%%%%%%%%%%%%%%%%%%%%%%%%%%%%%%%%%%%%%%%%%%%%%%%%%%%%%%%%%%%%%%%%
		
The setup is illustrated in Fig. \ref{FIG:soft_covering_superpos}, where inner and outer layer codewords are uniformly chosen and passed through a DMC to produce an output sequence. The induced distribution of the output should be asymptotically indistinguishable from a product distribution. The approximation is in terms of relative entropy, which is shown to converge to 0 exponentially quickly with high probability. The negligible probability is doubly-exponentially small with the blocklength $n$.

Fix $p_{U,V,W}\in\mathcal{P}(\mathcal{U}\times\mathcal{V}\times\mathcal{W})$ and let $I$ and $J$ be two independent random variables uniformly distributed over $\mathcal{I}_n\triangleq\big[1:2^{nR_1}\big]$ and $\mathcal{J}_n\triangleq\big[1:2^{nR_2}\big]$, respectively. Furthermore, let $\mathsf{B}^{(n)}_U \triangleq\big\{\mathbf{U}(i)\big\}_{i\in\mathcal{I}_n}$ be a random inner layer codebook, which is a set of random vectors of length $n$ that are i.i.d. according to $p_U^n$. A realization of $\mathsf{B}^{(n)}_U$ is denoted by $\mathcal{B}^{(n)}_U\triangleq\big\{\mathbf{u}(i)\big\}_{i\in\mathcal{I}_n}$. 

To describe the outer layer codebook, fix $\mathcal{B}^{(n)}_U$, and for every $i\in\mathcal{I}_n$, let $\mathsf{B}^{(n)}_V(i)\triangleq\big\{\mathbf{V}(i,j)\big\}_{j\in\mathcal{J}_n}$ be a collection of i.i.d. random vectors of length $n$ with distribution $p^n_{V|U=\mathbf{u}(i)}$. A random outer layer codebook (with respect to an inner codebook $\mathcal{B}^{(n)}_U$) is defined as $\mathsf{B}^{(n)}_V\triangleq\Big\{\mathsf{B}^{(n)}_V(i)\Big\}_{i\in\mathcal{I}_n}$. An outcome of $\mathsf{B}^{(n)}_V(i)$, for $i\in\mathcal{I}_n$ is denoted by  $\mathcal{B}^{(n)}_V(i)\triangleq\big\{\mathbf{v}(i,j,m)\big\}_{j\in\mathcal{J}_n}$. We also use $\mathcal{B}_V^{(n)}$ to denote an outcome of $\mathsf{B}_V^{(n)}$. A random superposition codebook  $\mathsf{B}_n\triangleq\Big\{\mathsf{B}^{(n)}_U,\mathsf{B}^{(n)}_V\Big\}$, while $\mathcal{B}_n\triangleq\Big\{\mathcal{B}^{(n)}_U,\mathcal{B}^{(n)}_V\Big\}$ denotes a fixed codebook. 

Letting $\mathfrak{B}_n$ be the set of all possible outcomes of $\mathsf{B}_n$, the above construction induces a distribution $\mu\in\mathcal{P}(\mathfrak{B}_n)$ over the codebook ensemble. For every $\mathcal{B}_n\in\mathfrak{B}_n$, we have
\begin{equation}
    \nu(\mathcal{B}_n)= \prod_{i\in\mathcal{I}_b}p^n_U\big(\mathbf{u}(i)\big) \prod_{\big(\hat{i},j\big)\in\mathcal{I}_n\times\mathcal{J}_n}p^n_{V|U}\Big(\mathbf{v}\big(\hat{i},j\big)\Big|\mathbf{u}(\hat{i})\Big).
\end{equation}

For a fixed superposition code $\mathcal{B}_n$, the output sequence $\mathbf{W}$ is generated by independently drawing $I$ and $J$ from $\mathcal{I}_n$ and $\mathcal{J}_n$, respectively, and feeding $\mathbf{u}(i)$ and $\mathbf{v}(i,j)$ into the DMC $p^n_{W|U,V}$. The induced distribution on $\mathcal{I}_n\times\mathcal{J}_n\times\mathcal{U}^n\times\mathcal{V}^n\times\mathcal{W}^n$ by $P^{(\mathcal{B}_n)}$ is\footnote{To simplify notation, from here on we assume that quantities of the form $2^{nR}$, where $n\in\mathbb{N}$ and $R\in\mathbb{R}_+$, are integers. Otherwise, simple modifications of some of the subsequent expressions using floor operations are required.}
\begin{equation}
	P^{(\mathcal{B}_n)}(i,j,\mathbf{u},\mathbf{v},\mathbf{w})=2^{-n(R_1+R_2)}\mathds{1}_{\big\{\mathbf{u}=\mathbf{u}(i)\big\}\cap\big\{\mathbf{v}=\mathbf{v}(i,j)\big\}}p^n_{W|U,V}(\mathbf{w}|\mathbf{u},\mathbf{v}).\label{EQ:induced_PMF_SCL}
\end{equation}
Accordingly, the induced output distribution is
\begin{equation}
    P^{(\mathcal{B}_n)}(\mathbf{w})=2^{-n(R_1+R_2)}\sum_{(i,j)\in\mathcal{I}_n\times\mathcal{J}_n}p^n_{W|U,V}\big(\mathbf{w}\big|\mathbf{u}(i),\mathbf{v}\big(i,j)\big)\label{EQ:induced_output_PMF_SCL}.
\end{equation}

We also set
\begin{equation}
    P(\mathcal{B}_n,i,j,\mathbf{u},\mathbf{v},\mathbf{w})\triangleq\mu(\mathcal{B}_n)\times P^{(\mathcal{B}_n)}(i,j,\mathbf{u},\mathbf{v},\mathbf{w}),
\end{equation}
and denote by $\mathbb{P}\triangleq \mathbb{P}_P$ the probability measure induced by $P$. This notation is used in the remainder of this section and in the proof of the strong SCL for superposition codes that is stated next. When switching to other probability measures, we do so in accordance with the notations defined in Section \ref{SEC:preliminaries}. 
		
\begin{lemma}[Strong Superposition SCL]\label{LEMMA:soft_covering}
For any $p_{U,V,W}$, where $|\mathcal{W}|<\infty$, and $(R_1,R_2)\in\mathbb{R}^2_+$ with
\begin{subequations}
    \begin{align}
	    R_1&>I(U;W)\\
	    R_1+R_2&>I(U,V;W),
	\end{align}
\end{subequations}
there exist $\gamma_1,\gamma_2 >0$, such that for $n$ large enough
\begin{equation}
	\mathbb{P}_\mu\bigg(\mathsf{D}\Big(P_\mathbf{W}^{(\mathsf{B}_n)}\Big|\Big|p_W^n\Big)> e^{-n\gamma_1}\bigg)\leq e^{- e^{n\gamma_2}}.\label{EQ:soft_covering}
\end{equation}
\begin{comment}
More precisely, for any $\delta_1 \in \big(0,R_1-I(U;W)\big)$ and $\delta_2 \in \big(0,R_1+R_2-I(U,V;W)\big)$ with $\delta_1<\delta_2<2\delta_1$ and $n$ sufficiently large
\begin{equation}
    \mathbb{P}_\mu\bigg(\mathsf{D}\Big(P^{(\mathsf{B}_n)}_{\mathbf{W}}\Big|\Big|p_W^n\Big)\geq c_{\delta_1,\delta_2} n2^{-n\gamma_{\delta_1,\delta_2}}\bigg)\leq 2^{nR_2}\cdot e^{-\frac{1}{3}2^{n\delta_1}} + |\mathcal{W}|^n\left[2^{nR_1}e^{-2^{n\frac{\delta_2}{2}}} +  e^{-\frac{1}{3}2^{n\frac{\delta_2-\delta_1}{2}}}\right],\label{EQ:soft_covering_precise}
\end{equation}
where
\begin{subequations}
	\begin{align}
		\gamma_{\delta_1,\delta_2} &= \sup_{\alpha > 1} \min\left\{\beta^{(1)}_{\alpha,\delta_1},\beta^{(2)}_{\alpha,\delta_2},\frac{\delta_1}{4}\right\},\label{EQ:soft_covering_exponent}\\
		\beta^{(1)}_{\alpha,\delta_1}&=\frac{\alpha-1}{2\alpha-1} \big(R_1-\delta_1-d_{\alpha}(p_{U,W},p_Up_W)\big),\label{EQ:soft_covering_exponent_beta1}\\
		\beta^{(2)}_{\alpha,\delta_2}&=\frac{\alpha-1}{2\alpha-1} \big(R_1+R_2-\delta_2-d_{\alpha}(p_{U,V,W},p_{U,V}p_W)\big),\label{EQ:soft_covering_exponent_beta2}\\			c_{\delta_1,\delta_2}&= 4\bigg(\log e+2\sup_{\alpha>1}\min\left\{\beta^{(1)}_{\alpha,\delta_1},\beta^{(2)}_{\alpha,\delta_2}\right\}\log2\bigg)+\log e+2 \log \left( \max_{w \in\supp(p_W)} \frac{1}{p_W(w)} \right),\label{EQ:soft_covering_coefficient1}
	\end{align}
\end{subequations}
and $d_{\alpha}(\mu,\nu)=\frac{1}{\alpha-1}\log_2\int d\mspace{2mu}\mu\left(\frac{d\mspace{2mu}\mu}{d\mspace{2mu}\nu}\right)^{1-\alpha}$ is the R\'{e}nyi divergence of order $\alpha$.
\end{comment}
\end{lemma}

The proof of the lemma is relegated to Appendix \ref{APPEN:soft_covering_proof}, where exact exponents of conversion can also be found.

\subsection{Strong Soft-Covering Implies Classic Soft-Covering}

The strong superposition SCL stated above implies the convergence to zero of the corresponding relative entropy's expected value \cite[Lemma 2]{Goldfeld_WTCII_semantic2015}. The expected value decay is used for SS analysis in the proof of Theorem \ref{TM:SDWTC_lower_bound}. For completeness, we next restate Lemma 2 from \cite{Goldfeld_WTCII_semantic2015}; the proof is omitted. 

\begin{lemma}[Stronger than Classic Soft-Covering]\label{LEMMA:soft_covering_stronger} Under the framework of the strong SCL for superposition codes from Subsection A, let $\gamma_1,\gamma_2>0$ be such that \eqref{EQ:soft_covering} holds for $n$ large enough. Then, for every such $n$ we have
\begin{equation}
\mathbb{E}_\mu\mathsf{D}\Big(P^{(\mathsf{B}_n)}_{\mathbf{W}}\Big|\Big|p_W^n\Big)\leq e^{-n\gamma_1}+n\log\left(\frac{1}{\mu_W}\right)e^{-e^{n\gamma_2}},\label{EQ:superpos_SCL_classic}
\end{equation}
where $\mu_W=\min_{w\in\supp(p_W)}p_W(w)>0$.
\end{lemma}

%%%%%%%%%%%%%%%%%%%%%%%%%%%%%%%%%%%%%%%%%%%%%%%%%%%%%%%%%%%%%%%%%%%%%%%%%%%%%%%%%%%%%%%%%%%%%%%%%%%%%%%%%%%%%%%%%%%
%%%%%%%%%%%%%%%%%%%%%%%%%%%%                                                         %%%%%%%%%%%%%%%%%%%%%%%%%%%%%%
%%%%%%%%%%%%%%%%%%%%%%%%%%%%       APPENDIX B - PROOF OF SUPERPOSITION SCL           %%%%%%%%%%%%%%%%%%%%%%%%%%%%%%
%%%%%%%%%%%%%%%%%%%%%%%%%%%%                                                         %%%%%%%%%%%%%%%%%%%%%%%%%%%%%%
%%%%%%%%%%%%%%%%%%%%%%%%%%%%%%%%%%%%%%%%%%%%%%%%%%%%%%%%%%%%%%%%%%%%%%%%%%%%%%%%%%%%%%%%%%%%%%%%%%%%%%%%%%%%%%%%%%%

\section{Proof of Lemma \ref{LEMMA:soft_covering}}\label{APPEN:soft_covering_proof}		
We state the proof in terms of arbitrary distributions (not necessarily discrete). When needed, we will specialize to the case where $\mathcal{W}$ is finite. For any fixed superposition codebook $\mathcal{B}_n$, let the Radon-Nikodym derivative of the induced distribution with respect to $p_W$ be denoted as
\begin{equation}
	\Delta_{\mathcal{B}_n}(\mathbf{w}) \triangleq \frac{d P^{(\mathcal{B}_n)}_{\mathbf{W}}}{d p_W^n}(\mathbf{w}).\label{EQ:RD_derivative}
\end{equation}
In the discrete case, $\Delta_{\mathcal{B}_n}$ is a ratio of PMFs. Accordingly, the relative entropy of interest, which is a function of the codebook $\mathcal{B}_n$, is given by
\begin{equation}
	\mathsf{D}\Big(P^{(\mathcal{B}_n)}_{\mathbf{W}}\Big|\Big|p_W^n\Big)=\int dP^{(\mathcal{B}_n)}_{\mathbf{W}}\log \Delta_{\mathcal{B}_n}.
\end{equation}

To describe the jointly-typical set over $u$-, $v$- and $w$-sequences, we first define information density $i_{p_{W|U}}:\mathcal{U}\times\mathcal{W}\to\mathbb{R}_+$ and $i_{p_{W|U,V}}:\mathcal{U}\times\mathcal{V}\times\mathcal{W}\to\mathbb{R}_+$ as
\begin{subequations}
	\begin{align}
		i_{p_{U,W}}(u,w)&\triangleq\log\left(\frac{dp_{W|U=u}}{dp_W}(w)\right)\label{EQ:information_density1}\\
		i_{p_{U,V,W}}(u,v,w)&\triangleq\log\left(\frac{dp_{W|U=u,V=v}}{dp_W}(w)\right).\label{EQ:information_density2}
	\end{align}\label{EQ:information_density}
\end{subequations}
In \eqref{EQ:information_density}, the arguments of the logarithms are the Radon-Nikodym derivatives of $p_{W|U=u}$ and $p_{W|U=u,V=v}$, respectively, with respect to $p_W$. Let $\epsilon_1,\epsilon_2\geq0$ be arbitrary, to be determined later, and define
\begin{equation}
	{\cal A}_{\epsilon_1,\epsilon_2} \triangleq \left\{ (\mathbf{u},\mathbf{v},\mathbf{w})\in\mathcal{U}^n\times\mathcal{V}^n\times\mathcal{W}^n \vast|\begin{array}{c}
		\frac{1}{n} i_{p^n_{U,W}}(\mathbf{u},\mathbf{w}) < I(U;W) + \epsilon_1\\
		\frac{1}{n} i_{p^n_{U,V,W}}(\mathbf{u},\mathbf{v},\mathbf{w}) < I(U,V;W) + \epsilon_2
	\end{array} \right\},\label{EQ:typical_set}
\end{equation}
and note that
\begin{subequations}
	\begin{align}
		i_{p^n_{U,W}}(\mathbf{u},\mathbf{w})&=\sum_{t=1}^ni_{p_{U,W}}(u_t,w_t)\label{EQ:information_density_product1}\\
		i_{p^n_{U,V,W}}(\mathbf{u},\mathbf{v},\mathbf{w})&=\sum_{t=1}^ni_{p_{U,V,W}}(u_t,v_t,w_t),\label{EQ:information_density_product2}
	\end{align}\label{EQ:information_density_product}
\end{subequations}
%    In expected value over the codebook distribution, $P_{V^n|{\cal C}}$ is unbiased with respect to the desired distribution:
%    \begin{align}
%        {\mathbf E} P_{V^n|{\cal C}} &= p_{V^n}.
%    \end{align}

We split $P^{(\mathcal{B}_n)}_{\mathbf{W}}$ into two parts, making use of the indicator function. For every $\mathbf{w}\in\mathcal{W}^n$, define
\begin{subequations}
	\begin{align}
		P_{\mathcal{B}_n,1}(\mathbf{v})&\triangleq 2^{-n(R_1+R_2)}\sum_{(i,j)\in\mathcal{I}_n\times\mathcal{J}_n}p_{W|U,V}^n\big(\mathbf{w}\big|\mathbf{u}(i),\mathbf{v}(i,j)\big)\mathds{1}_{\big\{\big(\mathbf{u}(i),\mathbf{v}(i,j),\mathbf{w}\big)\in\mathcal{A}_{\epsilon_1,\epsilon_2}\big\}}, \\
		P_{\mathcal{B}_n,2}(\mathbf{v})&\triangleq2^{-n(R_1+R_2)}\sum_{(i,j)\in\mathcal{I}_n\times\mathcal{J}_n}p_{W|U,V}^n\big(\mathbf{w}\big|\mathbf{u}(i),\mathbf{v}(i,j)\big)\mathds{1}_{\big\{\big(\mathbf{u}(i),\mathbf{v}(i,j),\mathbf{w}\big)\notin\mathcal{A}_{\epsilon_1,\epsilon_2}\big\}}.
	\end{align}
\end{subequations}
The measures $P_{\mathcal{B}_n,1}$ and $P_{\mathcal{B}_n,2}$ on the space $\mathcal{W}^n$ are not probability measures, but $P_{\mathcal{B}_n,1}+P_{\mathcal{B}_n,2}=P^{(\mathcal{B}_n)}_{\mathbf{W}}$ for each codebook $\mathcal{B}_n$. For every $\mathbf{w}\in\mathcal{W}^n$, we also define
\begin{equation}
	\Delta_{\mathcal{B}_n,j}(\mathbf{w})\triangleq\frac{dP_{\mathcal{B}_n,j}}{dp_W^n}(\mathbf{w}),\quad j=1,2.\label{EQ:delta_def}
\end{equation}
\indent With respect to the above definitions, Lemma \ref{LEMMA:soft_covering_UB} states an upper bound on the relative entropy of interest.
\begin{lemma}\label{LEMMA:soft_covering_UB}
	For every fixed superposition codebook $\mathcal{B}_n$, we have
	\begin{equation}
		\mathsf{D}\Big(P^{(\mathcal{B}_n)}_{\mathbf{W}}\Big|\Big|p_W^n\Big)\leq h\left(\int dP_{\mathcal{B}_n,1}\right)
		+ \int dP_{\mathcal{B}_n,1}\log\Delta_{\mathcal{B}_n,1}+\int dP_{\mathcal{B}_n,2}\log\Delta_{\mathcal{B}_n,2}, \numberthis\label{EQ:expanded divergence bound}
	\end{equation}
	where $h(\cdot)$ is the binary entropy function.
\end{lemma}
The proof of the lemma is omitted as it follows the same steps as in the proof of \cite[Lemma 3]{Goldfeld_WTCII_semantic2015} (see Appendix B therein for details). Based on Lemma \ref{LEMMA:soft_covering_UB}, to prove Lemma \ref{LEMMA:soft_covering} it suffices to show that the probability (with respect to a random superposition codebook) of the RHS of \eqref{EQ:expanded divergence bound} not vanishing exponentially fast to 0 as $n\to\infty$, is double-exponentially small.

%Based on Lemma \ref{LEMMA:soft_covering_UB}, if the relative entropy of interest does not decay exponentially fast, then the same is true for the terms on the RHS of \eqref{EQ:expanded divergence bound}. Therefore, to establish Lemma \ref{LEMMA:soft_covering}, its suffices to show that the probability (with respect to a random superposition codebook) of the RHS not vanishing exponentially fast to 0 as $n\to\infty$, is double-exponentially small.
%By Jensen's inequality (or the data processing inequality) we can upper bound the relative entropy of interest:

Notice that $P_{\mathcal{B}_n,1}$ usually contains almost all of the probability mass. That is, for any fixed $\mathcal{B}_n$, we have%denoting the complement of $\mathcal{A}_{\epsilon_1,\epsilon_2}$ by $\mathcal{A}_{\epsilon_1,\epsilon_2}^c$, we have
\begin{align*}
	\int dP_{\mathcal{B}_n,2}&=1-\int dP_{\mathcal{B}_n,1}\\&=2^{-n(R_1+R_2)}\sum_{(i,j)\in\mathcal{I}_n\times\mathcal{J}_n} \mathbb{P}_{p_{W|U,V}^n}\Big(\big(\mathbf{u}(i),\mathbf{v}(i,j,m),\mathbf{W}\big)\notin\mathcal{A}_{\epsilon_1,\epsilon_2}\Big|\mathbf{U}=\mathbf{u}(i),\mathbf{V}=\mathbf{v}(i,j)\Big).\numberthis\label{EQ:chernoff_pre_rvs1}
\end{align*}
For a random codebook, \eqref{EQ:chernoff_pre_rvs1} becomes
\begin{equation}
    \int dP_{\mathsf{B}_n,2}=2^{-n(R_1+R_2)}\sum_{(i,j)\in\mathcal{I}_n\times\mathcal{J}_n} \mathbb{P}_{p_{W|U,V}^n}\Big(\big(\mathbf{U}(i),\mathbf{V}(i,j),\mathbf{W}\big)\notin\mathcal{A}_{\epsilon_1,\epsilon_2}\Big|\mathbf{U}=\mathbf{U}(i),\mathbf{V}=\mathbf{V}(i,j)\Big),\label{EQ:chernoff_rvs1}
\end{equation}
where the RHS is an average of exponentially many i.i.d. random variables bounded between 0 and 1. Furthermore, the expected value of each one is the exponentially small probability of correlated sequences being atypical:
\begin{align*}
	&\mathbb{E}_\mu\mathbb{P}_{p_{W|U,V}^n}\Big(\big(\mathbf{U}(i),\mathbf{V}(i,j),\mathbf{W}\big)\notin\mathcal{A}_{\epsilon_1,\epsilon_2}\Big|\mathbf{U}=\mathbf{U}(i),\mathbf{V}=\mathbf{V}(i,j)\Big)\\&=\mathbb{P}_{p_{U,V,W}^n}\Big(\big(\mathbf{U},\mathbf{V},\mathbf{W}\big)\notin\mathcal{A}_{\epsilon_1,\epsilon_2}\Big)\\
	&= \mathbb{P}_{p_{U,V,W}^n} \left(\left\{\sum_{t=1}^n i_{p_{U,W}}(U_t,p_t) \geq n \big(I(U;W) + \epsilon_1\big)\right\}\bigcup\left\{\sum_{t=1}^n i_{p_{U,V,W}}(U_t,V_t,p_t) \geq n \big(I(U,V;W) + \epsilon_2\big)\right\} \right) \\
	&\leq \mathbb{P}_{p_{U,V,W}^n}\mspace{-2mu} \left( 2^{\lambda\sum_{t=1}^n i_{p_{U,W}}(U_t,p_t)} \mspace{-2mu}\geq \mspace{-2mu}2^{n\lambda(I(U;W) + \epsilon_1)}\right)\mspace{-4mu}+\mspace{-4mu}\mathbb{P}_{p_{U,V,W}^n}\mspace{-2mu} \left( 2^{\lambda\sum_{t=1}^n i_{p_{U,V,W}}(U_t,V_t,p_t)}\mspace{-2mu} \geq \mspace{-2mu}2^{n\lambda(I(U,V;W) + \epsilon_2)}\right)\numberthis\label{EQ:atypical_expectation_UB_step1},
\end{align*}
where the last inequality uses the union bound and is true for any $\lambda\geq0$. We further bound the two probability terms from the RHS of \eqref{EQ:atypical_expectation_UB_step1} by exponentially decaying functions of $n$ as follows. For the first term, consider:
\begin{align*}
	\mathbb{P}_{p_{U,V,W}^n} \left( 2^{\lambda\sum_{t=1}^n i_{p_{U,W}}(U_t,p_t)} \geq 2^{n\lambda(I(U;W) + \epsilon_1)}\right)&\stackrel{(a)}\leq \frac{\mathbb{E}_{p_{U,W}^n} 2^{\lambda \sum_{t=1}^ni_{p_{U,W}}(U_t,p_t)}}{2^{n\lambda (I(U;W) + \epsilon_1)}} \\
	&= \left( \frac{\mathbb{E}_{p_{U,W}} 2^{\lambda i_{p_{U,W}}(U,W)}}{2^{\lambda (I(U;W) + \epsilon_1)}} \right)^n \\
	&\stackrel{(b)}= 2^{n \lambda \left( \frac{1}{\lambda} \log_2 \mathbb{E}_{p_{U,W}} \big[2^{\lambda i_{p_{U,W}}(U;W)}\big] - I(U;W) - \epsilon_1 \right)} \\
	&\stackrel{(c)}= 2^{n \lambda \big( d_{\lambda+1}(p_{U,W},p_U p_W) - I(U;W) - \epsilon_1 \big)},\numberthis\label{EQ:atypical_expectation_UB1}
\end{align*}
where (a) is Markov's inequality, (b) follows by restricting $\lambda$ to be strictly positive, while (c) is from the definition of the R\'{e}nyi divergence of order $\lambda+1$. We use units of bits for mutual information and R\'{e}nyi divergence to coincide with the base two expression of rate. Similarly, the second term from the RHS of \eqref{EQ:atypical_expectation_UB_step1} is upper bounded by
\begin{equation}
	\mathbb{P}_{p_{U,V,W}^n} \left( 2^{\lambda\sum_{t=1}^n i_{p_{U,V,W}}(U_t,V_t,p_t)} \geq 2^{n\lambda(I(U,V;W) + \epsilon_2)}\right)\leq 2^{n \lambda \big( d_{\lambda+1}(p_{U,V,W},p_{U,V} p_W) - I(U,V;W) - \epsilon_2 \big)}.\label{EQ:atypical_expectation_UB2}
\end{equation}
			
Now, substituting $\alpha = \lambda+1$ into \eqref{EQ:atypical_expectation_UB1}-\eqref{EQ:atypical_expectation_UB2} gives
\begin{equation}
	\mathbb{E}_\mu\mathbb{P}_{p_{W|U,V}^n}\Big(\big(\mathbf{U}(i),\mathbf{V}(i,j),\mathbf{W}\big)\notin\mathcal{A}_{\epsilon_1,\epsilon_2}\Big|\mathbf{U}=\mathbf{U}(i),\mathbf{V}=\mathbf{V}(i,j)\Big)\leq 2^{-n\beta^{(1)}_{\alpha,\epsilon_1}}+2^{-n\beta^{(2)}_{\alpha,\epsilon_2}},\label{EQ:atypical probability expectation}
\end{equation}
where
\begin{subequations}
	\begin{align}
		\beta^{(1)}_{\alpha,\epsilon_1}&=(\alpha - 1) \big( I(U;W) + \epsilon_1 - d_{\alpha}(p_{U,W}, p_U p_W) \big),\label{EQ:chernoff_rvs1_prop_beta1}\\
		\beta^{(2)}_{\alpha,\epsilon_2}&=(\alpha - 1) \big( I(U,V;W) + \epsilon_2 - d_{\alpha}(p_{U,V,W}, p_{U,V} p_W) \big),\label{EQ:chernoff_rvs1_prop_beta2}
	\end{align}\label{EQ:chernoff_rvs1_prop}
\end{subequations}
\vspace{-6mm}

\noindent for every $\alpha>1$ and $\epsilon_1,\epsilon_2\geq 0$, over which we may optimize. The optimal choices of $\epsilon_1$ and $\epsilon_2$ are apparent when all bounds of the proof are considered together (some yet to be derived), but the formula may seem arbitrary at the moment.  Nevertheless, fix $\delta_1 \in \big(0,R_1-I(U;W)\big)$ and $\delta_2 \in \big(0,R_1+R_2-I(U,V;W)\big)$, as found in the theorem statement, and for any $\alpha>1$ set
\begin{subequations}
	\begin{align}
		\epsilon^{(1)}_{\alpha,\delta_1}&= \frac{ \frac{1}{2} (R_1-\delta_1) + (\alpha - 1) d_{\alpha} (p_{U,W},p_U p_W) }{ \frac{1}{2} + (\alpha- 1) } - I (U;W),\label{EQ:optimized_epsilon1}\\
		\epsilon^{(2)}_{\alpha,\delta_2}&= \frac{ \frac{1}{2} (R_1+R_2-\delta_2) + (\alpha - 1) d_{\alpha} (p_{U,V,W},p_{U,V} p_W) }{ \frac{1}{2} + (\alpha- 1) } - I (U,V;W).\label{EQ:optimized_epsilon2}
	\end{align}\label{EQ:optimized_epsilon}
\end{subequations}
Substituting into $\beta^{(1)}_{\alpha,\epsilon_1}$ and $\beta^{(2)}_{\alpha,\epsilon_2}$ gives
\begin{subequations}
	\begin{align}
		\beta^{(1)}_{\alpha,\delta_1}&\triangleq\beta^{(1)}_{\alpha,\epsilon^{(1)}_{\alpha,\delta_1}}=\frac{\alpha - 1}{2 \alpha - 1} \big(R_1-\delta_1 - d_{\alpha} (p_{U,W},p_U p_W)\big),\label{EQ:optimized_beta1}\\
		\beta^{(1)}_{\alpha,\delta_2}&\triangleq\beta^{(2)}_{\alpha,\epsilon^{(2)}_{\alpha,\delta_2}}=\frac{\alpha - 1}{2 \alpha - 1} \big(R_1+R_2-\delta_2 - d_{\alpha} (p_{U,V,W},p_{U,V} p_W)\big)\label{EQ:optimized_beta2}.
	\end{align}\label{EQ:optimized_beta}
\end{subequations}
Observe that $\epsilon^{(1)}_{\alpha,\delta_1}$ and $\epsilon^{(2)}_{\alpha,\delta_2}$ in \eqref{EQ:optimized_epsilon} are nonnegative. For example, $\epsilon^{(1)}_{\alpha,\delta_1}\geq 0$ due to the assumption that $R_1-\delta_1> I(U;W)$, because $\alpha>1$ and $d_{\alpha} (p_{U,W},p_W p_V) \geq d_{1} (p_{W,V},p_U p_W) = I(U;W)$.

Furthermore, the properties of R\'{e}nyi divergence imply the existence of an $\alpha>1$, for which 
\eqref{EQ:optimized_beta1} and \eqref{EQ:optimized_beta2} are strictly positive.

\begin{lemma}[Strictly Positive Exponents]\label{LEMMA:positive_exponents}
	There exists an $\alpha>1$ such that $\beta^{(j)}_{\alpha,\delta_j}> 0$, for $j=1,2$.
\end{lemma}
Lemma \ref{LEMMA:positive_exponents} is proven in Appendix \ref{APPEN:positive_exponents_proof} and shows that the RHS of \eqref{EQ:atypical probability expectation} can be made an exponentially decaying function of $n$. To bound the probability (with respect to a random superposition codebook) of \eqref{EQ:chernoff_rvs1} not producing this exponential decay, we use one of the Chernoff bounds stated in the following lemma.

\begin{lemma}[Chernoff Bound]\label{LEMMA:Chernoff}
	Let $\big\{X_m\big\}_{m=1}^M$ be a collection of i.i.d. random variables with $X_m\in[0,B]$ and $\mathbb{E}X_m\leq\mu\neq 0$ for all $m\in[1:M]$. Then, for any $c$ with $\frac{c}{\mu} \geq 1$
	\begin{subequations}
		\begin{equation}
			\mathbb{P} \left( \frac{1}{M} \sum_{m=1}^M X_m \geq c \right) \leq e^{-\frac{M \mu}{B} \Big( \frac{c}{\mu}\left(\ln\frac{c}{\mu} - 1 \right)+1\Big)}.\label{EQ:Chernoff1}
		\end{equation}
		Furthermore, if $\frac{c}{\mu}\in[1,2]$, then
		\begin{equation}
			\mathbb{P} \left( \frac{1}{M} \sum_{m=1}^M X_m \geq c \right) \leq e^{-\frac{M \mu}{3 B} \left( \frac{c}{\mu} - 1 \right)^2}.\label{EQ:Chernoff2}
		\end{equation}
	\end{subequations}
\end{lemma}
For the proof of the bounds see \cite[Appendix C]{Goldfeld_WTCII_semantic2015}. Having Lemma \ref{LEMMA:Chernoff}, we show that $\int d P_{\mathsf{B}_n,2}$ is exponentially small with a probability doubly-exponentially close to 1. To demonstrate this we exploit the fact that for any $j\in\mathcal{J}_n$, the structure of the superposition code implies that the collection $\big\{\big(\mathbf{U}(i),\mathbf{V}(i,j)\big)\big\}_{i\in\mathcal{I}_n}$ comprises i.i.d. pairs of random variables. Consequently, denoting
\begin{equation}
	f(\mathbf{u},\mathbf{v})\triangleq \mathbb{P}_{p^n_{W|U,V}}\Big((\mathbf{u},\mathbf{v},\mathbf{W})\notin\mathcal{A}_{\epsilon^{(1)}_{\alpha,\delta_1},\epsilon^{(2)}_{\alpha,\delta_2}}\Big|\mathbf{U}=\mathbf{u},\mathbf{V}=\mathbf{v}\Big),
\end{equation}
we have that $\big\{f\big(\mathbf{U}(i),\mathbf{V}(i,j)\big)\big\}_{i\in\mathcal{I}_n}$ are i.i.d. for any $j\in\mathcal{J}_n$, and that 
\begin{equation}
	\mathbb{E}_\mu f\big(\mathbf{U}(i),\mathbf{V}(i,j)\big)\leq2^{-n\beta^{(1)}_{\alpha,\delta_1}}+2^{-n\beta^{(2)}_{\alpha,\delta_2}},\quad\forall(i,j)\in\mathcal{I}_n\times\mathcal{J}_n.
\end{equation} 

For any $c\in\mathbb{R}_+$ consider now the following:
\begin{align*}
	\mathbb{P}_\mu\left(\int d P_{\mathsf{B}_n,2}\geq c\right)&=\mathbb{P}_\mu\left(2^{-n(R_1+R_2)}\sum_{(i,j)\in\mathcal{I}_n\times\mathcal{J}_n}f\big(\mathbf{U}(i),\mathbf{V}(i,j)\big)\geq c\right)\\
	&\leq\mathbb{P}_\mu\left(\bigcup_{j\in\mathcal{J}_n}\left\{2^{-n(R_1+R_2)}\sum_{i\in\mathcal{I}_n}f\big(\mathbf{U}(i),\mathbf{V}(i,j)\big)\geq c\cdot 2^{-nR_2}\right\}\right)\\
	&\leq\sum_{j\in\mathcal{J}_n}\mathbb{P}_\mu\left(2^{-nR_1}\sum_{i\in\mathcal{I}_n}f\big(\mathbf{U}(i),\mathbf{V}(i,j)\big)\geq c\right)\numberthis\label{EQ:atypical_chenoff_preliminary}
\end{align*}
where the last inequality is the union bound. Using \eqref{EQ:Chernoff2} on each of the summands from the RHS of \eqref{EQ:atypical_chenoff_preliminary} with $M = 2^{nR_1}$, $\mu =2^{-n\beta^{(1)}_{\alpha,\delta_1}}+2^{-n\beta^{(2)}_{\alpha,\delta_2}}$, $B=1$, and $\frac{c}{\mu} = 2$, gives
\begin{align*}
	\mathbb{P}_\mu\left(2^{-nR_1}\sum_{i\in\mathcal{I}_n}f\big(\mathbf{U}(i),\mathbf{V}(i,j)\big)\geq 2 \cdot \left(2^{-n\beta^{(1)}_{\alpha,\delta_1}}+2^{-n\beta^{(2)}_{\alpha,\delta_2}}\right) \right)&\leq e^{-\frac{1}{3}2^{nR_1}\left(2^{-n\beta^{(1)}_{\alpha,\delta_1}}+2^{-n\beta^{(2)}_{\alpha,\delta_2}}\right)}\\
	&\leq e^{-\frac{1}{3}2^{n\left(R_1-\beta^{(1)}_{\alpha,\delta_1}\right)}}.\numberthis\label{EQ:atypical_chernoff_perj}
\end{align*}
Inserting \eqref{EQ:atypical_chernoff_perj} into \eqref{EQ:atypical_chenoff_preliminary}, we have
\begin{equation}
	\mathbb{P}_\mu\left(\int d P_{\mathsf{B}_n,2}\geq 2 \cdot \left(2^{-n\beta^{(1)}_{\alpha,\delta_1}}+2^{-n\beta^{(2)}_{\alpha,\delta_2}}\right)\right)\leq 2^{nR_2}\cdot e^{-\frac{1}{3}2^{n\left(R_1-\beta^{(1)}_{\alpha,\delta_1}\right)}},\label{EQ:atypical_double_exp_bound}
\end{equation}
for which $\alpha>1$ can be chosen to produce a double-exponential convergence to 0 of the RHS because
\begin{equation}
	R_1-\beta^{(1)}_{\alpha,\delta_1}=\frac{\alpha R_1+(\alpha-1)\big(\delta_1+d_\alpha(p_{U,W},p_Up_W)\big)}{2\alpha-1}>0,\quad\forall\alpha>1.
\end{equation}

We next treat the random variables $\Delta_{\mathsf{B}_n,1}(\mathbf{w})$, where $\mathbf{w}\in\mathcal{W}^n$, and show that it also decays exponentially fast with a probability doubly-exponentially close to 1. To simplify notation, for each $\mathbf{w}\in\mathcal{W}^n$, let $g_\mathbf{w}:\mathcal{U}^n\times\mathcal{V}^n\to\mathbb{R}_+$ be a function specified by 
\begin{equation}
	g_\mathbf{w}(\mathbf{u},\mathbf{v})=\frac{d p_{W|U=\mathbf{u},V=\mathbf{v}}}{d p_W^n} (\mathbf{w}) \mathds{1}_{\left\{\big(\mathbf{u},\mathbf{v},\mathbf{w}\big)\in\mathcal{A}_{\epsilon^{(1)}_{\alpha,\delta_1},\epsilon^{(2)}_{\alpha,\delta_2}}\right\}}.
\end{equation}
Accordingly, note that
\begin{align*}
	\Delta_{\mathsf{B}_n,1}(\mathbf{w})=2^{-n(R_1+R_2)}\mspace{-14mu}\sum_{(i,j)\in\mathcal{I}_n\times\mathcal{J}_n}\mspace{-5mu}g_\mathbf{w}\big(\mathbf{U}(i),\mathbf{V}(i,j)\big)=2^{-nR_1}\sum_{i\in\mathcal{I}_n}\left[2^{-nR_2}\sum_{j\in\mathcal{J}_n}g_\mathbf{w}\big(\mathbf{U}(i),\mathbf{V}(i,j)\big)\right]\mspace{-3mu},\numberthis\label{EQ:chernoff_rvs2}
\end{align*}
where the RHS is an average of $2^{nR_1}$ i.i.d. random variables due to the structure of the superposition codebook. Next, for any $c'\in\mathbb{R}_+$ and $i\in\mathcal{I}_n$ define
\begin{subequations}
	\begin{equation}
		\mathcal{D}_i(c')=\left\{2^{-nR_2}\sum_{j\in\mathcal{J}_n}g_\mathbf{w}\big(\mathbf{U}(i),\mathbf{V}(i,j)\big)\geq c'\cdot 2^{n\left(I(U;W)+\epsilon^{(1)}_{\alpha,\delta_1}\right)}\right\},
	\end{equation}
	and set
	\begin{equation}
		\mathcal{D}(c')=\bigcup_{i\in\mathcal{I}_n}\mathcal{D}_i(c').
	\end{equation}
\end{subequations}
%Specializing to the case where $|\mathcal{U}|<\infty$, 
Consider the following upper bound on the probability that $\Delta_{\mathsf{B}_n,1}(\mathbf{w})$ is lower bounded by some constant $c\in\mathbb{R}_+$. For any $\mathbf{w}\in\mathcal{W}^n$, we have
\begin{align*}
	&\mathbb{P}_\mu\Big( \Delta_{\mathsf{B}_n,1}(\mathbf{w})\geq c\Big)\\
	&=\mathbb{P}_\mu\left(2^{-n(R_1+R_2)}\mspace{-14mu}\sum_{(i,j)\in\mathcal{I}_n\times\mathcal{J}_n}\mspace{-5mu}g_\mathbf{w}\big(\mathbf{U}(i),\mathbf{V}(i,j)\big) \geq c\right)\\
	&\leq\mathbb{P}_\mu\Big(\mathcal{D}(c')\Big)+\mathbb{P}_\mu\left(2^{-n(R_1+R_2)}\mspace{-14mu}\sum_{(i,j)\in\mathcal{I}_n\times\mathcal{J}_n}\mspace{-5mu}g_\mathbf{w}\big(\mathbf{U}(i),\mathbf{V}(i,j)\big) \geq c\ \vasti| \mathcal{D}(c')^c\right)\\
	&\begin{multlined}[b][.83\columnwidth]\leq \sum_{i\in\mathcal{I}_n}\mathbb{P}_\mu\left(2^{-nR_2}\sum_{j\in\mathcal{J}_n}g_\mathbf{w}\big(\mathbf{U}(i),\mathbf{V}(i,j)\big)\geq c'\cdot 2^{n\left(I(U;W)+\epsilon^{(1)}_{\alpha,\delta_1}\right)}\right)\\+\mathbb{P}_\mu\left(2^{-n(R_1+R_2)}\mspace{-14mu}\sum_{(i,j)\in\mathcal{I}_n\times\mathcal{J}_n}\mspace{-5mu}g_\mathbf{w}\big(\mathbf{U}(i),\mathbf{V}(i,j)\big) \geq c\ \vasti| \mathcal{D}(c')^c\right)\end{multlined}\\
	&\leq \sum_{i\in\mathcal{I}_n}\mspace{-5mu}\int\limits_{\mspace{8mu}\mathbf{u}\in\mathcal{U}^n}\mspace{-10mu}d\mspace{3mu}\mathbb{P}_\mu\Big(\mathbf{U}(i)=\mathbf{u}\Big)\underbrace{\mathbb{P}_\mu\mspace{-3mu}\left(2^{-nR_2}\mspace{-3mu}\sum_{j\in\mathcal{J}_n}g_\mathbf{w}\big(\mathbf{U}(i),\mathbf{V}(i,j)\big)\mspace{-3mu}\geq\mspace{-2mu} c'\mspace{-3mu}\cdot\mspace{-3mu} 2^{n\left(I(U;W)+\epsilon^{(1)}_{\alpha,\delta_1}\right)}\vasti|\mathbf{U}(i)\mspace{-3mu}=\mspace{-3mu}\mathbf{u}\right)}_{P_1(i,\mathbf{u})}\\
	&\mspace{220mu}+\underbrace{\mathbb{P}_\mu\left(2^{-nR_1}\sum_{i\in\mathcal{I}_n}\left[2^{-nR_2}\sum_{j\in\mathcal{J}_n}g_\mathbf{w}\big(\mathbf{U}(i),\mathbf{V}(i,j)\big)\right] \geq c\ \vast| \forall i\in\mathcal{I},\ \mathcal{D}_i(c')^c\right)}_{P_2}.\numberthis\label{EQ:typical_UB_preliminary}
\end{align*}
			
	To invoke the Chernoff bound from \eqref{EQ:Chernoff1} on $P_1(i,\mathbf{u})$, where $i\in\mathcal{I}_n$ and $\mathbf{u}\in\mathcal{U}^n$, first note that conditioned on $\mathbf{U}(i)=\mathbf{u}$, $\left\{g_\mathbf{w}\big(\mathbf{U}(i),\mathbf{V}(i,j)\big)\right\}_{j\in\mathcal{J}_n}$ are i.i.d. Furthermore, each random variable $g_\mathbf{w}\big(\mathbf{U}(i),\mathbf{V}(i,j)\big)$ is upper bounded by $2^{n\left(I(U,V;W)+\epsilon^{(2)}_{\alpha,\delta_2}\right)}$ with probability 1, and has an expectation that is upper bounded as
	\begin{align*}
		\mathbb{E}_\mu\Big[g_\mathbf{w}\big(\mathbf{U}(i,\mathsf{B}_U),\mathbf{V}(i,j)\big)\Big|\mathbf{U}(i)=\mathbf{u}\Big]&=\mathbb{E}_\mu\left[\frac{d p_{W|U=\mathbf{u},V=\mathbf{V}(i,j)}}{d p_W^n} (\mathbf{w}) \mathds{1}_{\left\{\big(\mathbf{u},\mathbf{V}(i,j),\mathbf{w}\big)\in\mathcal{A}_{\epsilon^{(1)}_{\alpha,\delta_1},\epsilon^{(2)}_{\alpha,\delta_2}}\right\}}\vast |\mathbf{U}(i)=\mathbf{u}\right]\\
		&\leq\mathds{1}_{\left\{\frac{dp^n_{W|U=\mathbf{u}}}{dp^n_W}(\mathbf{w})\leq 2^{n\left(I(U;W)+\epsilon^{(1)}_{\alpha,\delta_1}\right)}\right\}}\frac{dp_{W|U=\mathbf{u}}}{d p_W^n} (\mathbf{w})\\
		&\leq 2^{n\left(I(U;W)+\epsilon^{(1)}_{\alpha,\delta_1}\right)}.\numberthis
\end{align*}
Using \eqref{EQ:Chernoff1} with  $M = 2^{nR_2}$, $\mu =2^{n\left(I(U;W)+\epsilon^{(1)}_{\alpha,\delta_1}\right)}$, $B=2^{n\left(I(U,V;W)+\epsilon^{(2)}_{\alpha,\delta_2}\right)}$, and $c=c'\cdot\mu$, for any $c'\geq \frac{1}{\mu}$, gives
\begin{equation}
	P_1(i,\mathbf{u})\leq e^{-2^{n\left(R_2-I(V;W|U)+\epsilon^{(1)}_{\alpha,\delta_1}-\epsilon^{(2)}_{\alpha,\delta_2}\right)}\big(c'(\ln c'-1)+1\big)},\quad \forall(i,\mathbf{u})\in\mathcal{I}_n\times\mathcal{U}^n.\label{EQ:P1iu_Chenoff_UB}
\end{equation}

Next, for $P_2$ we have that $\left\{2^{-nR_2}\sum_{j\in\mathcal{J}_n}g_\mathbf{w}\big(\mathbf{U}(i),\mathbf{V}(i,j)\big)\right\}_{i\in\mathcal{I}_n}$ are i.i.d. by the codebook construction. The conditioning on $\mathcal{D}(c')^c$ implies that each random variable $2^{-nR_2}\sum_{j\in\mathcal{J}_n}g_\mathbf{w}\big(\mathbf{U}(i),\mathbf{V}(i,j)\big)$, for $i\in\mathcal{I}_n$, is bounded between 0 and $c'\cdot 2^{n\left(I(U;W)+\epsilon^{(1)}_{\alpha,\delta_1}\right)}$ with probability 1. The expected value of each term with respect to the codebook is bounded above by one, which is observed by removing the indicator function from $g_\mathbf{w}\big(\mathbf{U}(i),\mathbf{V}(i,j)\big)$. Setting $M = 2^{nR_1}$, $\mu = 1$, $B = 2^{n\left(I(U;W) + \epsilon^{(1)}_{\alpha,\delta_1}\right)}$, and any $c\in[1,2]$ into \eqref{EQ:Chernoff2}, gives
\begin{equation}
	P_2\leq e^{-\frac{1}{3}2^{n\left(R_1-I(U;W) -\epsilon^{(1)}_{\alpha,\delta_1}\right)}(c-1)^2}.\label{EQ:P2_Chenoff_UB}
\end{equation}

Inserting \eqref{EQ:P1iu_Chenoff_UB} and \eqref{EQ:P2_Chenoff_UB} into \eqref{EQ:typical_UB_preliminary}, we have that for any $\mathbf{w}\in\mathcal{W}^n$, $c\in[1,2]$ and $c'\geq 2^{-n\left(I(U;W)+\epsilon^{(1)}_{\alpha,\delta_1}\right)}$
\begin{equation}
	\mathbb{P}_\mu\Big( \Delta_{\mathsf{B}_n,1}(\mathbf{w})\geq c\Big)\leq 2^{nR_1}e^{-2^{n\left(R_2-I(V;W|U)+\epsilon^{(1)}_{\alpha,\delta_1}-\epsilon^{(2)}_{\alpha,\delta_2}\right)}\big(c'(\ln c'-1)+1\big)}+e^{-\frac{1}{3}2^{n\left(R_1-I(U;W) -\epsilon^{(1)}_{\alpha,\delta_1}\right)}\frac{(c-1)^2}{c'}}.\label{EQ:typical_UB_constants}
\end{equation} 
Our next step is to choose $c$ and $c'$ to get a doubly-exponentially decaying function on the RHS of \eqref{EQ:typical_UB_constants}. Let
\begin{equation}
   c'=2^{n\left(I(V;W|U)-R_2-\epsilon^{(1)}_{\alpha,\delta_1}+\epsilon^{(2)}_{\alpha,\delta_2}+2\beta^{(2)}_{\alpha,\delta_2}+\frac{\delta_2}{2}\right)}-1,\label{EQ:c'_choice}
\end{equation}
and note that the exponent is strictly positive since
\begin{align*}
	I(V;W|U)-R_2-\epsilon^{(1)}_{\alpha,\delta_1}+\epsilon^{(2)}_{\alpha,\delta_2}+2\beta^{(2)}_{\alpha,\delta_2}+\frac{\delta_2}{2}&\stackrel{(a)}=R_1-I(U;W)-\frac{\delta_2}{2}-\epsilon^{(1)}_{\alpha,\delta_1}\\
	&=\frac{2(\alpha-1)\Big[R_1-d_\alpha(p_{U,W},p_Up_W)-\delta_1\Big]+\frac{2\alpha-1}{2}(2\delta_1-\delta_2)}{2\alpha-1}\\
	&>0
\end{align*}
where (a) is because $\epsilon^{(2)}_{\alpha,\delta_2}+2\beta^{(2)}_{\alpha,\delta_2}=R_1+R_2-I(U,V;W)-\delta_2$ and the positivity is by the choice of $\alpha$ from Lemma \ref{LEMMA:positive_exponents} and since $\delta_2<2\delta_1$. Consequently, $c'\to\infty$ as $n\to\infty$, and, therefore, $c'\geq 2^{-n\left(I(U;W)+\epsilon^{(1)}_{\alpha,\delta_1}\right)}$ for sufficiently large $n$. Since $c'$ is unbounded (as a function of $n$), for $n$ large enough we also have $\ln c' -1\geq 1$, which simplifies the RHS of \eqref{EQ:typical_UB_constants} as
\begin{align*}
   2^{nR_1}e^{-2^{n\left(R_2-I(V;W|U)+\epsilon^{(1)}_{\alpha,\delta_1}-\epsilon^{(2)}_{\alpha,\delta_2}\right)}\big(c'(\ln c'-1)+1\big)}&\leq 2^{nR_1}e^{-2^{n\left(R_2-I(V;W|U)+\epsilon^{(1)}_{\alpha,\delta_1}-\epsilon^{(2)}_{\alpha,\delta_2}\right)}(c'+1)}\\
   &=2^{nR_1}e^{-2^{n\left(2\beta^{(2)}_{\alpha,\delta_2}+\frac{\delta_2}{2}\right)}},\numberthis\label{EQ:typical_term1_doubleUB}
\end{align*}
which decays doubly-exponentially quickly to 0.

Setting $c=1+2^{-n\frac{\delta_1}{4}}$, we upper bound the second term from the RHS of \eqref{EQ:typical_UB_constants} by
\begin{equation}
   e^{-\frac{1}{3}2^{n\left(R_1-I(U;W) -\epsilon^{(1)}_{\alpha,\delta_1}\right)}\frac{(c-1)^2}{c'}}\leq e^{-\frac{1}{3}2^{n\left(R_1-I(U;W) -\epsilon^{(1)}_{\alpha,\delta_1}\right)}\frac{(c-1)^2}{(c'+1)}}= e^{-\frac{1}{3}2^{n\frac{\delta_2-\delta_1}{2}}},\label{EQ:typical_term2_doubleUB}
\end{equation}
which also converges to 0 with double-exponential speed because $\delta_1<\delta_2$.

Concluding, \eqref{EQ:typical_UB_constants}, \eqref{EQ:typical_term1_doubleUB} and \eqref{EQ:typical_term2_doubleUB} upper bound the probability of interest as
\begin{equation}
    \mathbb{P}_\mu\Big( \Delta_{\mathsf{B}_n,1}(\mathbf{w})\geq 1+2^{-n\frac{\delta_1}{4}}\Big)\leq 2^{nR_1}e^{-2^{n\left(2\beta^{(2)}_{\alpha,\delta_2}+\frac{\delta_2}{2}\right)}} +  e^{-\frac{1}{3}2^{n\frac{\delta_2-\delta_1}{2}}}.\label{EQ:typical_UB_double_final}
\end{equation}

At this point, we specialize to $\mathcal{W}$ being a finite set. Consequently, $\Delta_{\mathsf{B}_n,2}$ is bounded as
\begin{equation}
      \Delta_{\mathsf{B}_n,2}(\mathbf{w}) \leq \left( \max_{w \in \supp(p_W)} \frac{1}{p_W(w)} \right)^n,\ \forall\mspace{3mu} \mathbf{w}\in\mathcal{W}^n,\label{EQ:atypical divergence bound}
\end{equation}
with probability 1. Notice that the maximum is only over the support of $p_W$, which makes this bound finite. The underlying reason for this restriction is that with probability one a conditional distribution is absolutely continuous with respect to any of its associated marginal distributions.

Having \eqref{EQ:atypical_double_exp_bound}, \eqref{EQ:typical_UB_double_final} and \eqref{EQ:atypical divergence bound}, we can now bound the probability that the RHS of \eqref{EQ:expanded divergence bound} is not exponentially small. Let $\mathcal{S}$ be the set of superposition codebooks $\mathcal{B}_n\in\mathfrak{B}_n$, such that all of the following are true:
\begin{subequations}
	\begin{align}
		\int d P_{\mathcal{B}_n,2}&<2 \cdot \left(2^{-n\beta^{(1)}_{\alpha,\delta_1}}+2^{-n\beta^{(2)}_{\alpha,\delta_2}}\right),\label{EQ:codebook_S1}\\	\Delta_{\mathcal{B}_n,1}(\mathbf{w})&<1+2^{-n\frac{\delta_1}{4}},\quad\forall\mathbf{w}\in\mathcal{W}^n,\label{EQ:codebook_S2}\\		\Delta_{\mathcal{B}_n,2}(\mathbf{w})&\leq\left(\max_{w\in\supp(p_W)}\frac{1}{p_W(w)} \right)^n,\quad\forall\mathbf{w}\in\mathcal{W}^n.\label{EQ:codebook_S3}
	\end{align}
\end{subequations}
First, we use the union bound, while taking advantage of the fact that the space $\mathcal{W}^n$ is only exponentially large, to show that the probability of a random codebook not being in $\mathcal{S}$ is double-exponentially small:
\begin{align*}
	\mathbb{P}_\mu\big(\mathsf{B}_n\notin \mathcal{S}\big)&\stackrel{(a)}\leq\mathbb{P}_\mu\bigg(\int dP_{\mathsf{B}_n,2}\geq 2\cdot2^{-n\beta_{\alpha,\delta}}\bigg)+\sum_{\mathbf{w}\in\mathcal{W}^n}\mathbb{P}_\mu\bigg(\Delta_{\mathsf{B}_n,1}(\mathbf{w})\geq1+2^{-\beta_{\alpha,\delta}  n}\bigg)\\&\mspace{330mu}+\sum_{\mathbf{w}\in\mathcal{W}^n}\mathbb{P}_\mu\Bigg(\Delta_{\mathsf{B}_n,2}(\mathbf{w})>\left(\max_{w\in\supp(p_W)}\frac{1}{p_W(w)} \right)^n\Bigg)\\
	&\stackrel{(b)}\leq 2^{nR_2}\cdot e^{-\frac{1}{3}2^{n\left(R_1-\beta^{(1)}_{\alpha,\delta_1}\right)}} + |\mathcal{W}|^n\left[2^{nR_1}e^{-2^{n\left(2\beta^{(2)}_{\alpha,\delta_2}+\frac{\delta_2}{2}\right)}} +  e^{-\frac{1}{3}2^{n\frac{\delta_2-\delta_1}{2}}}\right],\numberthis\label{EQ:notin_S_probability_UB}
\end{align*}
where (a) is the union bound, and (b) uses \eqref{EQ:atypical_double_exp_bound}, \eqref{EQ:typical_UB_double_final} and \eqref{EQ:atypical divergence bound}.
		
Next, we claim that for every codebook in $\mathcal{S}$, the RHS of (\ref{EQ:expanded divergence bound}) is exponentially small. Let $\mathcal{B}_n\in\mathcal{S}$ and consider the following. For every $x\in[0,1]$, $h(x) \leq x \log \frac{e}{x}$, which, using \eqref{EQ:codebook_S1}, implies that
\begin{align*}
	h\left( \int dP_{\mathcal{B}_n,1} \right)& = h\left( \int d P_{\mathcal{B}_n,2} \right)\\& < 2\left[\log e-\log2\cdot\log\left(2^{-n\beta^{(1)}_{\alpha,\delta_1}}+2^{-n\beta^{(2)}_{\alpha,\delta_2}}\right)\right]\left(2^{-n\beta^{(1)}_{\alpha,\delta_1}}+2^{-n\beta^{(2)}_{\alpha,\delta_2}}\right)\\
	&\stackrel{(a)}\leq 4\big(\log e+2\beta_{\alpha,\delta_1,\delta_2}\log2\big)n2^{-n\beta_{\alpha,\delta_1,\delta_2}},\numberthis\label{EQ:codebook_S_UB1}
\end{align*}
where (a) follows by setting $\beta_{\alpha,\delta_1,\delta_2}\triangleq\min\big\{\beta^{(1)}_{\alpha,\delta_1},\beta^{(2)}_{\alpha,\delta_2}\big\}$. Furthermore, by \eqref{EQ:codebook_S2}, we have
\begin{equation}
	\int d P_{\mathcal{B}_n,1} \log \Delta_{\mathcal{B}_n,1}< \int dP_{\mathcal{B}_n,1}\log \left(1+2^{-n\frac{\delta_1}{4}}\right)= \log \left(1+2^{-n\frac{\delta_1}{4}}\right)\stackrel{(a)}\leq 2^{-n\frac{\delta_1}{4}}\log e,\numberthis\label{EQ:codebook_S_UB2}
\end{equation}
where (a) is since $\log(1+x)\leq x\log e$, for every $x>0$. Finally, using \eqref{EQ:codebook_S3} and the definition of $\beta_{\alpha,\delta_1,\delta_2}$, we obtain
\begin{equation}
	\int d P_{\mathcal{B}_n,2} \log \Delta_{\mathcal{B}_n,2}\leq \int d P_{\mathcal{B}_n,2} \log \left( \max_{w \in\supp(p_W)} \frac{1}{p_W(w)} \right)^n< 2 \log \left( \max_{w \in\supp(p_W)} \frac{1}{p_W(w)} \right) n2^{-n\beta_{\alpha,\delta_1,\delta_2}}.\label{EQ:codebook_S_UB3}
\end{equation}

Combining \eqref{EQ:codebook_S_UB1}-\eqref{EQ:codebook_S_UB3}, while setting $\gamma_{\alpha,\delta_1,\delta_2}\triangleq\min\left\{\beta_{\alpha,\delta_1,\delta_2},\frac{\delta_1}{4}\right\}$, yields
\begin{align*}
	h\left(\int dP_{\mathcal{B}_n,1}\right)&+\int dP_{\mathcal{B}_n,1}\log\Delta_{\mathcal{B}_n,1}+\int dP_{\mathcal{B}_n,2}\log\Delta_{\mathcal{B}_n,2}\\
	&<\Bigg(4\big(\log e+2\beta_{\alpha,\delta_1,\delta_2}\log2\big)+\log e+2 \log \left( \max_{w \in\supp(p_W)} \frac{1}{p_W(w)} \right)\Bigg)n2^{-n\gamma_{\alpha,\delta_1,\delta_2}}\\
	&\stackrel{(a)}=c_{\alpha,\delta_1,\delta_2}n2^{-n\gamma_{\alpha,\delta_1,\delta_2}},\numberthis
\end{align*}
where (a) comes from setting 
\begin{equation}
	c_{\alpha,\delta_1,\delta_2}\triangleq 4\big(\log e+2\beta_{\alpha,\delta_1,\delta_2}\log2\big)+\log e+2 \log \left( \max_{w \in\supp(p_W)} \frac{1}{p_W(w)} \right).\label{EQ:c_alphadelta}
\end{equation}
This implies that
\begin{align*}
	\mathbb{P}_\mu\bigg(\mathsf{D}\Big(P^{(\mathsf{B}_n)}_{\mathbf{W}}\Big|\Big|p_W^n\Big)&\geq c_{\alpha,\delta_1,\delta_2}n2^{-n\gamma_{\alpha,\delta_1,\delta_2}}\bigg)\\
	&\leq\mathbb{P}_\mu\Bigg(h\left(\int dP_{\mathsf{B}_n,1}\right)+\int dP_{\mathsf{B}_n,1}\log\Delta_{\mathsf{B}_n,1}+\int dP_{\mathsf{B}_n,2}\log\Delta_{\mathsf{B}_n,2}\geq c_{\alpha,\delta}n2^{-n\beta_{\alpha,\delta}}\Bigg)\\
	&\leq \mathbb{P}_\mu\big(\mathsf{B}_n\notin\mathcal{S}\big)\\
	&\stackrel{(a)}\leq 2^{nR_2}\cdot e^{-\frac{1}{3}2^{n\left(R_1-\beta^{(1)}_{\alpha,\delta_1}\right)}} + |\mathcal{W}|^n\left[2^{nR_1}e^{-2^{n\left(2\beta^{(2)}_{\alpha,\delta_2}+\frac{\delta_2}{2}\right)}} +  e^{-\frac{1}{3}2^{n\frac{\delta_2-\delta_1}{2}}}\right]\\
	&\stackrel{(b)}\leq 2^{nR_2}\cdot e^{-\frac{1}{3}2^{n\delta_1}} + |\mathcal{W}|^n\left[2^{nR_1}e^{-2^{n\frac{\delta_2}{2}}} +  e^{-\frac{1}{3}2^{n\frac{\delta_2-\delta_1}{2}}}\right],\numberthis\label{EQ:bad_divergence_final_UB}
\end{align*}
where (a) follows from \eqref{EQ:notin_S_probability_UB}, while (b) is because $\beta^{(1)}_{\alpha,\delta_1}\leq\frac{1}{2}(R_1-\delta_1)$ and $\beta^{(2)}_{\alpha,\delta_2}\geq 0$. Denoting $c_{\delta_1,\delta_2}\triangleq \sup_{\alpha> 1}c_{\alpha,\delta_1,\delta_2}$, \eqref{EQ:bad_divergence_final_UB} further gives
\begin{equation}
	\mathbb{P}_\mu\bigg(\mathsf{D}\Big(P^{(\mathsf{B}_n)}_{\mathbf{W}}\Big|\Big|p_W^n\Big)\geq c_{\delta_1,\delta_2} n2^{-n\gamma_{\alpha,\delta_1,\delta_2}}\bigg)\leq 2^{nR_2}\cdot e^{-\frac{1}{3}2^{n\delta_1}} + |\mathcal{W}|^n\left[2^{nR_1}e^{-2^{n\frac{\delta_2}{2}}} +  e^{-\frac{1}{3}2^{n\frac{\delta_2-\delta_1}{2}}}\right].\label{EQ:bad_divergence_final_UB_final}
\end{equation}
Since \eqref{EQ:bad_divergence_final_UB_final} holds for all $\alpha>1$ (the interesting values of $\alpha$ are those from Lemma \ref{LEMMA:positive_exponents}, but the derivation is valid for all $\alpha>1$), it must also be true, with strict inequality in the LHS, when replacing $\gamma_{\alpha,\delta_1,\delta_2}$ with $\gamma_{\delta_1,\delta_2}\triangleq\sup_{\alpha> 1}\gamma_{\alpha,\delta_1,\delta_2}$, which is the exponential rate of convergence we derive for the strong SCL for superposition codes. 

Concluding, if $R_1 > I(U;W)$, $R_1+R_2>I(U,V;W)$, then for any $\delta_1\in\big(0,R_1-I(U;W)\big)$ and $\delta_2\in\big(0,R_1+R_2-I(U,V;W)\big)$ with $\delta_1<\delta_2<2\delta_1$ we get exponential convergence of the relative entropy at rate $O(2^{-n\gamma_{\delta_1,\delta_2}})$ with doubly-exponential certainty. Discarding the precise exponents of convergence and coefficients, we state that there exist $\gamma_1,\gamma_2>0$, such that, for $n$ large enough,
\begin{equation}
	\mathbb{P}_\mu\bigg(\mathsf{D}\Big(P^{(\mathsf{B}_n)}_{\mathbf{W}}\Big|\Big|p_W^n\Big)>e^{-n\gamma_1}\bigg)\leq e^{-e^{n\gamma_2}},
\end{equation}
as needed. 

%%%%%%%%%%%%%%%%%%%%%%%%%%%%%%%%%%%%%%%%%%%%%%%%%%%%%%%%%%%%%%%%%%%%%%%%%%%%%%%%%%%%%%%%%%%%%%%%%%%%%%%%%%%%%%%%%%%
%%%%%%%%%%%%%%%%%%%%%%%%%             APPENDIX C - POSITIVE EXPONENTS             %%%%%%%%%%%%%%%%%%%%%%%%%%%%%%%%%
%%%%%%%%%%%%%%%%%%%%%%%%%%%%%%%%%%%%%%%%%%%%%%%%%%%%%%%%%%%%%%%%%%%%%%%%%%%%%%%%%%%%%%%%%%%%%%%%%%%%%%%%%%%%%%%%%%%
	
\section{Proof of Lemma \ref{LEMMA:positive_exponents}}\label{APPEN:positive_exponents_proof}
	
The proof uses several basic properties of R\'{e}nyi divergence (see, e.g., \cite{vanEvren_Reyni_Div2014}). First, recall that for fixed measures $\mu$ and $\nu$, $d_{\alpha}(\mu,\nu)$ is monotone non-decreasing in $\alpha$. Furthermore, if $\mu\ll\nu$ then $d_{\alpha}(\mu,\nu)$ is continuous in $\alpha\in(1,\infty]$. Since a joint PMF is always absolutely continuous with respect to the product of its marginals and by the choices of $\delta_1$ and $\delta_2$, there exists $\alpha_1,\alpha_2>1$ such that
\begin{subequations}
    \begin{align}
		R_1-\delta_1&>d_{\alpha_1}(Q_{U,W},Q_U,Q_W)\geq d_1(Q_{U,W},Q_U,Q_W)= I(U;W),\\
		R_1+R_2-\delta_2&>d_{\alpha_2}(Q_{U,V,W},Q_{U,V},Q_W)\geq d_1(Q_{U,V,W},Q_{U,V},Q_W)= I(U,V;W).
	\end{align}\label{EQ:alpha1_alpha2}
\end{subequations}
On account of \eqref{EQ:alpha1_alpha2}, by setting $\alpha=\min\{\alpha_1,\alpha_2\}$, we conclude that $\beta^{(j)}_{\alpha,\delta_j}> 0$, for $j=1,2$.

%%%%%%%%%%%%%%%%%%%%%%%%%%%%%%%%%%%%%%%%%%%%%%%%%%%%%%%%%%%%%%%%%%%%%%%%%%%%%%%%%%%%%%%%%%%%%%%%%%%%%%%%%%%%%%%%%%%
%%%%%%%%%%%%%%%%%%%%%               APPENDIX D - ALTERNATIVE ACHIEVABILITY               %%%%%%%%%%%%%%%%%%%%%%%%%%
%%%%%%%%%%%%%%%%%%%%%%%%%%%%%%%%%%%%%%%%%%%%%%%%%%%%%%%%%%%%%%%%%%%%%%%%%%%%%%%%%%%%%%%%%%%%%%%%%%%%%%%%%%%%%%%%%%%
	
\section{Proof of Proposition \ref{PROP:alternative_rate}}\label{APPEN:alternative_achievability_proof}

For the first direction, i.e., that $R_\mathsf{A}\leq R_\mathsf{A}^\mathsf{Alt}$, note that the two first rate bounds in $R_\mathsf{A}^\mathsf{Alt}$ (see \eqref{EQ:SDWTC_lower_bound_prob}) are the same as those defining $R_\mathsf{A}$, while the third bound in $R_\mathsf{A}^\mathsf{Alt}$ is obtained by adding the first bound from $R_\mathsf{A}$ and the quantity $I(U;Y)-I(U;S)$, which we know is non-negative by \eqref{EQ:SDWTC_capacity_alt_lower_bound}. 

For the opposite direction consider the following. Let $p_{U,V,X|S}^\star:\mathcal{S}\to\mathcal{P}(\mathcal{U}\times\mathcal{V}\times\mathcal{X})$ be such that $R_\mathsf{A}^\mathsf{Alt}=R_\mathsf{A}^\mathsf{Alt}(p^\star_{U,V,X|S})>0$, i.e., $R_\mathsf{A}^\mathsf{Alt}$ is strictly positive (otherwise there is nothing to prove) and it is achieved by the input distribution $p_{U,V,X|S}^\star$. Recall that the mutual information terms in $R_\mathsf{A}^\mathsf{Alt}(p^\star_{U,V,X|S})$ are taken with respect to $p^\star\triangleq p_Sp^\star_{U,V,X|S}p_{Y,Z|X,S}$. First, note that if $p^\star_{U,V,X|S}$ is such that $I(U;Y)- I(U;S)\geq 0$, then $R_\mathsf{A}^\mathsf{Alt}\leq R_\mathsf{A}(p^\star_{U,V,X|S})= R_\mathsf{A}$ and the desired inequality holds.

Otherwise, i.e., if $p^\star_{U,V,X|S}$ induces $I(U;Y)- I(U;S)<0$, and let $U'=(U,\tilde{V})$ and $V'=V$, where $\tilde{V}$ is $V$ passed through an erasure channel, with erasures independent of all the other random variables. Denoting the probability of an erasure by $\epsilon\in[0,1]$, the joint distribution of $(S,U,V,X,Y,Z,\tilde{V},U',V')$ is given by
\begin{equation}
    p_{S,U,V,X,Y,Z,\tilde{V},U',V'}=p_Sp^\star_{U,V,X|S}p_{Y,Z|X,S}p_{\tilde{V}|V}\mathds{1}_{\big\{U'=(U,\tilde{V}),V'=V\big\}},\label{EQ:alt_proof_dist}
\end{equation}
where $p_{\tilde{V}|V}:\mathcal{V}\to\mathcal{V}\cup\{?\}$ with $?\notin\mathcal{V}$, is the transition probability of a $\mathsf{BEC}(\epsilon)$. The exact value of $\epsilon$ is to be specified later. All subsequent information measures in this proof are taken with respect to the distribution from \eqref{EQ:alt_proof_dist} or its appropriate marginals. 

We first show that by a proper choice of $\epsilon\in[0,1]$, the conditional marginal distribution $p_{U',V',X|S}$ is a valid input distribution in $R_\mathsf{A}$, i.e., that it satisfies
\begin{equation}
    I(U';Y)-I(U',S)\geq 0.\label{EQ:alt_proof_constraint}
\end{equation}
Consider
\begin{align*}
I(U';Y)-I(U';S)&=I(U;Y)-I(U;S)+I(\tilde{V};Y|U)-I(\tilde{V};S|U)\\
               &=I(U;Y)-I(U;S)+\bar{\epsilon}\Big[I(V;Y|U)-I(V;S|U)\Big],\numberthis\label{EQ:alt_proof_ineq1_p}
\end{align*}
where $\bar{\epsilon}=1-\epsilon$. Notice that when $\epsilon=1$ this quantity is negative by assumption, while $\epsilon=0$ gives 
\begin{equation}
    I(U';Y)-I(U';S)=I(U,V;Y)-I(U,V;S)>0
\end{equation}
by the second rate bound in $R_\mathsf{A}^\mathsf{Alt}$. We set $\epsilon\in[0,1]$ at the value that produces $I(U';Y)-I(U';S)=0$, thus satisfying \eqref{EQ:alt_proof_constraint}.  

Being an appropriate input distribution in $R_\mathsf{A}$, we next evaluate $R_\mathsf{A}(p_{U',V',X|S})$. Starting from the  second one rate bound, we have
\begin{equation}
I(U',V';Y)-I(U',V';S)=I(U,V,\tilde{V};Y)- I(U,V,\tilde{V};S)\stackrel{(a)}=I(U,V;Y)-I(U,V;S)\geq R_\mathsf{A}^\mathsf{Alt},\label{EQ:alt_proof_ineq1}
\end{equation}
where (a) uses the Markov chain $(S,U,X,Y,Z)-V-\tilde{V}$, which follows because $\tilde{V}$ is a noisy version of $V$.

For the first rate bound, note that
\begin{align*}
I(V';Y|U')-I(V';Z|U')&=I(V;Y|U,\tilde{V})-I(V;Z|U,\tilde{V})\\
                     &\stackrel{(a)}=I(V;Y|U)-I(V;Z|U)-\Big[I(\tilde{V};Y|U)-I(\tilde{V};Z|U)\Big]\\
                     &\stackrel{(b)}=I(V;Y|U)-I(V;Z|U)-\bar{\epsilon}\Big[I(V;Y|U) - I(V;Z|U)\Big]\\
                     &=\epsilon\Big[I(V;Y|U)-I(V;Z|U)\Big],\numberthis\label{EQ:alt_proof_justification1}
\end{align*}
where, as before, (a) and (b) follow by Markovity. A similar derivation also gives
\begin{equation}
I(V';Y|U')-I(V';S|U')=\epsilon\Big[I(V;Y|U)-I(V;S|U)\Big].\label{EQ:alt_proof_justification2}
\end{equation}

We complete the proof by considering two cases. First, if $I(V;S|U) \geq I(V;Z|U)$, we obtain
\begin{align*}
I(V';Y|U')-I(V';Z|U')&\stackrel{(a)}=\epsilon\Big[I(V;Y|U) - I(V;Z|U)\Big]\\
                     &\stackrel{(b)}\geq \epsilon\Big[I(V;Y|U) - I(V;S|U)\Big]\\
                     &\stackrel{(c)}= I(V';Y|U')-I(V';S|U')\\
                     &\stackrel{(d)}= I(U',V';Y')-I(U',V';S)\\
                     &\stackrel{(e)}\geq R_\mathsf{A}^\mathsf{Alt},\numberthis
\end{align*}
where (a) is \eqref{EQ:alt_proof_justification1}, (b) follows by the assumption that $I(V;S|U) \geq I(V;Z|U)$, (c) is \eqref{EQ:alt_proof_justification2}, (d) is by choosing $\epsilon$ to satisfy $I(U';Y)-I(U';S)=0$, while (e) uses \eqref{EQ:alt_proof_ineq1}.

Finally, assuming $I(V;S|U) < I(V;Z|U)$ produces:
\begin{align*}
I(V';Y|U')-I(V';Z|U')&\stackrel{(a)}=\epsilon\Big[I(V;Y|U) - I(V;Z|U)\Big]\\
        &=I(V;Y|U)-I(V;Z|U)-\bar{\epsilon}\Big[I(V;Y|U)-I(V;Z|U)\Big]\\
        &\stackrel{(b)}>I(V;Y|U)-I(V;Z|U)-\bar{\epsilon}\Big[I(V;Y|U)-I(V;S|U)\Big]\\
        &\stackrel{(c)}=I(V;Y|U)-I(V;Z|U)+I(U;Y)-I(U;S)\\
        &\stackrel{(d)}\geq R_\mathsf{A}^\mathsf{Alt},\numberthis
\end{align*}
where (a) is \eqref{EQ:alt_proof_justification1}, as before, (b) is by the assumption in the second case, (c) uses \eqref{EQ:alt_proof_ineq1_p} with $I(U';Y)-I(U';S)=0$, and, finally, (d) follows by the third rate bound in $R_\mathsf{A}^\mathsf{Alt}$.

Concluding, we see that 
\begin{equation}
    R_\mathsf{A}\geq R_\mathsf{A}(p_{U',V',X|S})=\min\Big\{I(V';Y|U')-I(V';Z|U'),I(U',V';Y)-I(U',V';S)\Big\}\geq R_\mathsf{A}^\mathsf{Alt},
\end{equation}
which completes the proof.

\section{Proof of Corollary \ref{CORR:LNWTC_capacity}}\label{APPEN:LNWTC_capacity_proof}

\subsection{Direct} We use Theorem \ref{TM:SDWTC_lower_bound} to establish the achievability of Corollary \ref{CORR:LNWTC_capacity}. For any $Q_{U,V,X|S}:\mathcal{S}\to\mathcal{U}\times\mathcal{V}\times\mathcal{X}$, replacing $Y$ and $Z$ in $R_\mathsf{A}\left(Q_{U,V,X|S}\right)$ with $(Y.S_1)$ and $(Z,S_2)$, respectively, gives that
\begin{align}
R^\mathrm{RLN}_\mathsf{A}(Q_{U,V,X|S})=\min\Big\{I(V;Y,S_1|U)-I(V;Z,S_2|U),I&(U,V;Y,S_1)-I(U,V;S)\nonumber\\&,I(U,V;Y,S_1)-I(U;S)-I(V;Z,S_2|U)\Big\}\label{EQ:LNWTC_RA_achievable_prob}
\end{align}
is achievable.

To properly define the choice of $Q_{U,V,X|S}$ that achieves \eqref{EQ:LNWTC_capacity}, recall the $P$ distribution stated after \eqref{EQ:LNWTC_capacity_prob} that factors as $p_SP_{A|S}P_{B|A}P_Xp_{S_1,S_2|S}p_{Y,Z|X}$ and let $\tilde{P}$ be a PMF over $\mathcal{S}\times\mathcal{A}\times\mathcal{B}\times\mathcal{X}\times\mathcal{Y}\times\mathcal{Z}\times\mathcal{S}_1\times\mathcal{S}_2\times\mathcal{B}\times\mathcal{X}$, such that
\begin{equation}
\tilde{P}_{S,A,B,X,S_1,S_2,Y,Z,\tilde{B},\tilde{X}}=P_{S,A,B,X,S_1,S_2,Y,Z}\mathds{1}_{\{\tilde{B}=B\}\cap\{\tilde{X}=X\}}.\label{EQ:LNWTC_tildeP}
\end{equation}

Now, fix $P_{S,A,B,X,S_1,S_2,Y,Z}$ and let $Q_{U,V,X|S}$ in \eqref{EQ:SDWTC_lower_bound_prob} be such that $V=(A,B)_{\tilde{P}}$, $U=(\tilde{B},\tilde{X})_{\tilde{P}}$ and $Q_{X|S,U,V}=\tilde{P}_X=P_X$, where the subscript $\tilde{P}$ means that the random variables on the RHS are distributed according to their marginal from \eqref{EQ:LNWTC_tildeP}. Consequently, $Q_{U,V,X|S}p_{S_1,S_2|S}p_{Y,Z|X}$ is equal to the RHS of \eqref{EQ:LNWTC_tildeP}. We next evaluate the mutual information term in $R_\mathsf{A}$ from \eqref{EQ:SDWTC_lower_bound_prob} and show it coincides with \eqref{EQ:LNWTC_capacity}. In doing so, we once again make use of the notation $I_Q$, $I_{\tilde{P}}$ and $I_P$ to indicated that a mutual information term is taken with respect to the PMF $Q$, $\tilde{P}$ or $P$, respectively. We have
\begin{align*}
    I_Q(V;Y,S_1|U)-I_Q(V;Z,S_2|U)&=I_{\tilde{P}}(A,B;Y,S_1|\tilde{B},\tilde{X})-I_{\tilde{P}}(A,B;Z,S_2|\tilde{B},\tilde{X})\\
    &\begin{multlined}[b][.63\textwidth]\stackrel{(a)}=I_P(A;S_1|B,X)+I_P(A;Y|B,X,S_1)-I_P(A;S_2|B,X)\\-I_P(A;Z|B,X,S_2)\end{multlined}\\
    &\stackrel{(b)}=I_P(A;S_1|B)-I_P(A;S_2|B),\numberthis\label{EQ:LNWTC_achieve_RB1}
\end{align*}
where (a) is because $\tilde{B}=B$ and $\tilde{X}=X$ with probability 1 and since $\tilde{P}_{S,A,B,X,S_1,S_2,Y,Z}=P_{S,A,B,X,S_1,S_2,Y,Z}$. Step (b) is because in $P$ the chain $(Y,Z)-X-(A,B,S_1,S_2)$ is Markov.

Next, consider 
\begin{align*}
    I_Q(U,V;Y,S_1)-I_Q(U,V;S)&=I_{\tilde{P}}(A,B,\tilde{B},\tilde{X};Y,S_1)-I_{\tilde{P}}(A,B,\tilde{B},\tilde{X};S)\\
    &\stackrel{(a)}=I_P(A,B,X;Y,S_1)-I_P(A,B,X;S)\\
    &\stackrel{(b)}=I_P(A,B,X;Y|S_1)-I_P(A,B;S|S_1)\\
    &\stackrel{(c)}=I_P(X;Y)-I_P(A;S|S_1),\numberthis\label{EQ:LNWTC_achieve_RB2}
\end{align*}
where:\\
(a) is for the same reason as step (a) in the derivation of \eqref{EQ:LNWTC_achieve_RB1};\\ 
(b) is because in $P$ we have the Markov chain $(A,B,X)-S-S_1$, since $X$ is independent of $(A,B,S,S_1)$ and due to the chain rule;\\
(c) follows because $(X,Y)$ is independent of $(A,B,S_1)$ and since  $I(B;S|S_1,A)=0$ as $B-A-(S,S_1)$ is also a Markov chain.

Finally, we shown that the third term from the RHS of \eqref{EQ:LNWTC_RA_achievable_prob} is redundant by establishing that $I_Q(V;S|U)\geq I_Q(V;Z,S_2|U)$ for the aforementioned choice of $Q_{U,V,X|S}$. Consider
\begin{align*}
I_Q(V;Z,S_2|U)&\stackrel{(a)}=I_P(A;S_2|B)\\
&\leq I_P(A;S,S_2|B)\\
&\stackrel{(b)}= I_P(A,B;S)-I(B;S)\\
&\stackrel{(c)}= I_P(A;S|B,X)\\
&\stackrel{(d)}= I_Q(A;S|B,X),\numberthis\label{EQ:LNWTC_achieve_RB3}
\end{align*}
where:\\
(a) is due to similar arguments as those justifying \eqref{EQ:LNWTC_achieve_RB1};\\
(b) is because $(A,B)-S-S_2$ forms a Markov chain in $P$;\\
(c) is by the independence of $(A,B,S)$ and $X$;\\
(d) follows from the definition of the $Q_{U,V,X|S}$ distribution.

Consequently, the third term in $R^\mathrm{RLN}_\mathsf{A}(Q_{U,V,X|S})$ is redundant due to \eqref{EQ:LNWTC_achieve_RB2}. Along with \eqref{EQ:LNWTC_achieve_RB1}, this  establishes the direct part of Corollary \ref{CORR:LNWTC_capacity}.

\subsection{Converse}\label{SUBSEC:LNWTC_capacity_proof_converse}

Let $\big\{c_{n}\big\}_{n\in\mathbb{N}}$ be a sequence of $(n,R)$ semantically-secure codes for the SD-WTC with a vanishing maximal error probability. Fix $\epsilon>0$ and let $n\in\mathbb{N}$ be sufficiently large so that \eqref{EQ:SDWTC_achievability} is satisfied. Since both \eqref{EQ:SDWTC_achievability_reliability} and \eqref{EQ:SDWTC_achievability_security} hold for any message distribution $P_M\in\mathcal{P}(\mathcal{M})$, in particular, they hold for a uniform $P^{(U)}_M$. All the following multi-letter mutual information and entropy terms are calculated with respect to the induced joint PMF from \eqref{EQ:SDWTC_induced_PMF}, where the channel $p_{Y,Z|X,S}$ is replaced with $p_{S_1,S_2,Y,Z|X,S}$ defined in Section \ref{SUBSUBSEC:LNWTC}. Fano's inequality gives
\begin{equation}
H(M|S_1^n,Y^n)\leq 1+n\epsilon R\triangleq n\epsilon_n,\label{EQ:LNWTC_converse_Fano}
\end{equation}
where $\epsilon_n=\frac{1}{n}+\epsilon R$.

The security criterion from \eqref{EQ:SDWTC_achievability_security} and the reversely less noisy property of the channel $p_{Y,Z|X}$ (that, respectively, justify the two following inequalities) further gives
\begin{align*}
\epsilon&\geq I(M;S_2^n,Z^n)\\
        &= I(M;S_2^n)+\sum_{\mathbf{s}_2\in\mathcal{S}_2^n}W^n_{S_2}(\mathbf{s}_2)I(M;Z^n|S_2^n=\mathbf{s}_2)\\
        &\geq I(M;S_2^n)+\sum_{\mathbf{s}_2\in\mathcal{S}_2^n}W^n_{S_2}(\mathbf{s}_2)I(M;Y^n|S_2^n=\mathbf{s}_2)\\
        &= I(M;S_2^n,Y^n).\numberthis\label{EQ:LNWTC_converse_security}
\end{align*}

Having \eqref{EQ:LNWTC_converse_Fano} and \eqref{EQ:LNWTC_converse_security}, we bound $R$ as
\begin{align*}
nR&=H(M)\\
  &\stackrel{(a)}\leq I(M;S_1^n,Y^n)-I(M;S_2^n,Y^n)+n\delta_n\\
  &= I(M;S_1^n|Y^n)-I(M;S_2^n|Y^n)+n\delta_n\\
  &\stackrel{(b)}=\sum_{i=1}^n\Big[I(M;S_1^i,S_{2,i+1}^n|Y^n)-I(M;S_1^{i-1},S_{2,i}^n|Y^n)\Big]+n\delta_n\\
  &=\sum_{i=1}^n\Big[ I(M;S_{1,i}|S_1^{i-1},S^n_{2,i+1},Y^n)-I(M;S_{2,i}|S_1^{i-1},S^n_{2,i+1},Y^n)\Big]+n\delta_n\\ 
  &\stackrel{(c)}=\sum_{i=1}^n\Big[I(M;S_{1,i}|B_i)-I(M;S_{2,i}|B_i)\Big]+n\delta_n\\
  &\stackrel{(d)}=n\sum_{i=1}^nP_T(i)\Big[I(M;S_{1,T}|B_T,T=i)-I(M;S_{2,T}|B_T,T=i)\Big]+n\delta_n\\
  &= n\Big[I(M;S_{1,T}|B_T,T)-I(M;S_{2,T}|B_T,T)\Big]+n\delta_n\\
  &\stackrel{(e)}= n\Big[I(A;S_1|B)-I(A;S_2|B)\Big]+n\delta_n,\numberthis\label{EQ:LNSWTC_converse_UB1}
\end{align*}
where:\\
(a) is by \eqref{EQ:LNWTC_converse_Fano} and \eqref{EQ:LNWTC_converse_security} while setting $\delta_n\triangleq\epsilon_n+\frac{\epsilon}{n}$;\\
(b) is a telescoping identity \cite[Eqs. (9) and (11)]{Kramer_telescopic2011};\\
(c) defines $B_i\triangleq(S_1^{i-1},S^n_{2,i+1},Y^n)$, for all $i\in[1:n]$;\\
(d) is by introducing a time-sharing random variable $T$ that is uniformly distributed over the set $[1:n]$ and is independent of all the other random variables in $P^{(c_n)}$;\\
(e) defines $S\triangleq S_T$, $S_1\triangleq S_{1,T}$, $S_2\triangleq S_{2,T}$, $X\triangleq X_T$, $Y\triangleq Y_T$,  $Z\triangleq Z_T$, $B\triangleq(B_T,T)$ and $A\triangleq(M,B)$.

Another way to bound $R$ is
\begin{align*}
nR&=H(M)\\
  &\stackrel{(a)}\leq I(M;S_1^n,Y^n)+n\epsilon_n\\
  &=I(M;S_1^n,Y^n,S^n)-I(M;S^n|S_1^n,Y^n)+n\epsilon_n\\
  &\stackrel{(b)}=I(M;Y^n|S_1^n,S^n)-I(M,Y^n;S^n|S_1^n)+I(S^n;Y^n|S_1^n)+n\epsilon_n\\
  &=I(M,S^n;Y^n|S_1^n)-I(M,Y^n;S^n|S_1^n)+n\epsilon_n\\
  &\stackrel{(c)}\leq I(M,S^n;Y^n)-I(M,Y^n;S^n|S_1^n)+n\epsilon_n\\
  &\stackrel{(d)}\leq I(X^n;Y^n)-I(M,Y^n;S^n|S_1^n)+n\epsilon_n\\
  &\stackrel{(e)}\leq \sum_{i=1}^n\Big[I(X_i;Y_i)-I(M,Y^n;S_i|S_1^n,S^{i-1})\Big]+n\epsilon_n\\
  &\stackrel{(f)}\leq \sum_{i=1}^n\Big[I(X_i;Y_i)-I(M,Y^n,S_1^{n\backslash i},S^{i-1};S_i|S_{1,i})\Big]+n\epsilon_n\\
  &\stackrel{(g)}\leq \sum_{i=1}^n\Big[I(X_i;Y_i)-I(M,B_i;S_i|S_{1,i})\Big]+n\epsilon_n\\
  &\stackrel{(h)}= n\sum_{i=1}^nP_T(i)\Big[I(X_T;Y_T|T=i)-I(M,B_T;S_T|S_{1,T},T=i)\Big]+n\epsilon_n\\
  &\stackrel{(i)}\leq n\Big[I(X_T;Y_T)-I(M,B_T,T;S_T|S_{1,T})\Big]+n\epsilon_n\\
  &\stackrel{(j)}\leq n\Big[I(X;Y)-I(A;S|S_1)\Big]+n\epsilon_n,\numberthis\label{EQ:LNSWTC_converse_UB2}
\end{align*}
where:\\
(a) is by \eqref{EQ:LNWTC_converse_Fano};\\
(b) uses the independence of $M$ and $(S_1^n,S^n)$ (1st term);\\
(c) is because conditioning cannot increase entropy and since $Y^n-(M,S^n)-S_1^n$ forms a Markov chain (1st term);\\
(d) uses the Markov relation $Y^n-X^n-(M,S^n)$;\\
(e) follows since conditioning cannot increase entropy and by the discrete and memoryless property of the WTC $p^n_{Y,Z|X}$;\\
(f) is because $P^{(c_n)}_{S^n,S_1^n,S_2^n}=p^n_{S,S_1,S_2}$, i.e., the marginal distribution of $(S^n,S_1^n,S_2^n)$ are i.i.d.;\\
(g) is by the definition of $B_i$;\\
(h) follows for the same reason as step (d) in the derivation of \eqref{EQ:LNSWTC_converse_UB1};\\
(i) is because conditioning cannot increase entropy and the Markov relation $Y_T-X_T-T$ (1st term), and because $\mathbb{P}\big(S_T=s,S_{1,T}=s_1,T=t\big)=p_{S,S_1}(s,s_1)P_T(t)$, for all $(s,s_1,t)\in\mathcal{S}\times\mathcal{S}_1\times [1:n]$ (2nd term);\\
(j) reuses the definition of the single-letter random variable from step (e) in the derivation of \eqref{EQ:LNSWTC_converse_UB1}.

The joint distribution of the defined random variables factors as
\begin{align*}
\mathbb{P}\big(S=s,S_1&=s_1,S_2=s_2,A=a,B=b,X=x,Y=y,Z=z\big)\\
&\begin{multlined}[b][.8\columnwidth]=p_S(s)p_{S_1,S_2|S}(s_1,s_2|s)\mathbb{P}\big(A=a\big|S=s,S_1=s_1,S_2=s_2\big)\mathbb{P}\big(B=b\big|A=a\big)
\\\times\mathbb{P}\big(X=x\big|S=s,S_1=s_1,S_2=s_2,A=a,B=b\big)p_{Y,Z|X}(y,z|x)\end{multlined},\numberthis
\end{align*}
where the equalities $\mathbb{P}\big(S=s,S_1=s_1,S_2=s_2\big)=p_S(s)p_{S_1,S_2|S}(s_1,s_2|s)$ and $\mathbb{P}\big(Y=y,Z=z\big|S=s,S_1=s_1,S_2=s_2,A=a,B=b,X=x\big)=p_{Y,Z|X}(y,z|x)$ are straightforward from the probabilistic relations in $P^{(c_n)}$ and the definition of the random variable $T$, while $\mathbb{P}\big(B=b\big|S=s,S_1=s_1,S_2=s_2,A=a\big)=\mathbb{P}\big(B=b\big|A=a\big)$ follows because $A=(M,B)$. Furthermore, for every $(s,s_1,s_2,a)\in\mathcal{S}\times\mathcal{S}_1\times\mathcal{S}_2\times\mathcal{A}$, it holds that $\mathbb{P}\big(A=a\big|S=s,S_1=s_1,S_2=s_2\big)=\mathbb{P}\big(A=a\big|S=s\big)$. To see this, for any $(s^n,s_1^n,s_2^n,y^n)\in\mathcal{S}^n\times\mathcal{S}_1^n\times\mathcal{S}_2^n\times\mathcal{Y}^n$, we define the corresponding realization of $A$ as $a=(t,m,b_t)$, where $(t,m)\in[1:n]\in\mathcal{M}_n$ and $b_t=\big(y^n,s_1^{t-1},s_{2,t+1}^n\big)$. For any $(s_t,s_{1,t},s_{2,t})\in\mathcal{S}\times\mathcal{S}_1\times\mathcal{S}_2$, we have
\begin{align*}
    \mathbb{P}\big(&A=a\big|S=s_t,S_1=s_{1,t},S_2=s_{2,t}\big)\\
    &\stackrel{(a)}=P_T(t)P^{(c_n)}\left(m,s_1^{t-1},s_{2,t+1}^n,y^n|s_t,s_{1,t},s_{2,t}\right)\\
    &=P_T(t)\sum_{(s^{n\backslash t},x^n)\in\mathcal{S}^{n-1}\times\mathcal{X}^n}P^{(c_n)}\left(s^{n\backslash t},x^n,m,s_1^{t-1},s_{2,t+1}^n,y^n\Big|s_t,s_1^t,s_{2,t}^n,m\right)\\
    &\begin{multlined}[b][.85\columnwidth]\stackrel{(b)}=P_T(t)P_M(m)\sum_{(s^{n\backslash t},x^n)\in\mathcal{S}^{n-1}\times\mathcal{X}^n}p_S^{n-1}\left(s^{n\backslash t}\right)p_{S_1|S}^{t-1}\left(s_1^{t-1}|s^{t-1}\right)p_{S_2|S}^{n-t}\left(s_{2,t+1}^n|s_{t+1}^n\right)\\
    \times f_n\left(x^n\big|m,s^n\right)p^n_{Y|X}\left(y^n|x^n\right)\end{multlined}\\
    &=P_T(t)P^{(c_n)}\left(m,s_1^{t-1},s_{2,t+1}^n,y^n|s_t\right)\\
    &=\mathbb{P}\big(A=a\big|S=s_t\big),\numberthis\label{EQ:LNSWTC_converse_markov}
\end{align*}
where (a) is because $T$ is independent of all the other random variables, while (b) uses the dependence relations in $P^{(c_n)}$ from \eqref{EQ:SDWTC_induced_PMF} with $p_{S_1,S_2|S}p_{Y,Z|X}$ in the role of the SDWTC.

Denoting $\mathbb{P}\big(A=a\big|S=s\big)\triangleq P_{A|S}(a|s)$, $\mathbb{P}\big(B=b\big|A=a\big)\triangleq P_{B|A}(b|a)$ and $\mathbb{P}\big(X=x\big|S=s,S_1=s_1,S_2=s_2,A=a,B=b\big)\triangleq P_{X|S,S_1,S_2,A,B}(x|s,s_1,s_2,a,b)$, we have the following bound on the achievable rate:
\begin{equation}
    R\leq \frac{\min\Big\{I(A;S_1|B)-I(A;S_2|B),I(X;Y)-I(A;S|S_1)\Big\}}{1-\epsilon}+\frac{1}{(1-\epsilon)n}+\frac{\epsilon}{1-\epsilon},\label{EQ:LNSWTC_converse_RB_final}
\end{equation}
where the mutual information terms are calculated with respect to the joint PMF $p_Sp_{S_1,S_2|S}P_{A|S}P_{B|A}P_{X|S,S_1,S_2,A,B}p_{Y,Z|X}$. However, noting that in none of the mutual information terms from \eqref{EQ:LNSWTC_converse_RB_final} do $X$ and $(S,S_1,S_2,A,B)$ appear together, we may replace $P_{X|S,S_1,S_2,A,B}$ with $P_X$ without affecting the expressions. Taking $\epsilon\to 0$ and $n\to\infty$ completes the proof of the converse.

%%%%%%%%%%%%%%%%%%%%%%%%%%%%%%%%%%%%%%%%%%%%%%%%%%%%%%%%%%%%%%%%%%%%%%%%%%%%%%%%%%%%%%%%%%%%%%%%%%%%%%%%%%%%%%%%%%%
%%%%%%%%%%%%%%%%%%%%%               APPENDIX C - SEMI-DETRMINISTIC CONVERSE               %%%%%%%%%%%%%%%%%%%%%%%%%%
%%%%%%%%%%%%%%%%%%%%%%%%%%%%%%%%%%%%%%%%%%%%%%%%%%%%%%%%%%%%%%%%%%%%%%%%%%%%%%%%%%%%%%%%%%%%%%%%%%%%%%%%%%%%%%%%%%%
	
\section{Converse Proof for Corollary \ref{CORR:Semi_SDWTC_capacity}}\label{APPEN:Semi_SDWTC_capacity_proof}

Let $\big\{c_{n}\big\}_{n\in\mathbb{N}}$ be a sequence of $(n,R)$ for the SD-WTC satisfying \eqref{EQ:SDWTC_achievability}. By similar arguments to those presented in the converse proof from Appendix \ref{SUBSEC:LNWTC_capacity_proof_converse}, we assume a uniform message distribution and note that all the following multi-letter mutual information and entropy terms are taken with respect to \eqref{EQ:SDWTC_induced_PMF}. By Fano's inequality, we have
\begin{equation}
H(M|Y^n)\leq 1+n\epsilon R\triangleq n\epsilon_n,\label{EQ:semi_det_converse_Fano}
\end{equation}
where $\epsilon_n=\frac{1}{n}+\epsilon R$.

\par First, we bound the rate $R$ as
\begin{align*}
nR&=H(M)\\
    &\stackrel{(a)}\leq I(M;Y^n)-I(M;Z^n)+n\epsilon'_n\\
    &\leq I(M;Y^n|Z^n)+n\epsilon'_n\\
    &\stackrel{(b)}\leq\sum_{i=1}^nH(Y_i|Z_i)+n\epsilon'_n,\numberthis\label{EQ:semi_det_converse_UB1}
\end{align*}
where (a) uses \eqref{EQ:SDWTC_achievability_security} and \eqref{EQ:semi_det_converse_Fano} and defines $\epsilon'_n\triangleq\epsilon_n+\frac{\epsilon}{n}$, and (b) follows by the chain rule and since conditioning cannot increase entropy.

Another way to bound $R$ is as follows:
\begin{align*}
    nR&=H(M)\\
    &\stackrel{(a)}\leq I(M;Y^n)-I(M;S^n)+n\epsilon_n\\
    &\leq I(M;Y^n|S^n)+n\epsilon'_n\\
    &\stackrel{(b)}\leq\sum_{i=1}^nH(Y_i|S_i)+n\epsilon_n,\numberthis\label{EQ:semi_det_converse_UB2}
\end{align*}
where (a) is due to \eqref{EQ:semi_det_converse_Fano} and because $M$ and $S^n$ are independent in \eqref{EQ:SDWTC_induced_PMF}, while (b) is justified similarly to step (b) in \eqref{EQ:semi_det_converse_UB1}. Having \eqref{EQ:semi_det_converse_UB1}-\eqref{EQ:semi_det_converse_UB2}, the converse is established by standard time-sharing argument (as in the proof of Corollary \ref{CORR:LNWTC_capacity} from Appendix \ref{APPEN:LNWTC_capacity_proof}).

\section{Proof of Lemma \ref{LEMMA:good_approximation}}\label{APPEN:good_approximation_proof}

First note that for any $\mathcal{C}_n\in\mathfrak{C}_n$ and $(i,j,m,\mathbf{s})\in\mathcal{I}_n\times\mathcal{J}_n\times\mathcal{M}_n\times\mathcal{S}^n$, we have
\begin{align*}
Q^{(\mathcal{C}_n)}(i,j|m,\mathbf{s})&=\frac{Q^{(\mathcal{C}_n)}(m,i,j,\mathbf{s})}{Q^{(\mathcal{C}_n)}(m,\mathbf{s})}\\                                                 &=\frac{\sum\limits_{(\mathbf{u},\mathbf{v})\in\mathcal{U}^n\times\mathcal{V}^n}\frac{1}{|\mathcal{M}_n||\mathcal{I}_n||\mathcal{J}_n|}\mathds{1}_{\big\{\mathbf{u}(i)=\mathbf{u}\big\}\cap\big\{\mathbf{v}(i,j,m)=\mathbf{v}\big\}}p^n_{S|U,V}(\mathbf{s}|\mathbf{u},\mathbf{v})}{\sum\limits_{\substack{(i',j',\mathbf{u}',\mathbf{v}')\\\in\mathcal{I}_n\times\mathcal{J}_n\times\mathcal{U}^n\times\mathcal{V}^n}}\frac{1}{|\mathcal{M}_n||\mathcal{I}_n||\mathcal{J}_n|}\mathds{1}_{\big\{\mathbf{u}(i')=\mathbf{u}'\big\}\cap\big\{\mathbf{v}(i',j',m)=\mathbf{v}'\big\}}p^n_{S|U,V}(\mathbf{s}|\mathbf{u}',\mathbf{v}')}\\
&=\frac{p^n_{S|U,V}\big(\mathbf{s}\big|\mathbf{u}(i),\mathbf{v}(i,j,m)\big)}{\sum\limits_{(i',j')\in\mathcal{I}_n\times\mathcal{J}_n}p^n_{S|U,V}\big(\mathbf{s}\big|\mathbf{u}(i'),\mathbf{v}(i',j',m)\big)}\\
&\stackrel{(a)}=P^{(\mathcal{C}_n)}(i,j|m,\mathbf{s}),\numberthis\label{EQ:approx_lemma_proof_equality1}
\end{align*}
where (a) is by the definition from \eqref{EQ:main_proof_likelihood_enc}. Having \eqref{EQ:approx_lemma_proof_equality1}, note that
\begin{align*}
    \Big|\Big|P^{(\mathcal{C}_n)}-Q^{(\mathcal{C}_n)}\Big|\Big|_{\mathsf{TV}}&\stackrel{(a)}=\sum_{m\in\mathcal{M}_n}\frac{1}{|\mathcal{M}_n|}\Big|\Big|P^{(\mathcal{C}_n)}_{\mathbf{S},I,J,\mathbf{U},\mathbf{V},\mathbf{X},\mathbf{Y},\mathbf{Z}|M=m}-Q^{(\mathcal{C}_n)}_{\mathbf{S},I,J,\mathbf{U},\mathbf{V},\mathbf{X},\mathbf{Y},\mathbf{Z}|M=m}\Big|\Big|_{\mathsf{TV}}\\
    &\stackrel{(b)}=\frac{1}{|\mathcal{M}_n|}\sum_{m\in\mathcal{M}_n}\Big|\Big|p_S^n-Q^{(\mathcal{C}_n)}_{\mathbf{S}|M=m}\Big|\Big|_{\mathsf{TV}},\numberthis\label{EQ:approx_lemma_proof_TV1}%\\
    %&\leq\max_{m\in\mathcal{M}_n}\Big|\Big|p_S^n-Q^{(\mathcal{C}_n)}_{\mathbf{S}|M=m}\Big|\Big|_{\mathsf{TV}}
\end{align*}
where (a) is because $Q^{(\mathcal{C}_n)}_M=P^{(\mathcal{C}_n)}_M=p^{(U)}_{\mathcal{M}_n}$, while (b) is based on the property of total variation that for any $p_X,q_X\in\mathcal{P}(\mathcal{X})$ and $p_{Y|X}:\mathcal{X}\to\mathcal{P}(\mathcal{Y})$ we have
$\big|\big|p_Xp_{Y|X}-q_Xp_{Y|X}\big|\big|_{\mathsf{TV}}=\big|\big|p_X-q_X\big|\big|_{\mathsf{TV}}$. Combining this with \eqref{EQ:approx_lemma_proof_equality1} and the relations 
\begin{subequations}
\begin{align}
&Q^{(\mathcal{C}_n)}_{\mathbf{U},\mathbf{V}|I,J,\mathbf{S},M=m}=\mathds{1}_{\big\{\mathbf{U}=\mathbf{u}(I)\big\}\cap\big\{\mathbf{V}=\mathbf{v}(I,J,m)\big\}}=P^{(\mathcal{C}_n)}_{\mathbf{U},\mathbf{V}|I,J,\mathbf{S},M=m}\label{EQ:good_approx_proof_Gamma_P_relations1}\\
&Q^{(\mathcal{C}_n)}_{\mathbf{X},\mathbf{Y},\mathbf{Z}|\mathbf{U},\mathbf{V},I,J,\mathbf{S},M=m}=p^n_{X|U,V,S}p^n_{Y,Z|X,S}=P^{(\mathcal{C}_n)}_{\mathbf{X},\mathbf{Y},\mathbf{Z}|\mathbf{U},\mathbf{V},I,J,\mathbf{S},M=m}\label{EQ:good_approx_proof_Gamma_P_relations2}
\end{align}\label{EQ:good_approx_proof_Gamma_P_relations}
\end{subequations}
justifies (b).

Now, consider%for any $\tilde{\alpha}>0$ and sufficiently large $n$ consider
\begin{align*}
\mathbb{E}_\mu\Big|\Big|P^{(\mathsf{C}_n)}-Q^{(\mathsf{C}_n)}\Big|\Big|_{\mathsf{TV}}&\stackrel{(a)}\leq\mathbb{E}_\mu \frac{1}{|\mathcal{M}_n|}\sum_{m\in\mathcal{M}_n}\Big|\Big|p_S^n-Q^{(\mathcal{C}_n)}_{\mathbf{S}|M=m}\Big|\Big|_{\mathsf{TV}}\\
&\stackrel{(b)}=\mathbb{E}_\mu \Big|\Big|p_S^n-Q^{(\mathcal{C}_n)}_{\mathbf{S}|M=1}\Big|\Big|_{\mathsf{TV}}\\
&\stackrel{(c)}\leq\mathbb{E}_\mu \sqrt{\frac{1}{2}\mathsf{D}\Big(Q^{(\mathcal{C}_n)}_{\mathbf{S}|M=1}\Big|\Big|p_S^n\Big)}\\
&\stackrel{(c)}\leq \sqrt{\frac{1}{2}\mathbb{E}_\mu\mathsf{D}\Big(Q^{(\mathcal{C}_n)}_{\mathbf{S}|M=1}\Big|\Big|p_S^n\Big)},\numberthis\label{EQ:good_approx_proof_TV_UB}%\\
%(\max_{m\in\mathcal{M}_n}\Big|\Big|p_S^n-Q^{(\mathsf{C}_n)}_{\mathbf{S}|M=m}\Big|\Big|_{\mathsf{TV}}> e^{-n\tilde{\alpha}}\bigg)\\
%&\stackrel{(b)}\leq\mathbb{P}\bigg(\max_{m\in\mathcal{M}_n}\mathsf{D}\Big(Q^{(\mathsf{C}_n)}_{\mathbf{S}|M=m}\Big|\Big|p_S^n\Big)> 2e^{-2n\tilde{\alpha}}\bigg)\\
%&\stackrel{(c)}\leq\sum_{m\in\mathcal{M}_n}\mathbb{P}\bigg(\mathsf{D}\Big(Q^{(\mathsf{C}_n)}_{\mathbf{S}|M=m}\Big|\Big|p_S^n\Big)> e^{-2n\tilde{\alpha}}\bigg),\numberthis\label{EQ:good_approx_proof_TV_UB}
\end{align*}
where (a) is due to \eqref{EQ:approx_lemma_proof_TV1}, (b) follws by symmetry of the codebook generation with respect to $m\in\mathcal{M}_n$, while (c) follows by Pinsker's Inequality (see \eqref{EQ:Pinsker_Inequality}), and (d) is Jensen's inequality.

To conclude the proof, note that the expected value inside the square root on the RHS of \eqref{EQ:good_approx_proof_TV_UB} falls within the framework of the SCL for superposition codes (Part 1 of Lemma \ref{LEMMA:soft_covering_stronger}), with respect to the DMC $p^n_{S|U,V}$. Therefore, taking $(R_1,R_2)$ as in \eqref{EQ:main_proof_approx_rate_bounds} implies that there exist $\tilde{\alpha}>0$ such that for any $n$ large enough
\begin{equation}
\mathbb{E}_\mu\mathsf{D}\Big(Q^{(\mathsf{C}_n)}_{\mathbf{S}|M=m}\Big|\Big|p_S^n\Big)\leq e^{-n\tilde{\alpha}}.\label{EQ:good_approx_proof_SCL_application}
\end{equation}
Combining this with \eqref{EQ:good_approx_proof_TV_UB} proves Lemma \ref{LEMMA:good_approximation} with $\alpha=\frac{\tilde{\alpha}}{2}$.
%The stronger result from Lemma \ref{LEMMA:good_approximation} (i.e., \eqref{EQ:main_proof_approx_soft_covering}) then follows from \eqref{EQ:good_approx_proof_TV_UB} and \eqref{EQ:good_approx_proof_SCL_application} for $\alpha_1=\frac{\gamma_1}{2}$ and $\alpha_2=\gamma_2$. To get \eqref{EQ:main_proof_approx_soft_covering_expect}, we use \cite[Lemma 2]{Goldfeld_WTCII_semantic2015}, where it is stated that the stronger version of the SCL indeed implies Wyner's original notion of soft-covering where the convergence is of the expected value. 

%%%%%%%%%%%%%%%%%%%%%%%%%%%%%%%%%%%%%%%%%%%%%%%%%%%%%%%%%%%%%%%%%%%%%%%%%%%%%%%%%%%%%%%%%%%%%%%%%%%%%%%%%%%%%%%%%%%
%%%%%%%%%%%%%%%%%%%%%                 APPENDIX H - SS UNDER P VS GAMMA                   %%%%%%%%%%%%%%%%%%%%%%%%%%
%%%%%%%%%%%%%%%%%%%%%%%%%%%%%%%%%%%%%%%%%%%%%%%%%%%%%%%%%%%%%%%%%%%%%%%%%%%%%%%%%%%%%%%%%%%%%%%%%%%%%%%%%%%%%%%%%%%
	
\section{Proof of Lemma \ref{LEMMA:ss_p_gamma}}\label{APPEN:ss_p_gamma_proof}

To simplify notation, throughout his proof we abbreviate $I_{P^{(\mathcal{C}_n)}}$ and $I_{Q^{(\mathcal{C}_n)}}$ as $I_P$ and $I_Q$, respectively. Consider the following:
\begin{align*}
    \Big|I_P(M;\mathbf{Z})-I_Q(M;\mathbf{Z})\Big|&=\Big|H_P(M)+H_P(\mathbf{Z})-H_P(M,\mathbf{Z})-H_Q(M)-H_Q(\mathbf{Z})+H_Q(M,\mathbf{Z})\Big|\\
    &\stackrel{(a)}\leq\Big|H_P(\mathbf{Z})-H_Q(\mathbf{Z})\Big|+\Big|H_Q(M,\mathbf{Z})-H_P(M,\mathbf{Z})\Big|\\
    &\stackrel{(b)}\leq\Big|\Big|P^{(\mathcal{C}_n)}_\mathbf{Z}-Q^{(\mathcal{C}_n)}_\mathbf{Z}\Big|\Big|_{\mathsf{TV}}\log\frac{|\mathcal{Z}^n|}{\Big|\Big|P^{(\mathcal{C}_n)}_\mathbf{Z}-Q^{(\mathcal{C}_n)}_\mathbf{Z}\Big|\Big|_{\mathsf{TV}}}\\&\mspace{150mu}+\Big|\Big|P^{(\mathcal{C}_n)}_{M,\mathbf{Z}}-Q^{(\mathcal{C}_n)}_{M,\mathbf{Z}}\Big|\Big|_{\mathsf{TV}}\log\frac{|\mathcal{M}_n|\cdot|\mathcal{Z}^n|}{\Big|\Big|P^{(\mathcal{C}_n)}_{M,\mathbf{Z}}-Q^{(\mathcal{C}_n)}_{M,\mathbf{Z}}\Big|\Big|_{\mathsf{TV}}}\\
    &\stackrel{(c)}\leq e^{-n\beta_1}\Big(n\log|\mathcal{Z}|+n\log\left(2^R|\mathcal{Z}|\right)\Big)-\Big|\Big|P^{(\mathcal{C}_n)}_\mathbf{Z}-Q^{(\mathcal{C}_n)}_\mathbf{Z}\Big|\Big|_{\mathsf{TV}}\log\Big|\Big|P^{(\mathcal{C}_n)}_\mathbf{Z}-Q^{(\mathcal{C}_n)}_\mathbf{Z}\Big|\Big|_{\mathsf{TV}}\\&\mspace{220mu}-\Big|\Big|P^{(\mathcal{C}_n)}_{M,\mathbf{Z}}-Q^{(\mathcal{C}_n)}_{M,\mathbf{Z}}\Big|\Big|_{\mathsf{TV}}\log\Big|\Big|P^{(\mathcal{C}_n)}_{M,\mathbf{Z}}-Q^{(\mathcal{C}_n)}_{M,\mathbf{Z}}\Big|\Big|_{\mathsf{TV}},\numberthis\label{EQ:ss_p_gamma_proof_UB1}
\end{align*}
where (a) is because $H_P(M)=H_Q(M)$ and due to the triangle inequality, (b) uses \cite[Theorem 17.3.3]{CovThom06}, while (c) follows by the hypothesis in \eqref{EQ:lemma_ss_if}.

The function $x\mapsto -x\log x$ is monotone increasing for $x\in\left[0,2^{-\frac{1}{\ln2}}\right]$ and, for large enough values of $n$, we have $e^{-n\beta_1}\in\left[0,2^{-\frac{1}{\ln2}}\right]$. Therefore, as $\Big|\Big|P^{(\mathcal{C}_n)}_\mathbf{Z}-Q^{(\mathcal{C}_n)}_\mathbf{Z}\Big|\Big|_{\mathsf{TV}}\leq \Big|\Big|P^{(\mathcal{C}_n)}_{M,\mathbf{Z}}-Q^{(\mathcal{C}_n)}_{M,\mathbf{Z}}\Big|\Big|_{\mathsf{TV}}\leq e^{-n\beta_1}$, we have that for sufficiently large $n$
\begin{equation}
-\Big|\Big|P^{(\mathcal{C}_n)}_\mathbf{Z}\mspace{-5mu}-Q^{(\mathcal{C}_n)}_\mathbf{Z}\Big|\Big|_{\mathsf{TV}}\mspace{-8mu}\log\Big|\Big|P^{(\mathcal{C}_n)}_\mathbf{Z}\mspace{-5mu}-Q^{(\mathcal{C}_n)}_\mathbf{Z}\Big|\Big|_{\mathsf{TV}}\mspace{-8mu}-\Big|\Big|P^{(\mathcal{C}_n)}_{M,\mathbf{Z}}\mspace{-5mu}-Q^{(\mathcal{C}_n)}_{M,\mathbf{Z}}\Big|\Big|_{\mathsf{TV}}\mspace{-8mu}\log\Big|\Big|P^{(\mathcal{C}_n)}_{M,\mathbf{Z}}\mspace{-5mu}-Q^{(\mathcal{C}_n)}_{M,\mathbf{Z}}\Big|\Big|_{\mathsf{TV}}\leq -2e^{-n\beta_1}\log e^{-n\beta_1}.\label{EQ:ss_p_gamma_proof_UB2}
\end{equation}
Inserting \eqref{EQ:ss_p_gamma_proof_UB2} into \eqref{EQ:ss_p_gamma_proof_UB1} gives
\begin{equation}
    \Big|I_P(M;\mathbf{Z})-I_Q(M;\mathbf{Z})\Big|\leq ne^{-n\beta_1}\left(2\log|\mathcal{Z}|+R+2\beta_1\frac{1}{\ln 2}\right),\label{EQ:ss_p_gamma_proof_UB_final}
\end{equation}
for the aforementioned values of $n$. This implies that \eqref{EQ:lemma_ss_then} holds and concludes the proof of Lemma \ref{LEMMA:ss_p_gamma}.

\newpage
\bibliographystyle{unsrt}
\bibliographystyle{IEEEtran}
\bibliography{ref}

\end{document}